\documentclass[prb,aps,nofootinbib,amssymb,twocolumn,superscriptaddress,10pt]{revtex4-2}
\usepackage{amsmath}
\usepackage{amssymb}
\usepackage{amsthm}
\usepackage{amsfonts}
\usepackage{enumerate}
\usepackage{latexsym}
\usepackage{color}
\usepackage{setspace} 
\usepackage{blindtext}
\usepackage{dsfont}
\usepackage{mathrsfs}
\usepackage[normalem]{ulem}

\usepackage{bm}
\usepackage{graphicx}
\usepackage{subfigure}

\usepackage{hyperref}

\newcommand{\beginsupplement}{
        \setcounter{table}{0}
        \renewcommand{\thetable}{S\arabic{table}}
        \setcounter{figure}{0}
        \renewcommand{\thefigure}{S\arabic{figure}}
        \setcounter{equation}{0}
        \renewcommand{\theequation}{S\arabic{equation}}
        \setcounter{section}{0}
        \renewcommand{\thesection}{Appendix \Alph{section}}
        \setcounter{subsection}{0}
        \renewcommand{\thesubsection}{\arabic{subsection}}
}

\begin{document}

\tolerance 10000

\newcommand{\vk}{\mathbf{k}}
\newcommand{\vp}{\mathbf{p}}
\newcommand{\vq}{\mathbf{q}}

\newcommand{\hc}{\hat{c}}
\newcommand{\hb}{\hat{b}}
\newcommand{\hd}{\hat{d}}
\newcommand{\ha}{\hat{a}}
\newcommand{\hf}{\hat{f}}

\title{TSTG II: Projected Hartree-Fock Study of Twisted Symmetric Trilayer Graphene}

\author{Fang Xie}
\affiliation{Department of Physics, Princeton University, Princeton, New Jersey 08544, USA}
\author{Nicolas Regnault}
\affiliation{Department of Physics, Princeton University, Princeton, New Jersey 08544, USA}
\affiliation{Laboratoire de Physique de l'Ecole normale superieure, ENS, Universit\'e PSL, CNRS, Sorbonne Universit\'e, Universit\'e Paris-Diderot, Sorbonne Paris Cit\'e, 75005 Paris, France}
\author{Dumitru C\u{a}lug\u{a}ru}
\affiliation{Department of Physics, Princeton University, Princeton, New Jersey 08544, USA}
\author{B. Andrei Bernevig}
\affiliation{Department of Physics, Princeton University, Princeton, New Jersey 08544, USA}
\affiliation{Donostia International Physics Center, P. Manuel de Lardizabal 4, 20018 Donostia-San Sebastian, Spain}
\affiliation{IKERBASQUE, Basque Foundation for Science, Bilbao, Spain}
\author{Biao Lian}
\affiliation{Department of Physics, Princeton University, Princeton, New Jersey 08544, USA}

\date{\today}

\begin{abstract}
The Hamiltonian of the magic-angle twisted symmetric trilayer graphene (TSTG) can be decomposed into a TBG-like flat band Hamiltonian and a high-velocity Dirac fermion Hamiltonian. 
We use Hartree-Fock mean field approach to study the projected Coulomb interacting Hamiltonian of TSTG developed in C\u{a}lug\u{a}ru \emph{et al}. [Phys. Rev. B {\bf 103}, 195411 (2021)] at integer fillings $\nu=-3, -2, -1$ and $0$ measured from charge neutrality. We study the phase diagram with $w_0/w_1$, the ratio of $AA$ and $AB$ interlayer hoppings, and the displacement field, which introduces an interlayer potential $U$ and hybridizes the TBG-like bands with the Dirac bands. 
At small $U$, we find the ground states at all fillings $\nu$ are in the same phases as the tensor products of a Dirac semimetal with the filling $\nu$ TBG insulator ground states, which are spin-valley polarized at $\nu=-3$, and fully (partially) intervalley coherent at $\nu=-2,0$ ($\nu=-1$) in the flat bands. An exception is $\nu=-3$ with $w_0/w_1 \gtrsim 0.7$, which possibly become a metal with competing orders at small $U$ due to charge transfers between the Dirac and flat bands.
At strong $U$ where the bandwidths exceed interactions, all the fillings $\nu$ enter a metal phase with small or zero valley polarization and intervalley coherence. Lastly, at intermediate $U$, semimetal or insulator phases with zero intervalley coherence may arise for $\nu=-2,-1,0$. 
Our results provide a simple picture for the electron interactions in TSTG systems, and reveal the connection between the TSTG and TBG ground states.
\end{abstract}

\maketitle

\section{Introduction}

% Review of TBG and TTG
The rich physics discovered in twisted bilayer graphene (TBG), including the correlated insulating phase at integer fillings and the superconducting phase with finite doping have attracted the attention of both experimental and theoretical communities \cite{LOP07,SUA10,BIS11,CAO18,CAO18a,CAO20,CAO20,CHE20a,LIU20a,LU19,LU20,PAR20a,POL19,SAI20,SAI21,SER20,STE20,WU20a,YAN19,CHO19,CHO20,KER19,NUC20,WON20,XIE19,JIA19,CHO21,KAN19,SEO19,BUL20,HEJ20,FER21,FER20,VEN18,POT21,ABO20,AHN19,BER20,BER20a,BER20b,BUL20a,CAO20b,CEA20,CHR20,CLA19,DA19,DA20,DAI16,DOD18,EFI18,EUG20,GON19,GUI18,GUO18,HEJ19,HEJ19a,HUA19,HUA20a,ISO18,JAI16,JUL20,KAN18,KAN20a,KEN18,KHA20,KON20,KOS18,LED20,LEW20,LIA19,LIA20,LIA20a,LIU12,LIU18,LIU19,LIU21,LIU21a,OCH18,PAD20,PEL18,PO18,PO19,REP20,REP20a,ROY19,SOE20,SON19,SON20b,TAR19,THO18,UCH14,VAF20,VEN18,WAN20,WIJ15,WIL20,WU18,WU19,WU19a,WU20b,XIE20,XIE20a,XIE20c,XIE20d,XU18,XU18b,YOU19,YUA18,ZHA20,ZOU18,KWAN2021, ZHANG2021QMC, Lee2021QMC}.
The progress on TBG systems has also inspired interest in other twisted moir\'e materials. 
Among the twisted multilayer graphene systems and motivated by theoretical proposals in Refs. \cite{SUA13,KHA19,MOR19,LI19,LOP20,CAR20,PAR20b,ZHU20,LEI20,WU20c}, the twisted symmetric trilayer graphene (TSTG) has recently been realized in experiments \cite{HAO20,PAR20,CAO21}. Correlated insulating states and superconducting states are also observed in TSTG. Similar to the twisted bilayer graphene, the electron density in TSTG is tunable via gate voltages. Moreover, an external displacement field perpendicular to the graphene sheets can be applied to the system, which makes the band structure also tunable by gate voltages. The experimental discoveries also triggered a deeper theoretical look at this system \cite{TSTGI,SHIN21,FIS21,ETHAN2021,QIN2021,CHOU2021}.

TSTG is made of three graphene sheets in AAA stacking, with the middle layer twisted by a small angle $\theta$ relative to the top and bottom sheets. This lattice structure is shown to be energetically stable \cite{CAR20}. In the absence of the external displacement field, the system has mirror symmetry, by reflection around the graphene middle layer. Therefore we are able to use the eigenstates of this mirror symmetry as the basis: the TSTG decouples into two sectors with $+1$ and $-1$ mirror eigenvalues, which correspond to a TBG-like Hamiltonian with the effective interlayer hopping enhanced by a $\sqrt2$ factor, and a Dirac cone Hamiltonian with a large unrenormalized Fermi velocity, respectively \cite{KHA19}. Similar to the pure TBG system, the TBG-like sector in TSTG exhibits flat bands at the TSTG magic angle $\theta_M\approx1.5^\circ$, which is $\sqrt{2}$ times of the TBG magic angle. The band dispersion also depends on the parameter $w_0/w_1\in[0,1]$, which is the ratio between interlayer in $AA$ and $AB$ hoppings. When an out-of-plane displacement field is turned on, these two mirror sectors will hybridize with each other. Equivalently, the out-of-plane displacement field can be captured by a interlayer potential $U$.
In Ref.~\cite{TSTGI}, we provided the perturbation schemes of the low energy bands in TSTG with and without the displacement field, derived the projected Hamiltonian for TSTG with a screened Coulomb interaction, and carefully analyzed the discrete symmetries and continuous symmetries of the TSTG Hamiltonian.
These provide the foundation of the TSTG projected Hamiltonian we study in this paper.

In this paper, we employ the Hartree Fock (HF) mean field theory to study numerically the ground states of the projected interacting Hamiltonian of magic angle TSTG with a screened Coulomb repulsive interaction derived in Ref.~\cite{TSTGI}. We focus on integer fillings $\nu=-3,-2,-1,0$, defined as the number of electrons per moir\'e unit cell relative to the charge neutrality, where insulating or semimetallic behaviors are observed experimentally \cite{HAO20,PAR20,CAO21}. Our numerical results show that at small $U$, the TSTG phases at all integer fillings $\nu$ are states that can adiabatically connect to the tensor product of a semimetal in the Dirac sector with the TBG sector ground states at flat band fillings $\nu$: the TBG sector flat bands are fully spin-valley polarized at $\nu=-3$, fully intervalley coherent at $\nu=-2$ and $0$, and partially intervalley coherent at $\nu=-1$. The only exception is the case of $\nu=-3$ with $w_0/w_1>0.7$, where the TSTG may enter a large Fermi surface metal phase with competing orders, including a potential translation symmetry breaking, due to the charge transfers between the Dirac and TBG sectors. At fillings $\nu=-2,-1,0$, as $U$ increases (at $w_0/w_1>0$), we find a universal first order transition into a phase with zero intervalley coherence, which either remains a semimetal ($\nu=-2,-1$) or may even become an insulator ($\nu=-1,0$). Lastly, at stronger $U$ for which the TSTG free bandwidth exceeds the Coulomb interaction energy scale, all the integer fillings enter a metallic phase with large Fermi surfaces and small or zero valley polarization and intervalley coherence.

% The organization of the paper
The rest of the paper is organized as follows. In Sec. \ref{sec:model}, we review the single body Hamiltonian of TSTG and its mirror symmetric basis. The projected Hamiltonian into the low energy bands being studied is also discussed. 
Sec. \ref{sec:HF} presents the Hartree-Fock mean field approximation to the TSTG projected Hamiltonian, the self consistent conditions, and the HF order parameters which characterize the physical properties of the mean field ground state. 
In Sec. \ref{sec:nu_-3}, we provide the HF numerical results at integer filling factor $\nu=-3$. The phase diagram and ground state properties are discussed. We have also calculated the HF band structure in different phases.
Similarly, the discussion of the HF numerical results at filling factors $\nu=-2, -1$ and $0$ are also presented in Secs.~\ref{sec:nu_-2}, \ref{sec:nu_-1} and \ref{sec:nu_0}, respectively.

\section{Interacting Model for TSTG}\label{sec:model}
We first briefly review the non-interacting Bistritzer-MacDonald Hamiltonian for mirror symmetric twisted trilayer graphene, which can be written as the sum of a TBG Hamiltonian \cite{BIS11} with renormalized interlayer hopping and an independent Dirac fermion Hamiltonian \cite{KHA19,TSTGI}. We also introduce a displacement field perpendicular to the graphene sheets that can couple the Dirac fermion and TBG fermion together. The interacting Hamiltonian projected into the low energy bands is also discussed in this section \cite{TSTGI}.

\subsection{Single particle Hamiltonian}
The twisted trilayer graphene geometry with mirror symmetry was introduced in Refs. \cite{KHA19,MOR19}. In this article we will use the notations of Ref. \cite{TSTGI,BER20,BER20a,BER20b,LIA20,SON19,SON20b,XIE20a} where the non-interacting model and its symmetries are discussed in detail. We use $\ha^\dagger_{\vp, \alpha, s, l}$ to represent the electron creation operator with momentum $\vp$ measured from the $\Gamma$ point of single layer graphene Brillouin zone, sublattice $\alpha=A, B$, spin $s=\uparrow, \downarrow$ and layer $l=1,2,3$. Similar to the derivation of Bistritzer-MacDonald model for twisted bilayer graphene, Dirac equation can be used to describe the low energy physics of each individual layer. We define $\mathbf{K}_+ = \mathbf{K}_1 = \mathbf{K}_3$ as the $K$ point of the bottom and the top layers, and $\mathbf{K}_- = \mathbf{K}_2$ for the middle layer. Here $|\mathbf{K}_\pm |= 1.073\textrm{\r{A}}^{-1}$. For convenience, we also define vectors $\vq_j = C_{3z}^{j-1}(\mathbf{K}_+ - \mathbf{K}_-)$. The reciprocal lattice of the moir\'e lattice $\mathcal{Q}_0$ is spanned by the basis vectors $\mathbf{b}_{M1} = \vq_3 - \vq_1$ and $\mathbf{b}_{M2} = \vq_3 - \vq_2$. Adding the vectors $\vq_i$ iteratively gives us momentum lattices $\mathcal{Q}_\pm = \mathcal{Q}_0 \pm \vq _1$, and they form the hexagon lattice in the momentum space. In order to describe the low energy physics, we introduce the electron operators $\ha_{\vk, \mathbf{Q}, \eta, \alpha, s, l} = \ha_{\eta\mathbf{K}_l + \vk -\mathbf{Q}, \alpha, s, l}$, where $\mathbf{Q}\in\mathcal{Q}_{\eta}$ if $l = 1,3$ or $\mathbf{Q}\in\mathcal{Q}_{-\eta}$ if $l=2$. Without the displacement field along $\hat{z}$ direction, the system is invariant under mirror symmetry $m_z$ which switches the first layer with the third layer, and leaves the middle layer invariant. Therefore, the Bistritzer-MacDonald model for TSTG can be simplified using the following basis transformation:
\begin{align}
	\hc^\dagger_{\vk, \mathbf{Q}, \eta, \alpha, s} = \left\{
		\begin{array}{ll}
			\frac{1}{\sqrt2} \left(\ha^\dagger_{\vk, \mathbf{Q}, \eta, \alpha, s, 1} + \ha^\dagger_{\vk, \mathbf{Q}, \eta, \alpha, s, 3}\right)	& \mathbf{Q} \in \mathcal{Q}_\eta\,,\\
			\ha^\dagger_{\vk, \mathbf{Q}, \eta, \alpha, s, 2} & \mathbf{Q} \in \mathcal{Q}_{-\eta}\,.
		\end{array}
	\right.\label{eqn:def_even_sector}
\end{align}
where $\vk$ belongs to the moir\'e Brillouin zone (MBZ). These operators (dubbed as \emph{TBG fermions}) have even eigenvalue under $m_z$ transformation. Fermion operators with odd $m_z$ eigenvalue (dubbed as \emph{Dirac fermions}) are given by:
\begin{equation}
	\hb^\dagger_{\vk, \mathbf{Q}, \eta, \alpha, s} = \frac{1}{\sqrt2} \left(\ha^\dagger_{\vk, \mathbf{Q}, \eta, \alpha, s, 1} - \ha^\dagger_{\vk, \mathbf{Q}, \eta, \alpha, 3}\right)~~~\mathbf{Q} \in \mathcal{Q}_{\eta}\,.
\end{equation}
Since the single body Hamiltonian commutes with $m_z$ transformation in the absence of the external displacement field, it can be written as a block diagonal form:
\begin{equation}
	\hat{H_0} = \hat{H}_{\rm TBG} + \hat{H}_D\,.
\end{equation}
It can be shown that the Hamiltonian in the mirror symmetric sector $\hat{H}_{\rm TBG}$ contains $\hc, \hc^\dagger$ operators and is identical to the ordinary TBG Hamiltonian \cite{BIS11,SON19}, with the interlayer hopping parameter multiplied by a factor of $\sqrt{2}$. It reads:
\begin{equation}
	\hat{H}_{\rm TBG} = \sum_{\substack{\vk\in\mathrm{MBZ} \\ \mathbf{QQ}'\in\mathcal{Q}_{\pm} \\ {\eta,s,\alpha,\eta}}}\left[h_{\mathbf{Q},\mathbf{Q}'}^{(\eta)}(\vk)\right]_{\alpha\beta}\hc^\dagger_{\vk, \mathbf{Q}, \eta, \alpha, s} \hc_{\vk,\mathbf{Q}',\eta, \beta, s}\,,
\end{equation}
in which the ``first quantized Hamiltonian'' of the $\eta=+$ valley is given by:
\begin{equation}\label{eqn:HTBG_valley+}
	h^{(+)}_{\mathbf{Q},\mathbf{Q}'}(\vk)=v_F\bm{\sigma}\cdot\left(\vk-\mathbf{Q}\right)\delta_{\mathbf{Q},\mathbf{Q}'} + \sum_{j=1,2,3}\sqrt2 T_j \delta_{\mathbf{Q} - \mathbf{Q}', \pm \vq_j}
\end{equation}
where $v_F = 6104.5\,\rm meV\cdot\textrm{\AA}$ is the Fermi velocity of single layer graphene, and interlayer hopping matrices $T_j$ are given by:
\begin{equation}
	T_j = w_0\sigma_0 + w_1\left[\cos\frac{2\pi(j-1)}{3}\sigma_x + \sin\frac{2\pi(j-1)}{3}\sigma_y\right]\,.
\end{equation}
Similar to the TBG Hamiltonian, $w_0$ and $w_1$ stand for the interlayer hopping strength around the $AA$ and $AB$ stacking regions, respectively. In this article we use $w_0$ as a tunable parameter, and keep the value of $w_1 = 110\,\rm meV$ fixed. Similar to ordinary TBG, we define $w_0=0$ as the chiral limit. In the realistic case we have $0 \leq w_0 < w_1$ due to lattice relaxation effects \cite{UCH14,WIJ15,JAI16,KOS18}. The $\sqrt2$ factor in Eq. (\ref{eqn:HTBG_valley+}) comes from the transformation in Eq. (\ref{eqn:def_even_sector}). Due to the fact that the effective interlayer hopping is stronger, the magic angle of TSTG where the bands around charge neutral point are flat will be around $\theta \approx 1.5^\circ$, which is bigger than the magic angle in TBG \cite{KHA19}. The Hamiltonian in valley $\eta = -$ can be obtained by applying $C_{2z}$ transformation to Eq. (\ref{eqn:HTBG_valley+}).

On the other hand, $\hat{H}_D$ only includes the contribution from mirror anti-symmetric sector. It is given by the following expression:
\begin{equation}\label{eqn:ham_dirac}
	\hat{H}_D = \sum_{\substack{\vk\in{\rm MBZ} \\ \eta, s, \alpha, \beta}}\sum_{\mathbf{Q}\in\mathcal{Q}_{\eta}}\left[h_{\mathbf{Q}}^{D,\eta}(\vk)\right]_{\alpha\beta}\hb^\dagger_{\vk, \mathbf{Q}, \eta, \alpha, s}\hb_{\vk, \mathbf{Q}, \eta, \beta, s}
\end{equation}
in which the first quantized Hamiltonian for Dirac cone reads:
\begin{align}
	h^{D,+}_\mathbf{Q}(\vk) &= v_F\bm{\sigma}\cdot(\vk - \mathbf{Q})\,,\\
	h^{D,-}_{\mathbf{Q}}(\vk) &= \sigma_x h^{D,+}_{-\mathbf{Q}}(-\vk)\sigma_x\,.
\end{align}

We can introduce an external displacement field perpendicular to the graphene sheets. When this external field is turned on, the mirror symmetry $m_z$ is broken, which will lead to mixing terms between the TBG fermions in the mirror symmetric sector and the Dirac fermions in the mirror anti-symmetric sector. We denote the potential difference between the top and bottom layer by $U$, and the Hamiltonian which describes the electric field can be written as:
\begin{equation}
	\hat{H}_U = \frac{U}{2} \sum_{\vk,\eta,s\alpha}\sum_{\mathbf{Q}\in\mathbf{Q}_\eta}\sum_{l=1,3}(l-2)\ha^\dagger_{\vk, \mathbf{Q}, \eta, \alpha, s, l}\ha_{\vk, \mathbf{Q}, \eta, \alpha, s,l}\,.
\end{equation}
This Hamiltonian can be rewritten using the Dirac and TBG fermions:
\begin{equation}
	\hat{H}_U = \frac{U}{2}\sum_{\vk,\eta,s\alpha}\sum_{\mathbf{Q}\in\mathbf{Q}_\eta}\left(\hb^\dagger_{\vk,\mathbf{Q},\eta, \alpha,s}\hc_{\vk, \mathbf{Q}, \eta, \alpha, s} + {\rm h.c.}\right)\,,\label{eq-HU-full}
\end{equation}
which couples the mirror symmetric and anti-symmetric sectors. In conclusion, the non-interacting Hamiltonian can be written as the summation of these three terms:
\begin{equation}\label{eqn:bm_hamiltonian}
	\hat{H}_0 = \hat{H}_{\rm TBG} + \hat{H}_D + \hat{H}_U\,.
\end{equation}

\subsection{Interaction and Projected Hamiltonian}
In this article we will assume that the interaction between electrons in TSTG system is given by the Coulomb potential screened by a top and bottom gate. The interaction Fourier transformation reads:
\begin{equation}\label{eq:Vq}
	V(\vq) = \pi\xi^2U_\xi\frac{\tanh(\xi q/2)}{\xi q/2}
\end{equation}
where $\xi \approx 10\,\rm nm$ is the distance between the top and bottom gates, and $U_\xi = e^2/\epsilon\xi \approx 24\,\rm meV$ is the strength of the Coulomb interaction with dielectric constant $\epsilon\sim 6$ \cite{CAO18,CAO18a,KAN18}. The interacting Hamiltonian can be written as \cite{KAN19,BER20a,TSTGI}:
\begin{equation}
	\hat{H}_I = \frac{1}{2N_M\Omega_c} \sum_{\vq\in{\rm MBZ}}\sum_{\mathbf{G}\in\mathcal{Q}_0}V(\vq + \mathbf{G}) \delta\rho_{\vq + \mathbf{G}} \delta\rho_{-\vq - \mathbf{G}}
\end{equation}
where $\Omega_c$ is the area of moir\'e unit cell, and $N_M$ is the number of moir\'e unit cells. $\delta\rho$ is the electron density at momentum $\vq + \mathbf{G}$ relative to the charge neutral point and can be written as:
\begin{align}
	\delta\rho_{\vq + \mathbf{G}} &= \delta\rho^{\hat{c}}_{\vq + \mathbf{G}} + \delta\rho^{\hat{b}}_{\vq + \mathbf{G}},\\
	\delta\rho^{\hat{c}}_{\vq + \mathbf{G}} &= \sum_{\substack{\vk,\eta, \alpha, s \\ \mathbf{Q}\in\mathcal{Q}_\pm}}\left(\hc^\dagger_{\vk+\mathbf{q},\mathbf{Q}-\mathbf{G},\eta,\alpha,s} \hc_{\mathbf{k},\mathbf{Q},\eta,\alpha,s} - \frac12 \delta_{\vq, 0}\delta_{\mathbf{G},0}\right),\\
	\delta\rho^{\hat{b}}_{\vq + \mathbf{G}} &= \sum_{\substack{\vk,\eta, \alpha, s \\ \mathbf{Q}\in\mathcal{Q}_\eta}}\left(\hb^\dagger_{\vk+\mathbf{q},\mathbf{Q}-\mathbf{G},\eta,\alpha,s} \hb_{\mathbf{k},\mathbf{Q},\eta,\alpha,s} - \frac12 \delta_{\vq, 0}\delta_{\mathbf{G},0}\right).
\end{align}

By projecting the system into the low energy bands, the dimension of Hamiltonian matrix in Hartree Fock calculation will be reduced dramatically, and therefore greatly improving the numerical calculations. By diagonalizing the single particle TBG Hamiltonian $h^{(\eta)}(\vk)$ and the Dirac Hamiltonian $h^{D,\eta}(\vk)$, we obtain the dispersion relation $\varepsilon^{\hat{f}}_{m,\eta}(\vk)$ and the single body wavefunctions $u_{\mathbf{Q}\alpha, m\eta}^{\hat{f}}(\vk)$ for the TBG and Dirac fermions ($\hat{f}=\hat{c},\hat{b}$). For each spin and valley, we project the kinetic Hamiltonian into the two bands which are closest to the charge neutral point for both $\hat{H}_{\rm TBG}$ and $\hat{H}_D$. Therefore, the kinetic part of the projected Hamiltonian can be written in the following form when $U=0$:
\begin{equation}
	H_{\rm TBG} + H_D = \sum_{\hat{f} = \hat{c},\hat{b}}\sum_{\vk, m=\pm1, \eta, s}\varepsilon_{m, \eta}^{\hat{f}}(\vk)\hf^\dagger_{\vk, m, \eta, s}\hf_{\vk, m, \eta, s} 
\end{equation}
where the creation operators in band indices are defined as $\hf^\dagger_{\vk,m,\eta,s} = \sum_{\mathbf{Q}\alpha}u^{\hat{f}}_{\mathbf{Q}\alpha, m\eta}\hf^\dagger_{\vk,\mathbf{Q}, \eta, \alpha, s}$. 
The Dirac fermions in the antisymmetric sector $\hb$ are degenerate on certain high symmetry lines between the projected bands and the bands above and below when folding over the MBZ, therefore there is an ambiguity of choosing its single-body wavefunction. 
We provide a careful discussion of this issue and how we solve it in \ref{app_sec:hamiltonian}.

As shown in Refs. \cite{BUL20,HEJ20,BER20a,LIA20}, by fixing the sewing matrix of $C_{2z}T$ symmetry to identity (where $C_{2z}$ is the 2-fold rotation about the $z$ axis, and $T$ is the time reversal), one can recombine the TBG flat energy band basis $\hc^\dagger_{\vk,m,\eta,s}$ into a Chern band basis
\begin{equation}\label{eq:chernbasis}
\hat{d}^\dagger_{\vk,e_Y,\eta,s}=\frac{\hc^\dagger_{\vk,+1,\eta,s}+ie_Y \hc^\dagger_{\vk,-1,\eta,s}}{\sqrt{2}}\ ,
\end{equation}
where $e_Y=\pm1$ gives the Chern number of the Chern band basis (which is also the eigenvalue of the Pauli matrix $\zeta_y$ in the space of TBG energy band index $m=\pm1$).

The displacement field term $\hat{H}_U$ in Eq. (\ref{eq-HU-full}) can also be written using band basis and projected into the lowest bands:
\begin{equation}\label{eq-HU}
	H_U = \frac{U}{2}\sum_{\vk, \eta, s}\sum_{m=\pm1}\sum_{n=\pm1}N^\eta_{mn}(\vk)\left(\hb^\dagger_{\vk, m, \eta, s}\hc_{\vk, n, \eta, s} + {\rm h.c.}\right)\,,
\end{equation}
where the displacement field overlap matrices are given by
\begin{equation}
	N^{\eta}_{mn}(\vk) = \sum_{\mathbf{Q}\in\mathcal{Q}_\eta, \alpha} u^{\hat{b}*}_{\mathbf{Q}\alpha, m\eta}(\vk)u^{\hat{c}}_{\mathbf{Q}\alpha, n\eta}\,.
\end{equation}
Thus the projected non-interacting Hamiltonian can be written as the following quadratic form:
\begin{align}
	H_0 =& H_{\rm TBG} + H_D + H_U \nonumber\\
	=& \sum_{\vk,\hf\hf',\eta\eta',ss'}\mathcal{H}^{(0)}_{\hat{f}m\eta s,\hat{f}'n\eta's'}(\vk)\hf^\dagger_{\vk,m,\eta,s}\hf'_{\vk,n,\eta',s'}\,,\label{eqn:non-interacting-quadratic}
\end{align}
in which the matrix $\mathcal{H}^{(0)}(\vk)$ is given by
\begin{align}
	&\mathcal{H}^{(0)}_{\hat{f}m\eta s,\hat{f}'n\eta's'}(\vk) = \varepsilon_{m,\eta}^{\hat{f}}(\vk) \delta_{\hat{f}\hat{f}'}\delta_{mn}\delta_{\eta\eta'}\delta_{ss'}\nonumber\\
	& + \frac{U}{2}(N^{\eta}_{mn}(\vk)\delta_{\hat{f}\hat{b}}\delta_{\hat{f}'\hat{c}} + N^{\eta*}_{mn}(\vk)\delta_{\hat{f}\hat{c}}\delta_{\hat{f}'\hat{b}})\delta_{\eta\eta'}\delta_{ss'}\,.\label{eqn:non-interacting-H0}
\end{align}
Here $\varepsilon_{m,\eta}^{\hf}(\vk)$ is the dispersion of TBG($\hf = \hc$) and Dirac($\hf = \hb$) fermions without displacement field. The eigenvalues of $\mathcal{H}^{(0)}(\vk)$ can give us the approximate dispersion of the non-interacting TSTG at non-zero displacement field.
The projected Hamiltonian can capture the band width of the bands around charge neutrality accurately \cite{TSTGI}. We also provide plots comparing the dispersion of the projected Hamiltonian $\mathcal{H}^{(0)}(\vk)$ and the band structure obtained from the unprojected BM Hamiltonian in \ref{app_sec:hamiltonian} Fig.~\ref{fig:projected} \cite{TSTGI}.

Similarly, the interacting Hamiltonian can also be projected into these bands:
\begin{equation}\label{eq:proj-HI}
	H_I = \frac{1}{2N_M\Omega_{c}}\sum_{\vq,\mathbf{G}\in\mathcal{Q}_0}V(\vq + \mathbf{G})\overline{\delta\rho}_{\vq + \mathbf{G}}\overline{\delta\rho}_{-\vq - \mathbf{G}}\,,
\end{equation}
in which the density operators after being projected are defined as:
\begin{align}
	\overline{\delta\rho}_{\vq + \mathbf{G}} =& \sum_{\hat{f} = \hat{c}, \hat{b}}\overline{\delta\rho}^{\hat{f}}_{\vq + \mathbf{G}}\\
	\overline{\delta\rho}^{\hat{f}}_{\vq + \mathbf{G}} =& \sum_{\vk,m,n,\eta,s}M^{\hat{f}, \eta}_{mn}(\vk, \vq + \mathbf{G})\nonumber\\
	&\left(\hf^\dagger_{\vk + \vq, m, \eta, s}\hf_{\vk, n, \eta, s} - \frac12 \delta_{\vq, 0}\delta_{mn} \right)\,,\\
	M^{\hat{f},\eta}_{mn}(\vk, \vq + \mathbf{G}) =& \sum_{\mathbf{Q}\alpha}u^{\hat{f}*}_{\mathbf{Q}-\mathbf{G}\alpha, m\eta}(\vk + \vq)u^{\hat{f}}_{\mathbf{Q}\alpha, n\eta}(\vk)\,.
\end{align}
The components of these form factors $M_{mn}^{\hat{f},\eta}(\vk, \vq + \mathbf{G})$ depend on the gauge choice of the single body wavefunctions. 
As mentioned in Eq. (\ref{eq:chernbasis}), we fix the gauge choice of the single-body wavefunction of the TBG fermions $u^{\hc}_{\mathbf{Q}\alpha,m\eta}(\vk)$ such that the sewing matrix of $C_{2z}T$ is the identity.

For convenience, we can rewrite the interacting Hamiltonian as the following form:
\begin{align}
	H_I =& \frac{1}{2\Omega_{\rm tot}} \sum_{\vk,\vk',\vq}\sum_{\eta\eta'ss'}\sum_{\hat{f},\hat{f}'=\hat{c},\hat{b}}\sum_{mnm'n'}\tilde{V}^{(\hat{f}\eta;\hat{f}'\eta')}_{mn;m'n'}(\vq;\vk,\vk')\nonumber\\
	& \times \left(\hf^\dagger_{\vk + \vq, m,\eta, s}\hf_{\vk, n,\eta, s} - \frac{1}{2}\delta_{\vq, 0}\delta_{mn}\right)\nonumber\\
	& \times\left(\hf'^\dagger_{\vk'-\vq, m',\eta', s'}\hf'_{\vk',n',\eta', s'} - \frac12 \delta_{\vq,0}\delta_{m'n'} \right)\,,
\end{align}
in which the matrix elements $\tilde{V}^{(\hat{f}\eta;\hat{h}\eta')}_{mn;m'n'}(\vq; \vk, \vk')$ are given by:
\begin{align}
	&\tilde{V}^{(\hat{f}\eta;\hat{h}\eta')}_{mn;m'n'}(\vq; \vk, \vk') \nonumber\\
	=& \sum_{\mathbf{G}}V(\vq + \mathbf{G})M^{\hat{f},\eta}_{mn}(\vk, \vq + \mathbf{G})M^{\hat{h},\eta'}_{m'n'}(\vk', -\vq - \mathbf{G})\,.
\end{align}
The mean field Hamiltonian will have a simpler form using this notation, as we will discuss in Sec. \ref{sec:HF}.

In this paper, we fix the twist angle to $\theta=1.51^\circ$, which is near the magic angle of TSTG and gives rise to flat bands in the mirror symmetric sector. 
Since both the band structure and the wavefunctions of the mirror symmetric sector depend on the parameter $w_0$, the projected Hamiltonian also depends on $w_0$. And by adding all the terms in kinetic energy and potential energy, we obtain the tunable Hamiltonian with parameters $w_0$ and $U$:
\begin{equation}\label{eqn:tunable_ham}
	H(w_0, U) = H_{\rm TBG}(w_0) + H_D + H_U(w_0, U) + H_I(w_0)\,.
\end{equation}
Similar to that in TBG, we define $w_0=0$ as the chiral limit, and $H_{\rm TBG}(w_0)=0$ (zero TBG bandwidth) as the flat (TBG band) limit. In these limits or their combinations, the symmetry of the TSTG is enhanced to a $U(4)$ symmetry in the combined spin and valley space \cite{TSTGI}. In this paper, we will not tune the bandwidth in the mirror symmetric (TBG) sector, therefore the non-interacting band structure will only depend on $w_0$ and $U$ (at the fixed twist angle $\theta=1.51^\circ$ and AB/BA interlayer hopping strength $w_1 = 110\,\rm meV$).

\section{Hartree-Fock Theory}\label{sec:HF}
We perform Hartree Fock (HF) mean field calculations for the projected Hamiltonian we obtained in Eq. (\ref{eqn:tunable_ham}), which is at fixed twist angle $\theta=1.51^\circ$. In \ref{app_sec:hartreefock}, we provide a more detailed discussion of the HF calculations. In this section, we focus on the assumption and the quantities that we will rely on in the rest of our paper. 

In Refs. \cite{KAN18, ZHA20, BUL20, LIA20,XIE20a}, it has been shown that the ground states of TBG at integer fillings (integer number of electrons per moir\'e unit cell, relative to the charge neutral point) around the chiral flat band limit (i.e. the value of $w_0/w_1$ is small and disregarding the flat band dispersion) are correlated insulator states (sometimes with non-zero Chern number) without translation symmetry breaking. This picture is expected to be valid till reasonably large physical $w_0/w_1$ values (depending on electron fillings) \cite{KAN20a,SOE20, XIE20a}. Meanwhile, the high Fermi velocity and vanishing Fermi surfaces of the Dirac fermions make them unlikely to contribute to translation symmetry breaking (which requires certain low energy Fermi surface nestings). 

Therefore, we assume there is no translation symmetry breaking in our HF calculation for TSTG (with a notable exception in \ref{app_subsec:-3} where we discuss the possible CDW order at $M_M$ point at $\nu=-3$ filling). This assumption simplifies our numerical calculation by reducing the number of HF mean field order parameters. For this reason, within the assumption, the HF mean field order parameter can be defined as the following $16\times 16$ matrix at each $\vk$:
\begin{equation}\label{eqn:def_order_parameter}
	\Delta_{\hat{f}m\eta s;\hat{f}'n\eta' s'}(\vk) = \Big{\langle} \hf^\dagger_{\vk,m,\eta, s}\hf'_{\vk,n,\eta', s'} - \frac12 \delta_{\hat{f}\hat{f}'}\delta_{mn}\delta_{\eta\eta'}\delta_{ss'} \Big{\rangle}\,,
\end{equation}
where $\hat{f}, \hat{f}'$ stand for the TBG and Dirac fermion operators. The matrix $\Delta(\vk)$ is the single-body density matrix at each momentum $\vk$.
As we explained, this assumption of no translation breaking is reasonable when $w_0/w_1$ is small (typically $w_0/w_1 \lesssim 0.7$), and it is possible that our assumption will be violated for large $w_0/w_1$ \cite{KAN18,XIE20a}. Therefore, the Hartree Fock result is less trustable when $w_0/w_1$ gets bigger.

For an arbitrary momentum $\vk$, the Hartree and Fock mean field Hamiltonians are given by the following:
 \begin{align}
	\mathcal{H}^{(H)}_{\hat{f}m\eta s,\hat{f}'n\eta's'}(\vk) =& \frac{1}{\Omega_{\rm tot}}\sum_{\vk', \hat{f}',m'n',\eta'', s''}\tilde{V}^{(\hat{f}\eta;\hat{f}'\eta'')}_{mn;m'n'}(0; \vk,\vk') \nonumber\\
	&\times\Delta_{\hat{f}'m'\eta''s'';\hat{f}'n'\eta''s''}(\vk') \delta_{\eta\eta'}\delta_{ss'} \\
	\mathcal{H}^{(F)}_{\hat{f}m\eta s,\hat{f}'n\eta's'}(\vk) =& -\frac{1}{\Omega_{\rm tot}}\sum_{\vk',m'n'}\tilde{V}^{(\hat{f}'\eta';\hat{f}\eta)}_{m'n;mn'}(\vk'-\vk;\vk,\vk')\nonumber\\
	&\times \Delta_{\hat{f}'m'\eta's'; \hat{f}n'\eta s}(\vk') 
\end{align}
Together with the non-interacting term $\mathcal{H}^{(0)}(\vk)$ defined in Eqs. (\ref{eqn:non-interacting-quadratic}) and (\ref{eqn:non-interacting-H0}), we obtain the Hartree Fock Hamiltonian $\mathcal{H}^{HF}(\vk) = \mathcal{H}^{(0)}(\vk) + \mathcal{H}^{(H)}(\vk) + \mathcal{H}^{(F)}(\vk)$. By diagonalizing the Hartree Fock Hamiltonian, we obtain the HF band structure $E_i(\vk)$, which is related to the dispersion of the charge excitations, and the corresponding wavefunction $\phi_{\hf m \eta s, i}(\vk)$:
\begin{equation}
	\sum_{\hf',n,\eta',s'}\mathcal{H}^{HF}_{\hat{f}m\eta s, \hat{f}'n\eta's'}(\vk) \phi_{\hat{f}'n\eta's', i}(\vk) = E_i(\vk)\phi_{\hat{f}m\eta s, i}(\vk)
\end{equation}
For a filling factor $\nu$, which is defined as the number of electrons per moir\'e unit cell relative to charge neutrality, the HF ground state is given by occupying the single particle states $E_i(\vk)$ (where $i=1,\cdots, 16$ at each $\vk$) from low to high up to filling $\nu$. For each single body state $E_i(\vk)$, valley polarization $v_i(\vk)$ can be defined as:
\begin{equation}\label{eqn:def_singlevp}
	v_i(\vk) = \sum_{\hf m s \eta\eta'}\phi^*_{\hf m \eta s, i}(\vk) (\tau_z)_{\eta\eta'}\phi_{\hf m\eta's, i}(\vk)\,,
\end{equation}
and the valley physics of the system can be captured by $v_i(\vk)$ of each individual occupied state at every $\vk$. 

The self-consistent condition also gives a relation between these wavefunctions and the order parameter:
\begin{align}
	\Delta_{\hat{f}m\eta s;\hf'n\eta's'}(\vk) =& \sum_{i\in{\rm occupied}}\Big{(}\phi^*_{\hat{f}m\eta s, i}(\vk)\phi_{\hat{f}'n\eta's', i}(\vk)\nonumber\\
	& - \frac12 \delta_{\hat{f}\hat{f}'}\delta_{mn}\delta_{\eta\eta'}\delta_{ss'}\Big{)}\,.\label{eq:ord-para}
\end{align}
For each integer filling factor $\nu$, we use various initial conditions in our HF calculation, and we choose the result with the lowest energy.
Detailed discussion about the choices of initial conditions at different filling factors can be found in \ref{app_sec:hartreefock}. In this article, the filling factor $\nu$ is measured from the charge neutrality, and it is related with the order parameter in Eq. (\ref{eq:ord-para}) by:
\begin{equation}
	\nu = \frac{1}{N_M}\sum_{\vk,\hf,m,\eta,s}\Delta_{\hf m \eta s; \hf m \eta s}(\vk)\,.
\end{equation}

Moreover, since the particle numbers of Dirac fermion and TBG fermion are conserved when the displacement field is turned off, we can define the filling factors (measured from the charge neutrality) for these fermion flavors separately:
\begin{align}
	\nu_{\rm TBG} &= \frac{1}{N_M}\sum_{\vk, m, \eta, s}\Delta_{\hc m \eta s; \hc m \eta s}(\vk)\,,\\
	\nu_D &= \frac{1}{N_M}\sum_{\vk, m, \eta, s}\Delta_{\hb m \eta s; \hb m \eta s}(\vk)\,.
\end{align}
The summation of these two quantities is the total filling factor: 
\begin{equation}
\nu = \nu_D + \nu_{\rm TBG}\,.
\end{equation}
For the projected bands we keep, the two filling factors range within $\nu_D\in[-4,4]$ and $\nu_{\rm TBG}\in [-4,4]$, respectively. We will be focusing on total integer fillings $\nu=-3,-2,-1,0$ in this paper. Since the physics at filling $-\nu$ is particle-hole symmetric to that at filling $\nu$ \cite{TSTGI}, it is sufficient to study fillings $\nu\le 0$.

Various physical quantities can be derived from $\Delta_{\hat{f}m\eta s;\hat{f}'n\eta' s'}(\vk)$, which can be used to describe the nature of the ground state, such as the intervalley coherence and valley polarization. As shown in Ref. \cite{LIA20}, the ground state at $\nu=\pm2$ filling in TBG has intervalley coherence when the system is non-chiral non-flat. In order to measure the coherence between the two valleys, we define the quantity $\mathcal{C}$ which is based on the norm of the off-diagonal block in valley space:
\begin{equation}\label{eqn:def_vc}
	\mathcal{C} = \frac{1}{N_M}\sum_{\vk\in{\rm MBZ}}\sum_{\hat{f}\hat{f}',mn,ss'}|\Delta_{\hat{f}m+s;\hat{f}'n-s'}(\vk)|^2 \,,
\end{equation}
where $N_M$ is the number of moir\'e lattice sites. This quantity includes both the contribution from the TBG flat bands and the Dirac fermions. Its value is 
\begin{equation}
    \mathcal{C} = \frac{n}{4}
\end{equation} if there are $n$ filled TBG flat bands which are fully intervalley coherent. 

The expectation value of any single-body quantity can be obtained from the Hartree-Fock order parameter $\Delta(\vk)$. 
In this article, we calculate three quantities that we will now define: the valley polarization $N_v$, the spins in each valley $S^\eta$ and the quantity $\mathrm{Ch}$ which provides information about the Chern number of the occupied TBG fermions.

The valley polarization $N_v$ is the electron number difference between the two valleys. This can also be obtained from the order parameter:
\begin{equation}\label{eqn:def_Nv}
	N_v = \sum_{\vk}\sum_{\hat{f}=\hat{c},\hat{b}}\sum_{\eta\eta' m s}(\tau_z)_{\eta\eta'}\Delta_{\hat{f}m\eta s; \hat{f}m \eta' s}(\vk)\,,
\end{equation}
where $\tau_z$ is the Pauli $z$ matrix in valley space.

Similarly, we can track the spin order. Due to the U(2)$\times$U(2) symmetry of the system, the total spin of the two valleys are conserved independently. For each valley $\eta$, the semi-classical total spin per moir\'e unit cell $\vec{S}^\eta$ can be obtained by the following equation:
\begin{equation}\label{eqn:def_Spm}
	2\vec{S}^\eta(\vk) = \frac{1}{N_M}\sum_{\vk}\sum_{\hat{f}=\hat{c},\hat{b}}\sum_{mss'}(\vec{s})_{ss'}\Delta_{\hat{f}m\eta s; \hat{f}m\eta s'}(\vk)\,,
\end{equation}
where $\vec{s}=(s_x, s_y, s_z)$ are the Pauli matrices in spin space. 

Finally, we can define a quantity within the TBG band sector:
\begin{equation}\label{eqn:def_xi_y}
	\mathrm{Ch} = \frac{1}{N_M} \sum_{\vk}\sum_{\eta s mn}(\zeta_y)_{mn}\Delta_{\hat{c}m\eta s; \hat{c}n\eta s}(\vk)\,,
\end{equation}
where $\zeta_y$ is the Pauli $y$ matrix in the space of the energy band index $m$. If the Dirac bands and the TBG bands in the HF Hamiltonian are decoupled (e.g. at $U=0$ and without $m_z$ breaking order parameters), $\mathrm{Ch}$ characterizes the Chern number in the TBG sector when the TBG sector is insulating, which can be seen by transforming $\mathrm{Ch}$ into the Chern band basis in Eq. (\ref{eq:chernbasis}). Generically (e.g., $U>0$), $\mathrm{Ch}$ is not necessarily an integer, but it is related with the Chern number of the (partially or fully) occupied TBG flat band basis. For example, this value is close to $\pm2$ if the two occupied TBG flat bands have the same Chern number. Similar to the filling factor for Dirac and TBG fermion flavors, this quantity is a useful characterization of the many-body state when $U$ is close to zero.

We perform the HF calculations on a $C_{3z}$ preserving $N_L\times N_L$ momentum lattice in the MBZ (see Fig. \ref{fig:mbz_dirac}), with $N_L$ up to $10$. As discussed in \ref{app_sec:hartreefock}, we are also able to obtain the band structure plot along high symmetry lines by using the HF order parameters we obtained on these $N_L\times N_L$ lattices. In the band structure plots, we use subscript $M$ to denote the high symmetry points in the moir\'e Brillouin zone. Our HF calculations are restricted within the pamameter ranges $0.1\le w_0/w_1\le 1$ and $U\ge0$. We do not discuss the HF calculation in the chiral limit $w_0=0$ in this paper, the convergence of which is difficult due to the enhanced symmetry and enlarged ground state degeneracy manifold. We note that the realistic TSTG is always away from the $w_0=0$ chiral limit.

\section{Numerical Results at filling factor \texorpdfstring{$\nu=-3$}{nu=-3}}\label{sec:nu_-3}

\begin{figure*}[t]
	\centering
	\includegraphics[width=\linewidth]{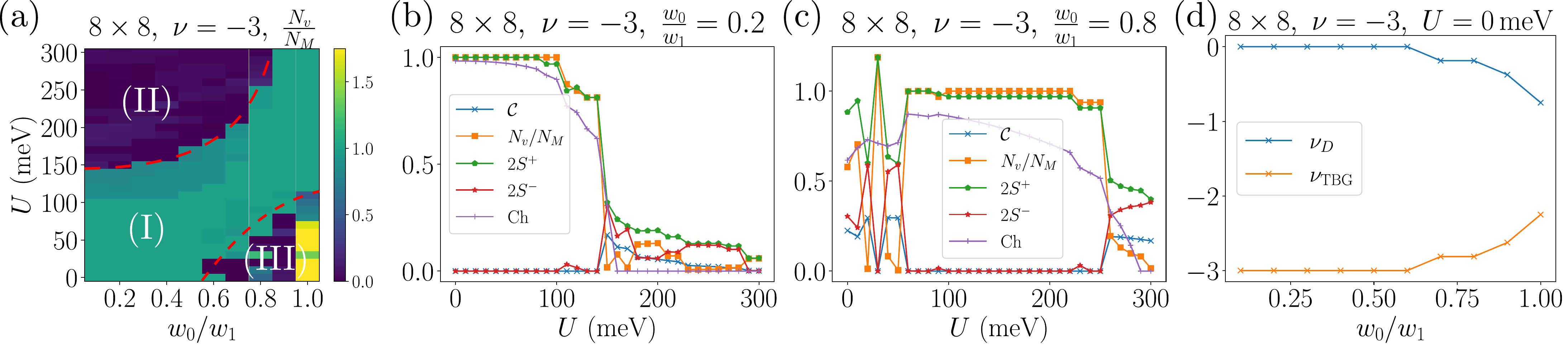}
	\caption{(a) The phase diagram at filling factor $\nu=-3$ obtained on $8\times 8$ momentum lattice in the $(w_0, U)$ plane. The color represents the valley polarization $N_v/N_M$ of the ground state. (b) The displacement field dependence of other quantities $\mathcal{C}, N_v, S^\pm$ and $\mathrm{Ch}$ on $8\times 8$ lattice at $w_0/w_1 = 0.2$. (c) Similar to sub-figure (b), the displacement field dependence of these quantities with $w_0/w_1 =0.8$. (d) The filling factors for Dirac fermions and TBG fermions as a function of $w_0$. In this plot, the interlayer potential is fixed to be $U=0\rm\,meV$. }
	\label{fig:nu=-3_phase}
\end{figure*}

\begin{figure*}[t]
	\centering
	\includegraphics[width=\linewidth]{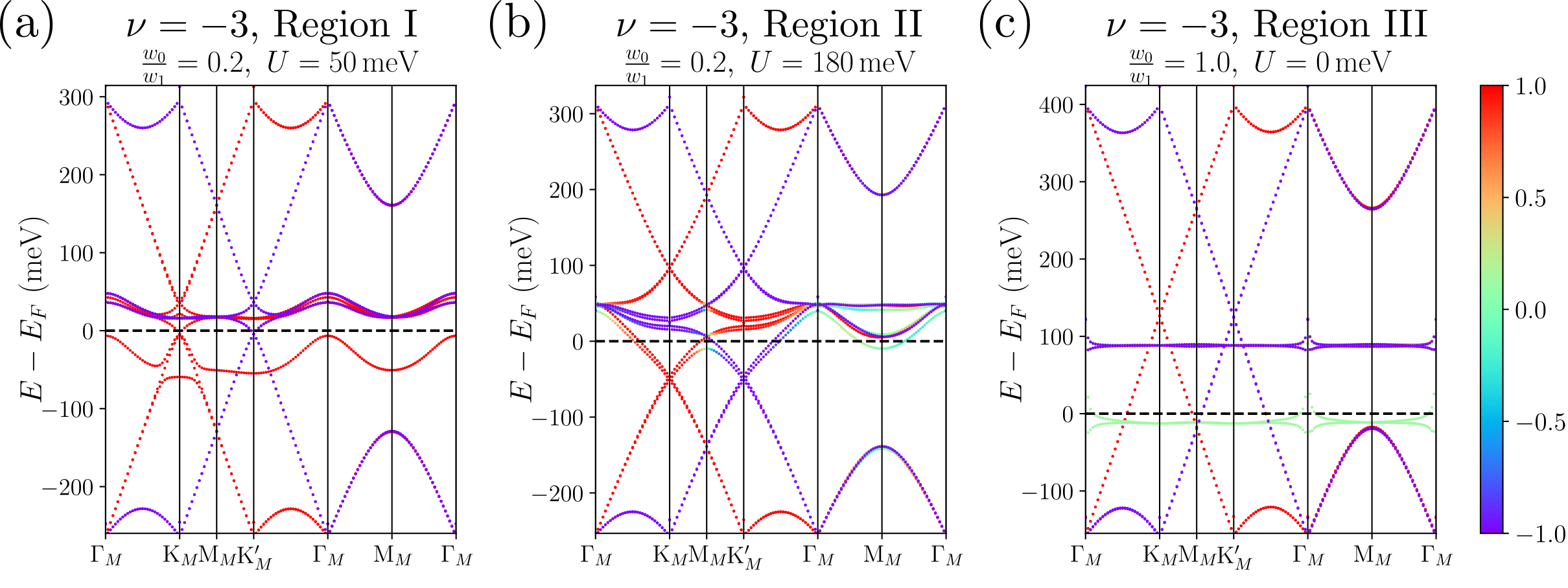}
	\caption{Some typical HF band structures illustrating the three regions of the phase diagram at filling factor $\nu=-3$ on $10\times 10$ momentum lattice. (a) The band structure in region I with $w_0/w_1 = 0.2$ and $U=50\rm\,meV$. (b) The band structure in region II with $w_0/w_1 = 0.2$ and $U = 180\rm\,meV$. (c) The band structure in region III with $w_0/w_1 = 1$ and $U = 0\,\rm meV$. The color of each point represents the valley polarization $v_i(\vk)$ of each single body state, which is defined in Eq. (\ref{eqn:def_singlevp}).}
	\label{fig:nu=-3_band}
\end{figure*}

We start our discussion about HF calculations for TSTG with filling factor $\nu=-3$. As a comparison, the ground state at $\nu=-3$ filling in TBG at small $w_0$ and small nonzero bandwidth is a spin and valley polarized Chern insulator state with Chern number $\pm1$, and may enter translation or rotation symmetry breaking phases at large $w_0$, which has been predicted in Refs.~\cite{ZHA20,BUL20,LIA20,XIE20a}. In this section, we will explore the HF ground states in TSTG at $\nu=-3$ in the parameter space of $w_0/w_1$ and $U$ (see Eq. (\ref{eqn:tunable_ham}) for definition).

Here we restrict the parameter ranges within $0.1\leq w_0/w_1 \leq 1$ and $0 \leq U \leq 300\,\rm meV$. The maximal value of $U$ is motivated by the experimental results \cite{PAR20}.
The valley polarization $N_v$ as a function of $w_0$ and $U$ is shown in Fig.~\ref{fig:nu=-3_phase}(a). We find the HF ground states show different behaviors in three different parameter regions, which are labeled by I, II and III in Fig.~\ref{fig:nu=-3_phase}(a). We also calculate other physical quantities, including $\mathcal{C}, N_v, S^\pm$ and $\mathrm{Ch}$, the values of which along certain line cuts in the parameter space are shown in Fig.~\ref{fig:nu=-3_phase}(b) and (c). Based on these quantities, we describe the TSTG phases in the three regions in details below.

{\it Region I}: we find $\mathcal{C}\approx 0, N_v/N_M\approx 1, 2S^+\approx 1, 2S^- \approx 0$ and $\mathrm{Ch}\approx 1$ throughout the whole region (Fig.~\ref{fig:nu=-3_phase}(b) and (c)). This indicates that the ground state is a spin-valley polarized state dominantly occupying one Chern band in the TBG sector (defined in Eq. (\ref{eq:chernbasis})) of a particular spin and valley. In particular, at $U=0$, where the electron numbers in the Dirac sector and the TBG sector are both conserved, we find $\nu_D=0$ and $\nu_{\rm TBG}=-3$ within region I (see $w_0/w_1<0.6$ in Fig.~\ref{fig:nu=-3_phase}(d)). Therefore, in region I, the $\nu=-3$ HF ground state at $U=0$ is the tensor product of the $\nu_{\rm TBG}=-3$ TBG spin-valley polarized Chern insulator and the Dirac fermion semimetal at charge neutrality $\nu_D=0$. The ground states at $U>0$ in region I are adiabatically in the same semimetal phase.
As an example, the band structure at $w_0/w_1 = 0.2$ and $U=50\,\rm meV$ is shown in Fig.~\ref{fig:nu=-3_band}(a), which is almost a Dirac semimetal. At $U>0$, where the Dirac and TBG sectors are hybridized, the gapless Dirac nodes are due to the $C_{2z}T$ symmetry within the empty valley-spin flavors, as shown in \ref{sec:Ggammasymmetry}. The color (from red to purple) indicates the valley polarization of of each band, and an occupied flat band can be seen clearly.

{\it Region II}: we find the valley polarization $N_v/N_M$ drops abruptly to small values near zero, and so do the other quantities as shown in Fig.~\ref{fig:nu=-3_phase}(b) in this region where the displacement field is large. Accordingly, the HF ground state can be understood as a metal with little spin/valley polarization or intervalley coherence.
A typical HF band structure in region II is shown in Fig.~\ref{fig:nu=-3_band}(b), which has a large Fermi surface around $K_M$ ($K_M'$) point in valley $\eta=+$ ($\eta=-$), showing that the system is a metal. A sharp phase boundary between region I and II can be identified in Fig.~\ref{fig:nu=-3_phase}(a), which is at $U\approx 150\rm\,meV$ when $w_0/w_1=0.2$, and at $U \approx 250\rm\,meV$ when $w_0/w_1=0.8$. The reason for such a metallic phase is that a large $U$ significantly hybridizes the Dirac sector and the TBG sector, and turns the flat bands near $K_M$ ($K_M'$) point of valley $+$ ($-$) into dispersive Dirac fermions with kinetic energies comparable to the interaction energies. This leads to a Fermi surface reconstruction, where electrons prefer to occupy the electron states near the $K_M$ and $K_M'$ points with lower kinetic energies and form a metal. We provide the non-interacting band width as a function of $w_0/w_1$ and $U$ in Figs.~\ref{fig:bandwidth}(a) and (b) of \ref{app_sec:hamiltonian}. The phase boundary between region I and II is close to an equal value contour in these figures, which also implies that the transition to the metallic phase happens as the non-interacting bandwidth exceeds a critical value around the order of the interaction energy scale.

{\it Region III}: we find that the HF ground state exhibit competing orders in this region which is located in the weak displacement field region with $w_0/w_1 \gtrsim 0.6$. In Fig.~\ref{fig:nu=-3_phase}(c) we plot the HF mean field quantities, e.g. $\mathcal{C}$, $S^\pm$ and $\mathrm{Ch}$, at $w_0/w_1 = 0.8$ with respect to $U$. When $U < 50\,\rm meV$ (region III), we see all the quantities are strongly oscillating. Moreover, we also notice strong size effect in this region, which can be seen by considering other system sizes at $w_0/w_1 = 0.8$, as discussed in \ref{app_sec:numerical_results}. In previous numerical studies in TBG systems \cite{KAN20a,SOE20,XIE20a} (which do not have the $U$ parameter), it has been shown that the translation symmetry of the TBG at filling $\nu=-3$ could be broken at large $w_0/w_1$ (typically $w_0/w_1 \gtrsim 0.7$). Therefore, we expect the ground states in region III not to be accurately captured by our HF calculation, which does not allow translation symmetry breaking. 
In \ref{app_subsec:-3}, we provide numerical evidence for a translation symmetry breaking phase via a modified HF calculation. 
Nevertheless, we provide some universal observation of our HF results in region III. In Fig.~\ref{fig:nu=-3_phase}(d), we plot $\nu_D$ and $\nu_{\rm TBG}=-3-\nu_D$ as a function of $w_0/w_1$ at $U = 0$. We find the Dirac electron filling $\nu_D$ is $0$ for $w_0/w_1<0.6$ (i.e., in region I), but begins to decrease as $w_0/w_1$ increases beyond $0.6$ (i.e., in region III). This indicates that electrons are transferred from the Dirac valence bands into the TBG flat bands in region III, making $\nu_D<0$ and $\nu_{\rm TBG}>-3$. For instance, when $w_0 = w_1$ at $U=0$, our HF calculation shows that $\nu_D \approx -1$ and $\nu_{\rm TBG} \approx -2$, the HF band structure of which is shown in Fig.~\ref{fig:nu=-3_band}(c). The Fermi level of this HF band structure in region III is far from the Dirac point energy, giving rise to a metal with large Fermi surfaces. Therefore, the ground states in region III are likely to be metals with competing orders, such as translation symmetry breaking.

In summary, at $\nu=-3$, we have identified three phases in three regions of Fig.~\ref{fig:nu=-3_phase}(a). In region I the ground state is almost a spin-valley polarized semimetal, in region II the ground state is a metal with little spin/valley polarization or intervalley coherence, while in region III the ground state may be a metal with competing orders.

\section{Numerical Results at filling factor \texorpdfstring{$\nu=-2$}{nu=-2}}\label{sec:nu_-2}
In this section, we study the HF results for TSTG at integer filling $\nu=-2$. By comparison, in TBG systems, the ground state at $\nu=-2$ at small $w_0$ and small bandwidth is given by an intervalley coherent insulator with Chern number $0$, which has been predicted in Refs. \cite{KAN18,BUL20,ZHA20,LIA20,XIE20a}. At large $w_0$, the TBG ground state may become a metal \cite{BER20b}. However, there is no evidence of translation symmetry breaking at $\nu=-2$ in TBG so far. Therefore, we also conjecture that translation breaking is less likely in the TSTG at $\nu=-2$, and thus regard our HF results as more reliable than at $\nu=-3$ in the large $w_0/w_1$ region.

\begin{figure*}[!htbp]
	\centering
	\includegraphics[width=0.75\linewidth]{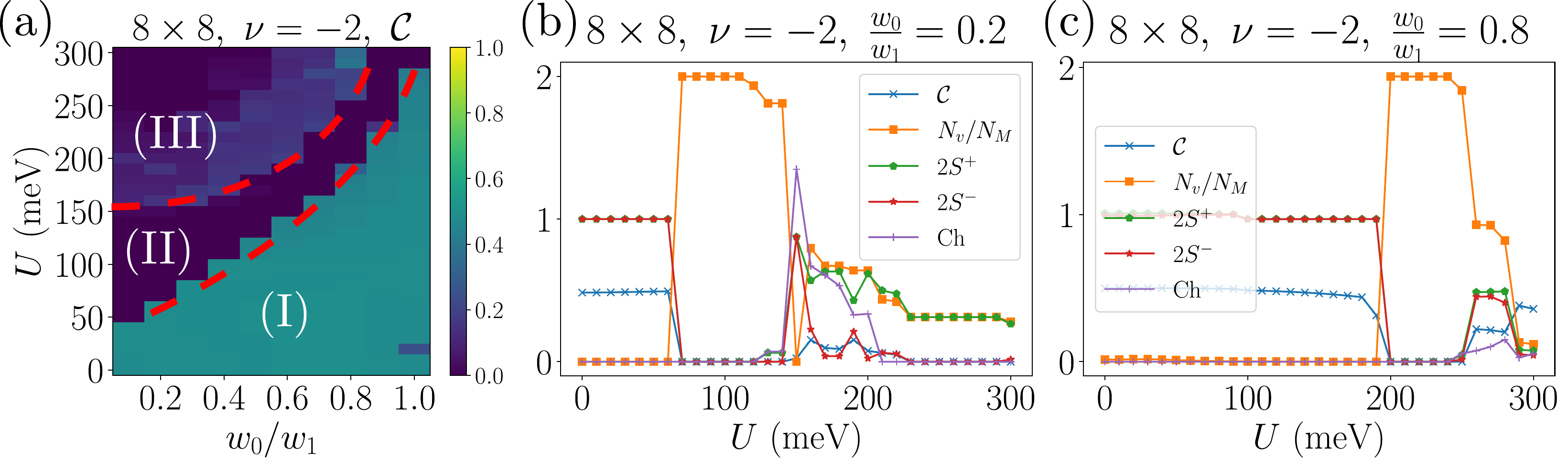}
	\caption{(a) The phase diagram at filling factor $\nu=-2$ obtained on a $8\times 8$ momentum lattice in the $(w_0, U)$ plane, and the color represents the intervalley coherence, which is defined in Eq. (\ref{eqn:def_vc}). (b) and (c) The displacement field dependence of physical quantities $\mathcal{C}, N_v, \mathrm{Ch}$ and $S^\pm$ on a $8\times 8$ at fixed $w_0/w_1 = 0.2$ (b) and $w_0/w_1 = 0.8$ (c). By considering the different HF parameters and band structure, we can define three different regions in the phase diagram, denoted I, II and III in (a).}
	\label{fig:nu_-2_phase}
\end{figure*}

Our HF results for TSTG at $\nu=-2$ identified 3 distinct regions I, II, III in the $w_0/w_1$ and $U$ parameter space as shown in Fig.~\ref{fig:nu_-2_phase}(a). In Fig.~\ref{fig:nu_-2_phase}(a), the color scale indicates the $\nu=-2$ ground state intervalley coherence $\mathcal{C}$, defined in Eq. (\ref{eqn:def_vc}) (note that this is different from the $\nu=-3$ phase diagram Fig.~\ref{fig:nu=-3_phase}(a), where valley polarization is shown by color, while intervalley coherence is near zero). Other HF quantities along certain constant $w_0/w_1$ line cuts are shown in Fig.~\ref{fig:nu_-2_phase}(b) and (c). From these quantities, we can see clear phase transitions between regions I and II, and between regions II and III. We now describe the HF ground states in the three regions, respectively.

\begin{figure*}[!htbp]
	\centering
	\includegraphics[width=\linewidth]{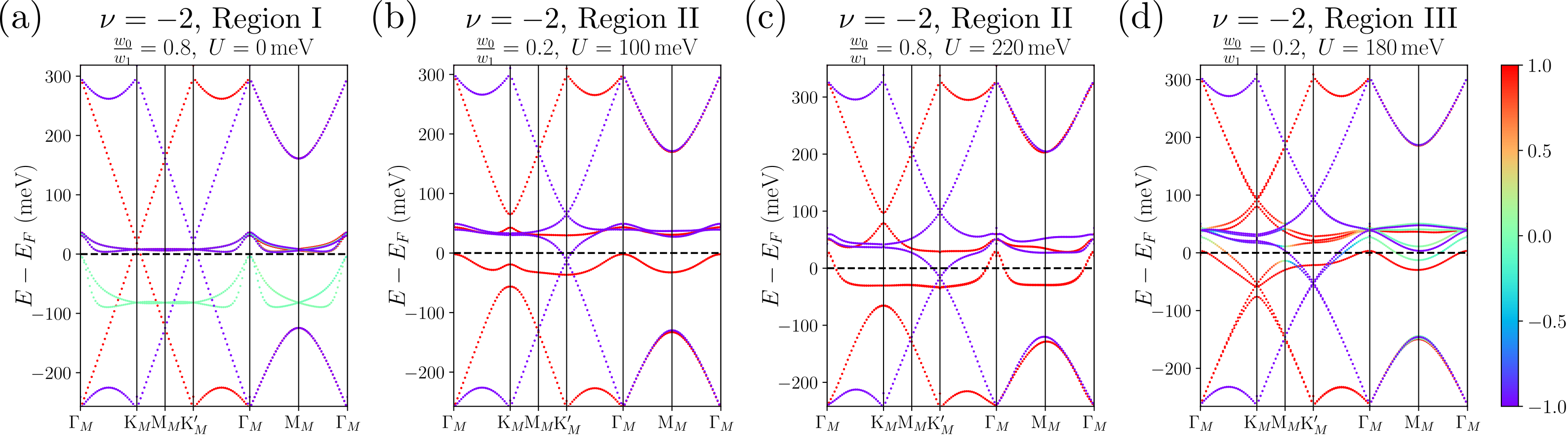}
	\caption{The HF band structure at $w_0/w_1 = 0.8$ for $U = 0$ (a), at $w_0/w_1 = 0.2$ for $U=100\,\rm meV$ (b), at $w_0/w_1=0.8$ for $U=220\rm\,meV$ (c) and at $w_0/w_1 = 0.2$ for $U = 180\,\rm meV$ (d) on a $10\times 10$ lattice at filling factor $\nu=-2$. The color represents the valley polarization $v_i(\vk)$ of each single body state defined in Eq. (\ref{eqn:def_singlevp}).}
	\label{fig:nu_-2_band}
\end{figure*}

{\it Region I}: this region contains the entire range of $w_0/w_1$ up to some $w_0$-dependent $U$ value. There we find $\mathcal{C}\approx 0.5$, $\mathrm{Ch}\approx 0$, $N_v/N_M\approx 0$ and $2S^\pm \approx 1$. This implies that there are two fully intervalley coherent flat bands occupied, which have the same spin and have zero total Chern number. This is the same as the TBG ground state at $\nu=-2$ filling. When $U=0$ in region I, the electron numbers in the Dirac sector and the TBG sector are conserved, respectively, and the HF ground state is almost the tensor product of the $\nu_{\rm TBG}=-2$ intervalley coherent TBG ground state predicted in Refs. \cite{KAN18,BUL20,ZHA20,LIA20,XIE20a} and the Dirac band ground state at charge neutrality $\nu_D=0$. A typical band structure in region I at $w_0/w_1=0.8$ and $U=0$ is given in Fig.~\ref{fig:nu_-2_band}(a), where the valley polarization values $v_i(\vk)$ of the occupied single body states (defined in Eq.~(\ref{eqn:def_singlevp})) are represented by color. One can see the valley polarization of the 2 occupied flat bands are approximately zero, consistent with an intervalley coherent state. The $\nu=-2$ ground state in region I is thus almost an intervalley coherent semimetal, in which the Dirac fermion is slightly doped away from the Dirac nodes. In particular, at $U>0$ where the Dirac and TBG sectors are hybridized, the gapless Dirac nodes are protected by a remaining anti-unitary symmetry $\mathcal{G}_\gamma$ ($\mathcal{G}_\gamma^2=1$), which is a combination of the $C_{2z}T$ and a relative intervalley phase rotation (see \ref{sec:Ggammasymmetry}).

{\it Region II}: the interlayer potential $U$ is intermediate, and we find $\mathcal{C}\approx 0, N_v/N_M \approx 2, \mathrm{Ch}\approx 0$ and $2S^\pm \approx 0$. This indicates that the ground state becomes a valley polarized state, and the two occupied TBG flat bands approximately have zero total Chern number.
We plot two typical HF band structures with different $w_0/w_1$ values in Fig.~\ref{fig:nu_-2_band}(b) and (c). In both of the band structure plots, the valley polarization values of occupied single body states in the flat bands are $v_i(\vk)\approx 1$. The occupied flat bands with smaller (larger) $w_0/w_1$ value has smaller (larger) band width. The band structures plots also show that there is a small electron pocket around $K_M'$ point, and a small hole pocket around $\Gamma_M$ point, indicating the system is almost a semimetal with a small Fermi surface.

{\it Region III}: the interlayer potential $U$ is further increased (e.g., $U \gtrsim 150\,\rm meV$ at $w_0/w_1=0.2$, and $U \gtrsim 280\,\rm meV$ at $w_0/w_1 = 0.8$), the valley polarization $N_v/N_M$ drops significantly, and the intervalley coherence slightly re-enters, as shown in Fig.~\ref{fig:nu_-2_phase}(b) and (c). In this case, the $\nu=-2$ TSTG enters a metallic phase with large Fermi surfaces. A HF band structure in this region is shown in Fig.~\ref{fig:nu_-2_band}(c). Similar to the region II phase at filling $\nu=-3$, the region III phase at $\nu=-2$ here is due to the change of flat bands into high energy dispersive Dirac bands near $K_M$ ($K_M'$) point of valley $+$ ($-$) at large $U$, yielding transitions into less valley polarized metal with large Fermi surfaces.

To summarize, the phase diagram at filling factor $\nu=-2$ can be roughly separated into three regions, as shown in Fig.~\ref{fig:nu_-2_phase}(a). In the small $U$ region I, the ground state is nearly an intervalley coherent semimetal and is adiabatically connected with the tensor product of the TBG ground state and a high velocity Dirac fermion at charge neutrality. In region II with intermediate $U$, the ground state is fully valley polarized and almost a semimetal. Finally, in region III with large $U$, the system enters a metal phase with partial valley polarization.

\section{Numerical Results at filling factor \texorpdfstring{$\nu=-1$}{nu=-1}}\label{sec:nu_-1}
In this section, we discuss the HF calculation results for TSTG at filling factor $\nu=-1$. We first recall that the ground state at $\nu=-1$ in nonchiral-nonflat TBG systems carries a Chern number $\nu_C = \pm 1$ and has two intervalley coherent bands and one valley polarized band occupied, as shown in Refs. \cite{LIA20,ZHA20}. Similar to filling $\nu=-3$ and $-2$, we expect the $\nu=-1$ TSTG ground state at small $w_0/w_1$ and $U=0$ to be the tensor product of the TBG ground state at this filling and the half filled Dirac fermion bands.

\begin{figure*}[!htbp]
	\centering
	\includegraphics[width=0.75\linewidth]{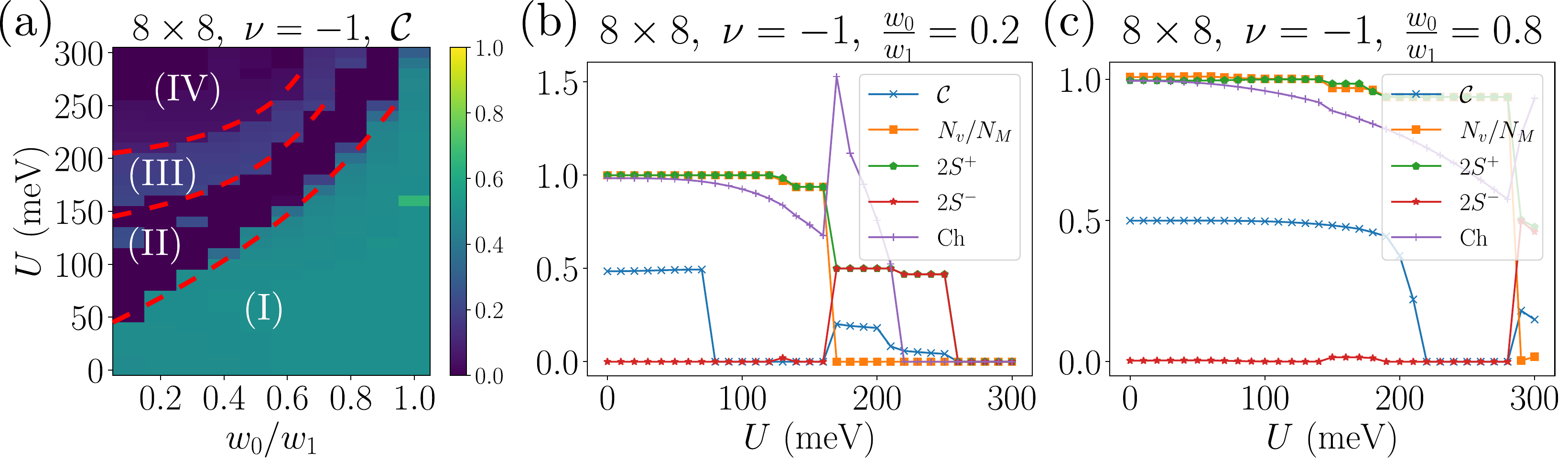}
	\caption{(a) The phase diagram at filling factor $\nu=-1$ obtained on a $8\times 8$ momentum lattice in the $(w_0, U)$ parameter space. The color represents the intervalley coherence $\mathcal{C}$. (b) and (c) The displacement field dependence of physical quantities $\mathcal{C}, N_v, \mathrm{Ch}$ and $S^\pm$ on $8\times 8$ at fixed $w_0/w_1 = 0.2$ (b) and $w_0/w_1 = 0.8$ (c).}
	\label{fig:nu_-1_phase}
\end{figure*}

The intervalley coherence $\mathcal{C}$ of the TSTG HF ground state at $\nu=-1$ as a function of $U$ and $w_0/w_1$ is represented by the color code in Fig.~\ref{fig:nu_-1_phase}(a). Other HF quantities at $w_0/w_1 = 0.2$ and $w_0/w_1 = 0.8$ are shown in Figs.~\ref{fig:nu_-1_phase}(b) and (c), respectively. Based on these quantities and the HF band structures, we are able to identify four different regions I, II, III and IV in $w_0/w_1$ and $U$ parameter space as shown in Fig.~\ref{fig:nu_-1_phase}(a). We now describe the HF mean field results in these regions.

\begin{figure*}[!htbp]
	\centering
	\includegraphics[width=\linewidth]{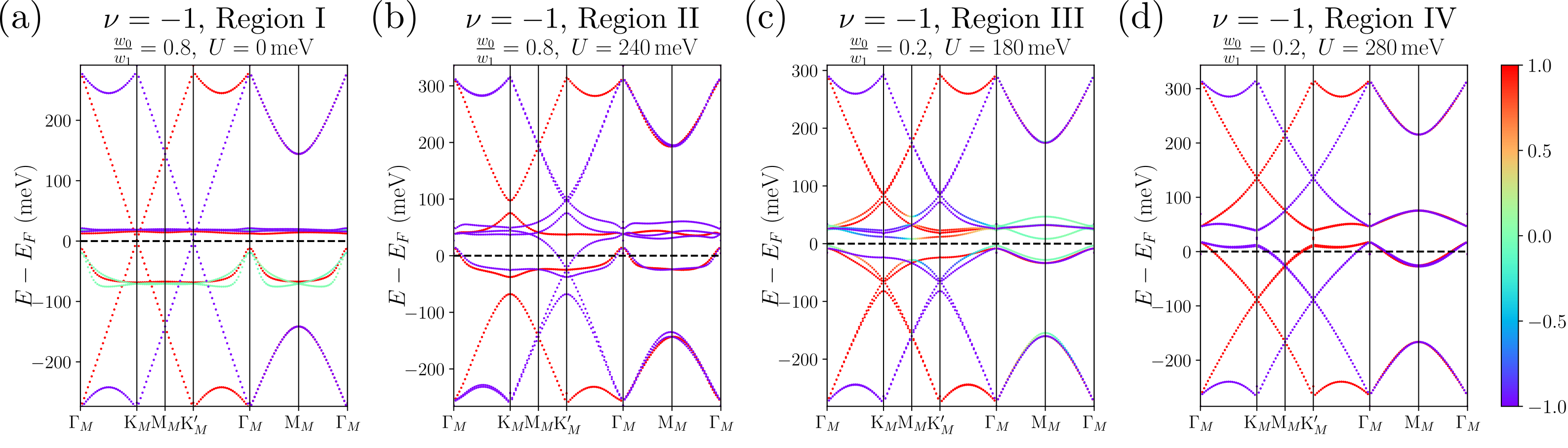}
	\caption{The HF band structure at $w_0/w_1 = 0.8$ for $U = 0$ in region I (a), $w_0/w_1 = 0.8$ for $U=240\rm\,meV$ in region II (b), $w_0/w_1 = 0.2$ for $U=180\rm\,meV$ in region III (c) and $w_0/w_1 = 0.2$ for $U = 280\,\rm meV$ in region IV (d) on a $10\times 10$ lattice at filling factor $\nu=-1$.  The color represents the valley polarization $v_i(\vk)$ of each single body state.}
	\label{fig:nu_-1_band}
\end{figure*}

{\it Region I}: this region encompasses the entire range of $w_0/w_1$, and up to certain $w_0/w_1$-dependent $U$ value, and we find that $\mathcal{C} \approx 0.5$, $\mathrm{Ch}\approx 1$, $N_v/N_M \approx 1$, $2S^+ \approx 1$ and $2S^- \approx 0$. The value of intervalley coherence indicates that among the three occupied TBG flat bands, two of them are intervalley coherent. These values also imply that the HF ground state at $U = 0$ is approximately equal to the tensor product of a $\nu_{\rm TBG} = -1$ intervalley coherent state \cite{ZHA20,LIA20} and a half-filled Dirac semimetal. Fig.~\ref{fig:nu_-1_band}(a) shows a typical HF band structure in region I at $w_0/w_1 = 0.8$ and $U = 0$. Among the three occupied flat bands in Fig.~\ref{fig:nu_-1_band}(a), two of them have zero valley polarization, while the other one is valley polarized, which agrees with the expected ground state in the TBG sector. The $U>0$ ground states of region I is adiabatically connected to the $U=0$ ground state. Therefore, region I is a semimetal phase with partially intervalley coherent flat bands. Similar to the $\nu=-3$ case, the gapless Dirac nodes at $U>0$ are protected by the $C_{2z}T$ symmetry within an empty valley-spin flavor, as shown in \ref{sec:Ggammasymmetry}.

{\it Region II}: the displacement field is intermediate in this region (e.g. $80{\rm\, meV} \lesssim U \lesssim 150 {\rm\, meV} $ at $w_0/w_1 = 0.2$, or $220{\rm\, meV} \lesssim U \lesssim 280 {\rm\, meV} $ at $w_0/w_1 = 0.8$). We find that the values of HF quantities $N_v/N_M$, $\mathrm{Ch}$ and $S^\pm$ are close to their values in region I. However, the intervalley coherence $\mathcal{C}$ vanishes abruptly in this region. We present a HF band structure at $w_0/w_1 = 0.8$ and $U=240\,\rm meV$ in Fig.~\ref{fig:nu_-1_band}(b). The valley polarization of the three occupied flat bands are $v_i(\vk) \approx \pm 1$. The band structure also shows small electron pocket around $K_M'$ point, and hole pocket around $\Gamma_M$ point, which means the system is also almost a semimetal without intervalley coherence.

{\it Region III}: the displacement field in this region (which is $160{\rm\,meV} \lesssim U \lesssim 220 {\rm\,meV}$ at $w_0/w_1 = 0.2$) is stronger than that in the region II. We find the valley polarization $N_v/N_M$ drops to zero, and the intervalley coherence slightly increases to $\mathcal{C}\approx 0.2$, as shown in Fig.~\ref{fig:nu_-1_phase}(b). The HF band structure in this region, which can be found in Fig.~\ref{fig:nu_-1_band}(c), shows that there is a direct band gap around the Fermi level. Therefore, we identify an insulating state at $\nu=-1$ filling with a non-zero displacement field in region III. Such a phase does not occur at $\nu=-3$ or $\nu=-2$ fillings. 

{\it Region IV}: the displacement field is further increased (e.g., $U \gtrsim 220\,\rm meV$ at $w_0/w_1 = 0.2$). Similar to the strong field phase at $\nu=-3$ and $\nu=-2$, the increased bandwidth of the non-interacting dispersion becomes comparable to or larger than the strength of the Coulomb interaction. 
Therefore, the electrons will first occupy the low energy states around $K_M$ and $K_M'$ at $E - E_F \approx -90\,\rm meV$ which can be seen in Fig.~\ref{fig:nu_-1_band}(d).
A large Fermi surface can also be observed in the band structure, which implies that region IV is a metallic phase. Both the valley polarization $N_v/N_M$ and the intervalley coherence $\mathcal{C}$ are nearly zero in this region.

In summary, there are four phases in the phase diagram at filling factor $\nu=-1$. When the displacement field is close to zero, i.e., in region I, the ground state is an intervalley coherent semimetal. As the displacement field increases into region II, the ground state becomes a semimetal without intervalley coherence. When the field further increases into region III, the HF band structure becomes gapped, and therefore the ground state is an insulator. We note that this phase does not occur at fillings $\nu=-3$ and $\nu=-2$. Finally in region IV with the strongest displacement field, the system becomes a metal, similar to the filling factors $\nu=-3$ and $\nu=-2$.

\section{Numerical Results at filling factor \texorpdfstring{$\nu=0$}{nu=0}}\label{sec:nu_0}
Lastly, we present our HF calculation results for TSTG at filling factor $\nu=0$. In comparison, in the TBG system the ground state at $\nu=0$ is an insulator state with four occupied intervalley coherent bands and zero total Chern number \cite{ZHA20,LIA20}. Similar to other integer fillings, we expect the ground state of TSTG at $\nu=0$ and $U=0$ to be the tensor product of a TBG intervalley coherent insulator ground state and half filled Dirac semimetal.

\begin{figure*}[!htbp]
	\centering
	\includegraphics[width=\linewidth]{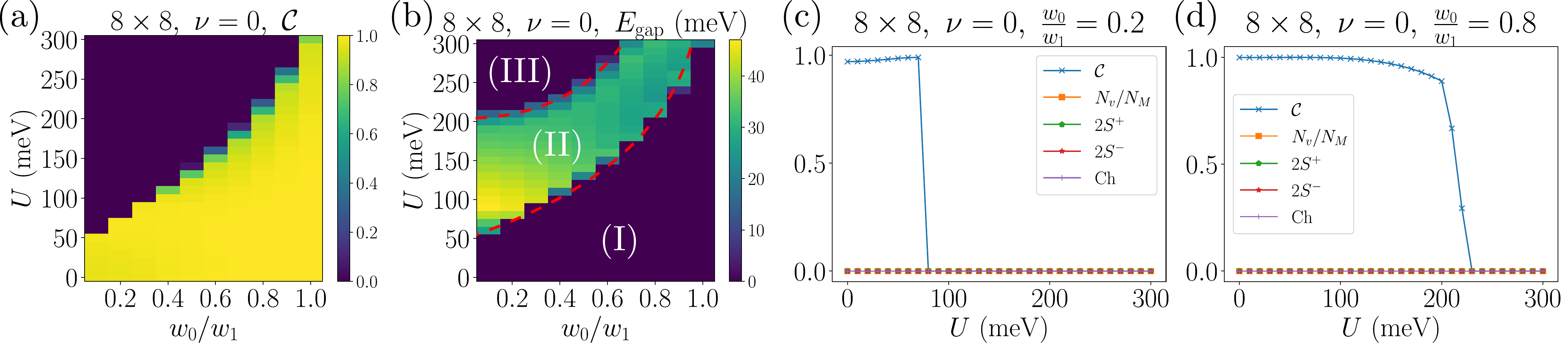}
	\caption{Phase diagrams at filling factor $\nu=0$. (a) The two dimensional phase diagram on $8\times 8$ momentum lattice in $(w_0, U)$ parameter space. It can be seen that in the weak $U$ phase, the intervalley coherence $\mathcal{C} \approx 1$ shows that there are four occupied intervalley coherent bands. (b) The energy gap along the high symmetry lines as a function of $w_0/w_1$ and $U$. Here we use the method discussed in \ref{app_subsec:analysis} to obtain the Hartree-Fock Hamiltonian along the high symmetry lines, therefore we are able to estimate the energy gap from the $8\times 8$ lattice. (c) and (d) The displacement field dependence of several quantities $\mathcal{C}, N_v, S^{\pm}$ and $\mathrm{Ch}$ on $8\times 8$ lattice with $w_0/w_1 = 0.2$ (c) and $w_0/w_1 = 0.8$ (d).}
	\label{fig:phase_0}
\end{figure*}

\begin{figure*}
	\centering
	\includegraphics[width=\linewidth]{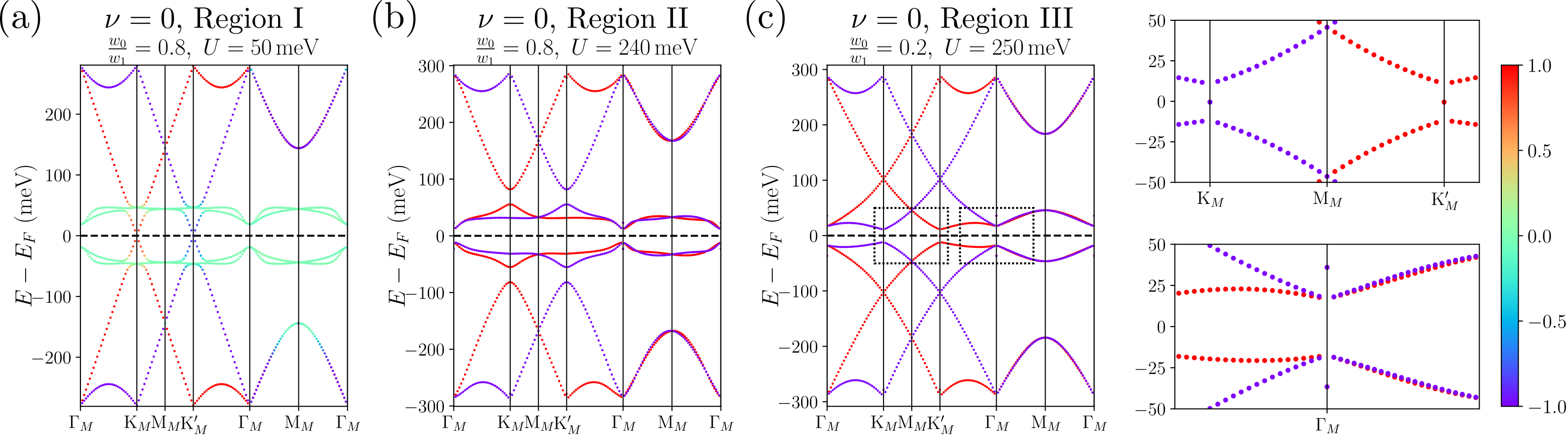}
	\caption{(a-c) The HF band structure on a $10\times 10$ lattice at filling factor $\nu=0$ at $w_0/w_1 = 0.8$ for $U=50\,\rm meV$ in region I (a), at $w_0/w_1 = 0.8$ for $U=200\rm\,meV$ in region II (b) and at $w_0/w_1 = 0.2$ for $U=250\,\rm meV$ in region III (c), respectively.  The color stands for the valley polarization $v_i(\vk)$ of each single body state. The zoom in band structures around $K_M$, $K_M'$ and $\Gamma_M$ points in the dashed boxes in subfigure (c) are also shown. It is visible that the HF band structure is discontinuous at these points, and it is also gapless at $K_M$ and $K_M'$ points.}
	\label{fig:band_0}
\end{figure*}

In Fig.~\ref{fig:phase_0}(a), we show the intervalley coherence $\mathcal{C}$ in the $w_0/w_1$ and $U$ parameter space at $\nu=0$. By using the same method as the HF band structure along the high symmetry lines, which is discussed in \ref{app_subsec:analysis}, we can estimate the HF Hamiltonian $\mathcal{H}^{HF}(\vk)$ at any momenta not included in the momentum lattice employed in our HF iterations.
Thus, the energy gap around the Fermi level along the high symmetry lines as a function of $w_0/w_1$ and $U$ can be calculated, which is shown in Fig.~\ref{fig:phase_0}(b).
We are able to identify three different regions I, II and III in the $w_0/w_1$ and $U$ parameter space, based on the valley coherence $\mathcal{C}$ and the energy gap. Other HF quantities at fixed $w_0/w_1 = 0.2$ and $w_0/w_1 = 0.8$ are also shown in Figs.~\ref{fig:phase_0}(c) and (d). We now use these quantities to describe the HF ground states in these regions.

{\it Region I}: this region is in the low displacement field regime, and we find the values of the HF quantities are $\mathcal{C}\approx 1$, $N_v/N_M \approx 0$, $\mathrm{Ch} \approx 0$ and $S^\pm \approx 0$. The value of the intervalley coherence $\mathcal{C}\approx 1$ shows that there are four occupied intervalley coherent bands and have zero total Chern number. Therefore, these values indicates that the HF ground state at $U=0$ can be well approximated by the tensor product of the insulating intervalley coherent TBG ground state at $\nu_{\rm TBG} = 0$ predicted in Refs.~\cite{ZHA20,LIA20}, and the ground state at $U > 0$ in region I is adiabatically connected to this tensor product state.
A typical HF band structure can be found in Fig.~\ref{fig:band_0}(a). The occupied flat bands have zero valley polarization, which agree with the intervalley coherent ground state. Therefore, the $\nu=0$ TSTG ground state is an intervalley coherent semimetal. As we show in \ref{sec:Ggammasymmetry}, the gapless Dirac nodes of this phase at $U>0$ is protected by a remaining anti-unitary symmetry $\mathcal{G}_\gamma$ ($\mathcal{G}_\gamma^2=1$), which is a combination of the $C_{2z}T$ and a relative phase rotation between the two valleys.

{\it Region II}: the displacement field is intermediate, and as seen in both Figs.~\ref{fig:phase_0}(b) and (c), the intervalley coherence $\mathcal{C}$ drops to zero in this region. 
Other HF parameters, including $N_v/N_M$, $\mathrm{Ch}$ and $S^\pm$ are equal to zero in region II. We also notice that there is another state with non-zero $\mathrm{Ch}$ values in region II, whose energy increment from the state with $\mathrm{Ch} = 0$ is within the machine precision when the parameters are around the boundary between regions II and III, showing a possible competing order. 
A typical HF ground state band structure in region II is shown in Fig.~\ref{fig:band_0}(b). 
The occupied flat bands have valley polarization values $v_i(\vk) \approx \pm1$, and there is a large direct gap around the Fermi level. This result indicates that region II is an insulating phase, akin to the region III at $\nu=-1$ filling.

{\it Region III}: here the interlayer potential $U$ is stronger, and the HF quantities $\mathcal{C}$, $N_v/N_M$, $\mathrm{Ch}$ and $S^\pm$ in this large $U$ region are the same as in region II. However, the band structures undergo an abrupt transition. As discussed in previous sections, the bandwidth of the low energy bands become large when $U$ is large, and therefore the effect of the interaction will be suppressed by the kinetic energy. A HF band structure in this region is shown in Fig.~\ref{fig:band_0}(c). The HF mean field band structure is similar to the non-interacting band dispersion, which has gapless Dirac points at $K_M$ and $K_M'$ points. The discontinuous dispersions in Fig.~\ref{fig:band_0}(c) at $K_M$ and $K_M'$ (see the zoom-in plots in Fig.~\ref{fig:band_0}(c)) are due to neglecting of the higher bands in the TSTG projected Hamiltonian, as explained in \ref{app_sec:hamiltonian}. 
From the HF band structure, we conclude that the large displacement field phase in region III at filling $\nu=0$ becomes a semimetal.

To summarize, there are three phases at filling factor $\nu=0$, as shown in Fig.~\ref{fig:phase_0}(b). Within the small $U$ region I, the HF ground state is an intervalley coherent semimetal. In region II with an intermediate $U$, the ground state is an insulator without intervalley coherence or valley polarization. Finally, in region III with a large $U$, the system becomes a semimetal with no valley polarization or intervalley coherence.

\section{Conclusion}

Through projected Hartree-Fock mean field calculations, our work unveiled the close relationship between TSTG at weak displacement field and TBG systems at integer fillings $\nu=-3, -2, -1$ and $0$. We show that at weak displacement fields, the TSTG ground states at integer fillings are almost semimetal states which are in the same phase as the tensor product of the TBG ground states at the same filling and a Dirac semimetal. Beyond the phases inherited from the TBG physics, the TSTG undergoes transitions into large Fermi surface metals or insulators as the displacement field increases. Besides, we generically find that the displacement field destabilizes the intervalley coherence of the flat bands.

For filling factor $\nu=-3$, we found three regions of different phases. At small displacement field, the TSTG ground state is a semimetal with an occupied spin-valley polarized flat band when $w_0/w_1 \lesssim 0.6$. At large displacement fields, the TSTG undergoes a first order phase transition into a metallic phase with large Fermi surfaces and zero valley polarization, due to the enlarged band width. When $w_0/w_1 \gtrsim 0.7$ and $U = 0$, we observed that the electrons transfer from the Dirac cones into the TBG flat bands, which yields a metallic phase with competing orders. Moreover, similar to pure TBG systems at $\nu=-3$, it is possible to have translation symmetry breaking, some evidence of which is shown in \ref{app_subsec:-3}. We leave the study of translation breaking TSTG phases in the future.

For filling factors $\nu=-2, -1$ and $0$, our HF numerical results show that the TSTG ground states at weak displacement fields are semimetals with intervalley coherent flat bands occupied.
At intermediate displacement fields, the intervalley coherence drops abruptly to zero, signaling a transition into phases without intervalley coherence, which are either semimetals (at $\nu=-2$ and $-1$) or insulators (at $\nu=-1$ and $\nu=0$). With a stronger displacement field, the dispersive energy bands will have bandwidths exceeding the energy scale of Coulomb interactions, which leads the system into a metallic state with little valley polarization or intervalley coherence.

Our work reveals two roles of the displacement field in TSTG with Coulomb interaction: destabilizing the intervalley coherence (if any), and increasing the flat band width and thus weakening the correlations due to interactions. Our results may provide guidance to the analytical studies of TSTG ground states in the future.

\begin{acknowledgments}
We are grateful to Zhi-Da Song for previous collaboration on related works and enlightening discussions. We thank Oskar Vafek, Pablo Jarillo-Herrero, and Dmitri Efetov for fruitful discussions. This work was supported primarily by the ONR No. N00014-20-1-2303, the Schmidt Fund for Innovative Research, Simons Investigator Grant No. 404513, the Packard Foundation, the Gordon and Betty Moore Foundation through Grant No. GBMF8685 towards the Princeton theory program, and a Guggenheim Fellowship from the John Simon Guggenheim Memorial Foundation. Further support was provided by the NSF-EAGER No. DMR 1643312, NSF-MRSEC No. DMR-1420541 and DMR-2011750, DOE Grant No. DE-SC0016239 , Gordon and Betty Moore Foundation through Grant GBMF8685 towards the Princeton theory program, BSF Israel US foundation No. 2018226, and the Princeton Global Network Funds. B.L. acknowledges support from the Alfred P. Sloan Foundation.
\end{acknowledgments}
\noindent
{\it Note added.}---During the final preparation of this manuscript, a recent preprint Ref.~\cite{christos2021correlated} appeared, with numerical results consistent with ours at even integer fillings.

\bibliographystyle{apsrev4-2}
\bibliography{TTGbib.bib}

\begin{widetext}

\tableofcontents

\beginsupplement

\section{Projected Hamiltonian}\label{app_sec:hamiltonian}

In order to simplify the numerical calculation, we project the Hamiltonian into the low energy bands. We start with solving the Hamiltonian in mirror symmetric (TBG fermions) and anti-symmetric (Dirac fermions) sectors in the absence of external displacement field.
By diagonalizing the TBG Hamiltonian $h^{(\eta)}(\vk)$ and the Dirac Hamiltonian $h^{D, \eta}(\vk)$, we obtained the band structure $\varepsilon_{m,\eta}^{\hat{f}}(\vk)$ and the single body wavefunctions $u^{\hat{f}}_{\mathbf{Q}\alpha,m\eta}(\vk)$, where $\hat{f} = \hat{c}, \hat{b}$. 
The single body wavefunction of TBG fermions can be gauge fixed as in Ref. \cite{XIE20a}, and thus the $C_{2z}T$ sewing matrix in the symmetric sector is identity. Therefore, the electron operators in energy band basis can be defined as $\hc^\dagger_{\vk,m,\eta,s} = \sum_{\mathbf{Q}\alpha}u^{\hat{c}}_{\mathbf{Q}\alpha,m\eta}(\vk)c^\dagger_{\vk,\mathbf{Q},\eta,\alpha,s}$. Moreover, by using this gauge fixing choice, we obtain the following electron operators $\hat{d}^\dagger_{\vk,e_Y,\eta,s}$ and its corresponding single body wavefunction $u^{\hat{d}}_{\mathbf{Q}\alpha,\eta e_Y}(\vk)$, which can form a band with Chern number $e_Y = \pm1$:
 \begin{align}
 	\hat{d}^\dagger_{\vk, e_Y, \eta, s} &= \frac{1}{\sqrt2}\left(\hc^\dagger_{\vk,1,\eta,s} + ie_Y \hc^\dagger_{\vk,-1,\eta,s} \right)\,\\
 	u^{\hat{d}}_{\mathbf{Q}\alpha,e_Y\eta}(\vk) &= \frac{1}{\sqrt2}\left(u^{\hat{c}}_{\mathbf{Q}\alpha, 1\eta}(\vk) + ie_Y u^{\hat{c}}_{\mathbf{Q}\alpha, -1\eta}(\vk) \right)\,,
 \end{align}
Indeed, these states are eigenstates of Pauli $y$ matrix $\zeta_y$ in energy band basis. 
For TBG fermions, we only keep the two bands which are closest to the charge neutral point, which are equivalent to the two narrow bands in TBG per spin and valley. The projected kinetic Hamiltonian for the TBG fermions is:
\begin{equation}
	H_{\rm TBG} = \sum_{\vk,m,\eta,s}\varepsilon^{\hc}_{m\eta}(\vk)\hc^\dagger_{\vk,m,\eta,s}\hc_{\vk,m,\eta,s}\,.
\end{equation}

For Dirac fermions, we also keep the two bands which are closest to the charge neutrality per spin and valley. As shown in Eq.~(\ref{eqn:ham_dirac}), the Hamiltonian of the Dirac fermion is block diagonal in $\mathbf{Q}$ basis. Therefore, for a general point in the MBZ, the wavefunction of a Dirac fermion state $u^{\hb}_{\mathbf{Q}\alpha,m\eta}(\vk) \neq 0$ for only one $\mathbf{Q}$. Since the wavefunction in the valley $\eta = -$ can be obtained by performing a $C_{2z}$ transformation to the wavefunctions in the valley $\eta = +$, we only discuss $\eta = +$ here (the spin degree of freedom can also be dropped). As seen in Fig.~\ref{fig:mbz_dirac}(a), there are slices of Dirac cones sitting on the three $K_M$ points \cite{TSTGI}, which are labeled by three different colors. For a given momentum $\vk$, there are two states which are closest to the charge neutrality, one has positive energy $+v_F|\vk - \mathbf{K}_M|$ and the other one has negative energy $-v_F|\vk - \mathbf{K}_M|$. Both of the states' wavefunction have non-zero components $u^{\hb}_{\mathbf{Q}\alpha,m +}(\vk)$ when $\mathbf{Q}$ is equal to its closest $K_M$ point. For example, the wavefunction of the Dirac fermion at momentum $\vk_1$ shown in Fig.~\ref{fig:mbz_dirac} has only non-zero components when $\mathbf{Q}$ is the $K_M$ point labeled by blue.

However, the distances between a momentum point along the $\Gamma_M$-$K_M'$ lines and two $K_M$ points are the same. For example, the $\vk_2$ point in Fig.~\ref{fig:mbz_dirac}(a) is at equal distance from the red and green $K_M$ points. This leads to some ambiguity in the choice of $u^{\hb}_{\mathbf{Q}\alpha,m+}(\vk)$. 
As seen in Fig.~\ref{fig:mbz_dirac}, there are three $\Gamma_M$-$K_M'$ lines in the MBZ. We choose the single body wavefunction, such that $u^{\hb}_{\mathbf{Q}\alpha,m+}(\vk) \neq 0$ only when $\mathbf{Q}$ is the $K_M$ point with the same color as the corresponding $\Gamma_M$-$K_M'$ line. As an example, the $\mathbf{Q}$ index of the only non-zero components of $u^{\hb}_{\mathbf{Q}\alpha,m+}(\vk_2)$ is equal to the $K_M$ point labeled by red. 
Thus, the wavefunctions along these high symmetry lines satisfy the $C_{3z}$ symmetry.
Moreover, at $\Gamma_M$ and $K_M'$ points, these bands are three-fold degenerate. At these points, we choose the state whose $C_{3z}$ eigenvalue is $1$. Indeed, choosing real $C_{3z}$ eigenvalues at $\Gamma_M$ and $K_M'$ leads to a more accurate approximation by the projected Hamiltonian when $U>0$ at these points as shown later in this appendix. Therefore, our choice of $u^{\hb}_{\mathbf{Q}\alpha,m\eta}(\vk)$ will satisfy the $C_{3z}$ symmetry. Similar to the TBG fermion, the kinetic Hamiltonian of Dirac fermions after the projection can be written as:
\begin{equation}\label{eqn:h_D_appendix}
	H_D = \sum_{\vk,m=\pm1,\eta,s}\varepsilon^{\hb}_{m\eta}(\vk)\hb^\dagger_{\vk,m,\eta,s}\hb_{\vk,m,\eta,s}\,,
\end{equation}
in which $\hb^\dagger_{\vk,m,\eta,s} = \sum_{\mathbf{Q}\alpha}u^{\hb}_{\mathbf{Q}\alpha,m\eta}(\vk)\hb^\dagger_{\vk,\mathbf{Q},\eta,\alpha,s}$, and the dispersion is given by $\varepsilon_{m\eta}^{\hb}(\vk) = m v_F |\vk - \mathbf{K}_M|$, where $\mathbf{K}_M$ is the closest to the $\vk$ point.

\begin{figure}
	\centering
	\includegraphics[width=\linewidth]{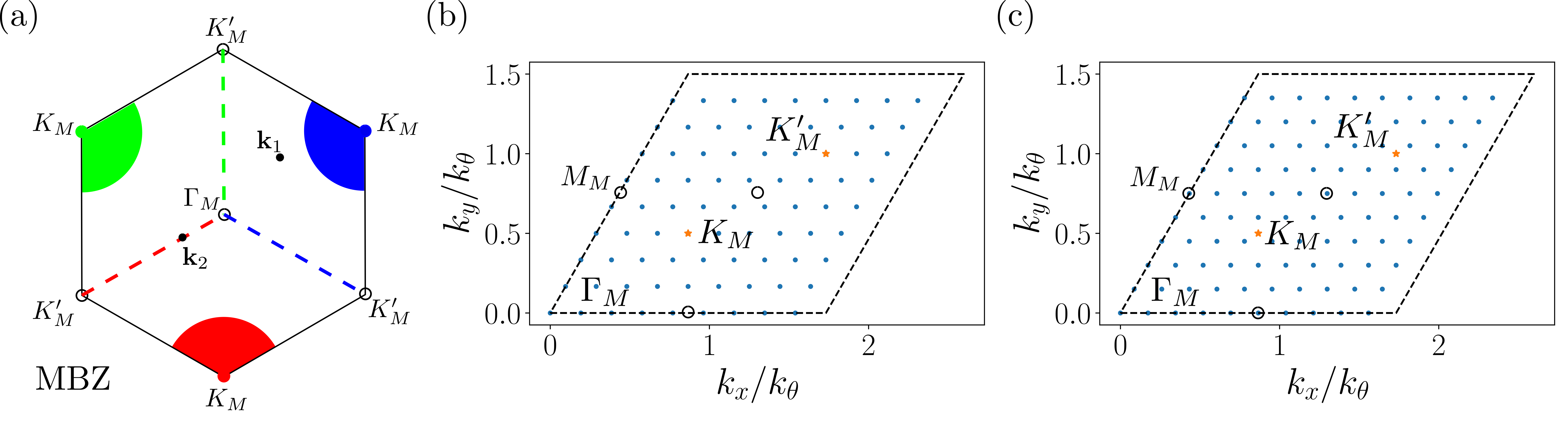}
	\caption{(a) The high symmetry lines in MBZ. The Dirac cones in valley $\eta = +$ are located on $K_M$ points, which are represented by the sectors with colors. For the momentum along $\Gamma_M$-$K_M'$ lines, we choose the wavefunction of the Dirac fermion $u^{\hb}_{\mathbf{Q}\alpha,m\eta}(\vk) \neq 0$, only when $\mathbf{Q}$ is the $K_M$ point, which is labeled by the same color as momentum $\vk$. For example, the point $\vk_2$ along the red high symmetry line implies choosing for $\mathbf{Q}$ the red $K_M$ point. This wavefunction choice preserves $C_{3z}$ symmetry. (b) and (c) The $9\times 9$ (b) and $10\times 10$ (c) momentum lattices in the first MBZ. Here we are using a parallelogram as the reciprocal unit cell. The high symmetry points are also labeled in these figures. Note that only lattices $3m\times 3m$ include the Dirac points $K_M$ and $K_M'$, such as Fig.~\ref{fig:mbz_dirac}(b).}
	\label{fig:mbz_dirac}
\end{figure}

Thus, the projected non-interacting Hamiltonian is given by:
\begin{equation}
	H_0 \big{|}_{U=0} = H_{\rm TBG} + H_{D} = \sum_{\vk}\sum_{\hat{f}=\hat{c},\hat{b}}\sum_{m=\pm1 }\sum_{\eta, s}\varepsilon_{m\eta}^{\hat{f}}(\vk) \hf^\dagger_{\vk,m,\eta,s} \hf_{\vk, m, \eta, s}\,,
\end{equation}
in which $\hf^\dagger_{\vk, m,\eta, s} = u^{\hat{f}}_{\mathbf{Q}\alpha, m\eta}(\vk)\hf^\dagger_{\vk,\mathbf{Q},\eta, \alpha, s}$ is the electron operator in energy band basis. Next, we project the displacement field term $\hat{H}_U$ into the Hilbert space spanned by the low energy states at $U=0$:
\begin{equation}
	H_U = \frac{U}{2}\sum_{\vk, \eta, s}\sum_{m=\pm1}\sum_{n=\pm1}N^\eta_{mn}(\vk)\left(\hb^\dagger_{\vk, m, \eta, s}\hc_{\vk, n, \eta, s} + {\rm h.c.}\right)\,,
\end{equation}
where the displacement field overlap matrices are defined by:
\begin{equation}
	N^{\eta}_{mn}(\vk) = \sum_{\mathbf{Q}\in\mathcal{Q}_\eta, \alpha} u^{\hat{b}*}_{\mathbf{Q}\alpha, m\eta}(\vk)u^{\hat{c}}_{\mathbf{Q}\alpha, n\eta}\,.
\end{equation}
The projected non-interacting Hamiltonian is then given by the summation of these terms:
\begin{equation}\label{eqn:def_projected_hamiltonian}
	H_0 = \sum_{\vk}\sum_{\hat{f}=\hat{c},\hat{b}}\sum_{m=\pm1 }\sum_{\eta, s}\varepsilon_{m\eta}^{\hat{f}}(\vk) \hf^\dagger_{\vk,m,\eta,s} \hf_{\vk, m, \eta, s} + \frac{U}{2}\sum_{\vk, \eta, s}\sum_{m=\pm1}\sum_{n=\pm1}N^\eta_{mn}(\vk)\left(\hb^\dagger_{\vk, m, \eta, s}\hc_{\vk, n, \eta, s} + {\rm h.c.}\right)\,.
\end{equation}
For convenience, this quadratic Hamiltonian can also be written as the following form:
\begin{align}
	H_0 &= \sum_{\vk,\hf\hf',\eta\eta',ss'}\mathcal{H}^{(0)}_{\hat{f}m\eta s,\hat{f}'n\eta's'}(\vk)\hf^\dagger_{\vk,m,\eta,s}\hf'_{\vk,n,\eta',s'}\\
	\mathcal{H}^{(0)}_{\hat{f}m\eta s,\hf'n\eta's'}(\vk) &= \varepsilon_{m,\eta}^{\hat{f}}(\vk) \delta_{\hat{f}\hf'}\delta_{mn}\delta_{\eta\eta'}\delta_{ss'} + \frac{U}{2}(N^{\eta}_{mn}(\vk)\delta_{\hat{f}\hat{b}}\delta_{\hf'\hat{c}} + N^{\eta*}_{mn}(\vk)\delta_{\hat{f}\hat{c}}\delta_{\hf'\hat{b}})\delta_{\eta\eta'}\delta_{ss'}\,.\label{eqn:kinetic_k}
\end{align}

\begin{figure}[!htbp]
    \centering
    \includegraphics[width=\linewidth]{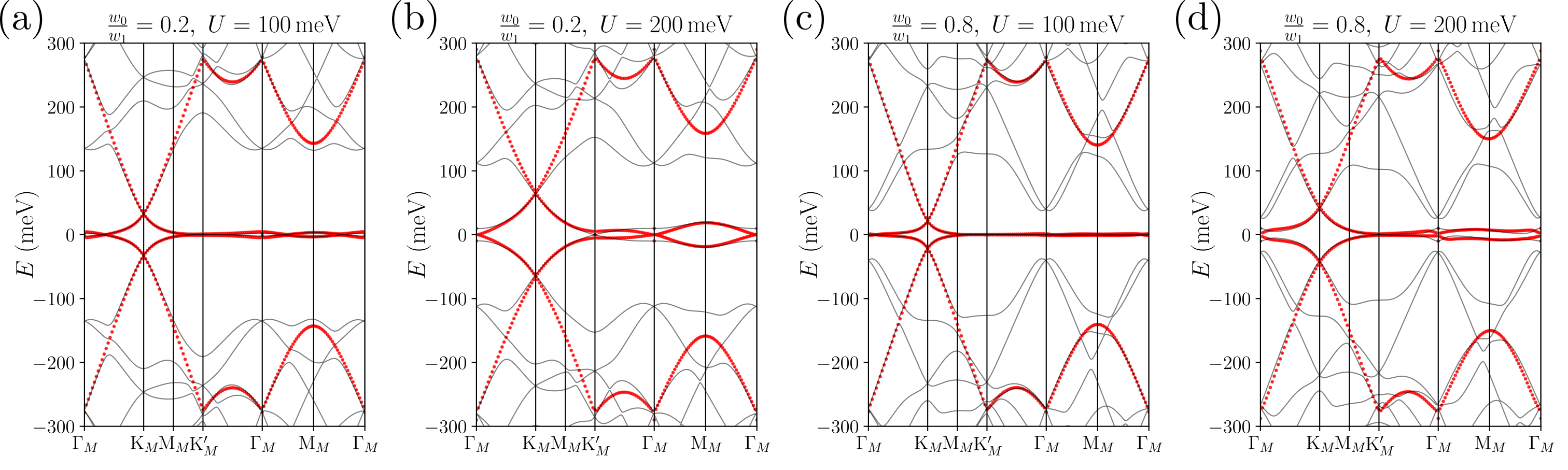}
    \caption{The dispersion of the projected kinetic Hamiltonian in Eq.~(\ref{eqn:kinetic_k}), represented by red dots. In these plots, only the $\eta = +$ valley bands are shown. The red dots represent the band structure of the projected Hamiltonian in Eq.~(\ref{eqn:def_projected_hamiltonian}) and black solid lines represent the dispersion of the BM model in Eq.~(\ref{eqn:bm_hamiltonian}).}
    \label{fig:projected}
\end{figure}

\begin{figure}
    \centering
    \includegraphics[width=\linewidth]{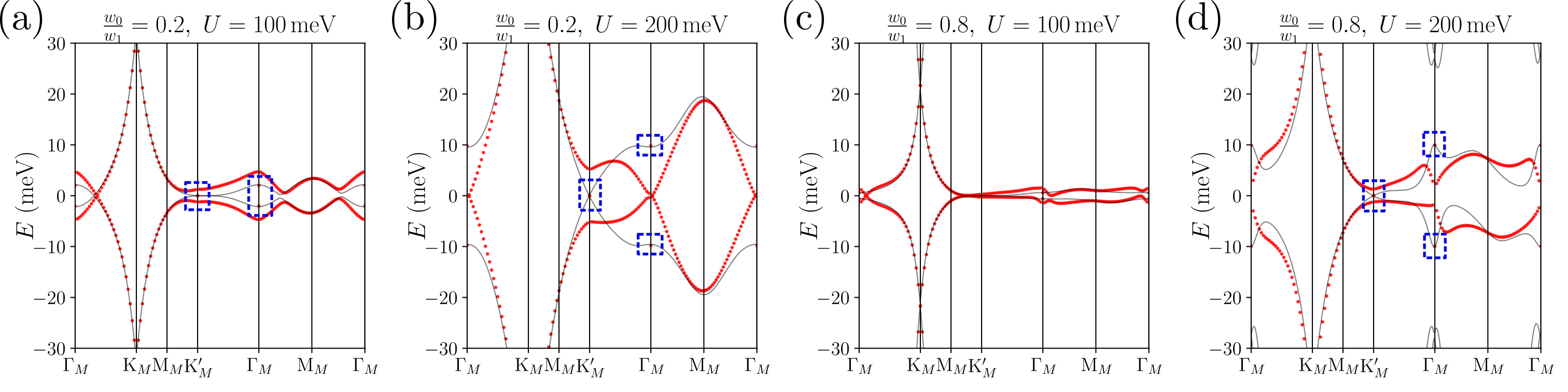}
    \caption{The dispersion of the projected kinetic Hamiltonian. The data is the same as in Fig.~\ref{fig:projected} but we provide a zoom close the zero energy. The red dots represents the dispersion of the projected Hamiltonian, and the black solid line represents the band structure of the BM model. There, we clearly see the discontinuities in the projected kinetic Hamiltonian at the $\Gamma_M$ and $K_M'$ with isolated energies (dots in the blue boxes) for the active bands. Note that the discontinuities are barely visible for (c), so we do not show any boxes there.}
    \label{fig:projected_zoomin}
\end{figure}

\begin{figure}
    \centering
    \includegraphics[width=\linewidth]{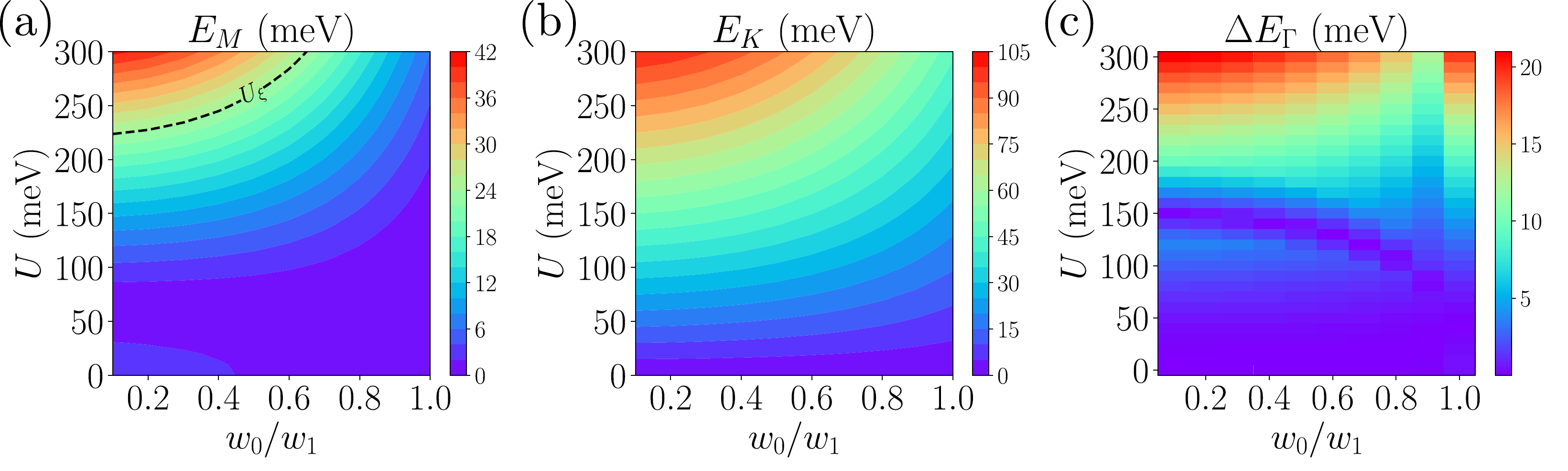}
    \caption{(a) The flat band energy at $M_M$ point as a function of $w_0/w_1$ and $U$. This value measures the band width of the flat bands. The value of the energy at $M_M$ point along black dashed line is equal to $U_\xi \approx 24\,\rm meV$, which measures the strength of the Coulomb interaction. (b) The flat band energy at $K_M$ point as a function of $w_0/w_1$ and $U$. It describes the energy shift of the Dirac cones at $K_M$ point, and it also captures the band width. (c) The energy jump of the flat bands at $\Gamma_M$ point. It measures the discontinuity of the projected Hamiltonian at high symmetry points.}
    \label{fig:bandwidth}
\end{figure}

The dispersion of the projected kinetic Hamiltonian $\mathcal{H}^{(0)}(\vk)$ with different $w_0/w_1$ and $U$ values are shown in Fig.~\ref{fig:projected}. In these non-interacting band structure plots, we find that the projected Hamiltonian can capture well the Dirac cone shift with non-zero $U$ around $K_M$ point. However, as shown in Fig.~\ref{fig:projected}(d), we can also find that the energy of the second bands of the TBG fermions, which are not included in the projected Hamiltonian, are comparable to the shifted Dirac cones in the projected bands when $w_0/w_1$ and $U$ are large. Thus, the HF results obtained in the large $w_0/w_1$ and $U$ region will be less reliable.

As shown in Fig.~\ref{fig:projected_zoomin}, the projected band structure is discontinuous at $\Gamma_M$ and $K_M'$ when the displacement field is strong (as shown in the blue dashed boxes). In particular, we see that the projected band energies at $\Gamma_M$ and $K_M'$ (red dots in the blue dashed boxes) agree quite well with the unprojected band energies (black lines).
As discussed in the paragraph above Eq.~(\ref{eqn:h_D_appendix}) and Fig.~\ref{fig:mbz_dirac}(a), the single-body wavefunction of the Dirac fermion at $\Gamma_M$ and $K_M'$ are chosen such that the states are $C_{3z}$ symmetric. Selecting the linear combination with the $C_{3z}$ eigenvalue $+1$ provides the most accurate energy for the projected Hamiltonian at $U>0$.
However, the Dirac fermion wavefunction of the neighborhood of $\Gamma_M$ and $K'_M$ points only has non-zero components on the nearest $K_M$ point, while the Dirac fermion wavefunction at $\Gamma_M$ (or $K'_M$) has an equal amplitude on the three nearest $K_M$ points.
Therefore, the projected Dirac wavefunction $u^{\hb}_{\mathbf{Q}\alpha,m\eta}(\vk)$ is not continuous at $\Gamma_M$ and $K'_M$. 
The projected band energies are immediately different from the dispersion of the BM model away from $\Gamma_M$ and $K_M'$ points, because of the abrupt change of the projected Dirac wavefunction $u^{\hb}_{\mathbf{Q}\alpha,m\eta}(\vk)$ and neglecting of the higher Dirac bands.
We note that in the HF bands where the Hartree and Fock energies are comparable to the kinetic energies, the discontinuities in the HF band dispersions are usually smeared out and barely noticeable, because of the summation over $\vk'$ in the HF mean field terms. However, when the interacting effects are weak (i.e., the HF mean field terms are small), this spurious discontinuity will be noticeable in the HF band dispersion (e.g., in Fig. \ref{fig:band_0}(c)).

Finally, we give the energy value of the non-interacting projected Hamiltonian in Eq.~(\ref{eqn:def_projected_hamiltonian}) at $M_M$ and $K_M$ points as a function of $w_0/w_1$ and $U$, which are shown in Figs.~\ref{fig:bandwidth}(a) and (b). The energy value closest to zero energy at $M_M$ point roughly captures the band width of the flat bands, and the energy shift of coupling with the Dirac cone is inferred from the energy closest to zero at $K_M$ point. In Fig.~\ref{fig:bandwidth}(c), we also provide the energy value jump of the non-interacting projected Hamiltonian at $\Gamma_M$ point, which describes the discontinuity of the projection.

\section{Hartree-Fock Mean Field Hamiltonian}\label{app_sec:hartreefock}
In this appendix, we give a short review of the HF mean field theory applied to the TSTG. We will also provide the initial conditions for our HF calculation and we will discuss the methodology used to plot the HF band structures along the high symmetry lines.
\subsection{Self-Consistent Mean Field Hamiltonian}
Assuming that there is no translation symmetry breaking, the HF order parameter can be defined as:
\begin{equation}
    \label{eq:ord_param_no_symbreak}
	\Delta_{\hat{f}m\eta s; \hat{f}'n\eta's'}(\vk) = \Big{\langle} \hf^\dagger_{\vk,m,\eta,s}\hf'_{\vk,n,\eta',s'} - \frac12 \delta_{\hat{f}\hat{f}'}\delta_{mn}\delta_{\eta\eta'}\delta_{ss'} \Big{\rangle}\,,
\end{equation}
in which $\hat{f},\hat{f}'$ stand for Dirac and TBG fermion operators. Therefore, by using the Hartree Fock mean field approximation, the interacting Hamiltonian can be written in the following form:
\begin{align}
	H^{(H)} &= \sum_{\vk, \hat{f}, mn, \eta, s}\mathcal{H}^{(H)}_{\hat{f}m\eta s,\hat{f}n\eta s}(\vk)\left(\hf^\dagger_{\vk,m,\eta,s}\hf_{\vk,n,\eta,s} - \frac{1}{2}\delta_{mn}\right)\\
	H^{(F)} &= \sum_{\vk,\hat{f}\hat{f}',mn,\eta\eta',ss'}\mathcal{H}^{(F)}_{\hat{f}m\eta s,\hat{f}'n\eta's'}(\vk)\left(\hf^\dagger_{\vk,m,\eta,s}\hf'_{\vk,n,\eta',s'} - \frac{1}{2}\delta_{\hat{f}\hat{f}'}\delta_{mn}\delta_{\eta\eta'}\delta_{ss'}\right)\,.
\end{align}
The matrices $\mathcal{H}^{(H)}(\vk)$ and $\mathcal{H}^{(F)}(\vk)$ are given by:
\begin{align}
	\mathcal{H}^{(H)}_{\hat{f}m\eta s,\hat{f}'n\eta's'}(\vk) =& \frac{1}{\Omega_{\rm tot}}\sum_{\vk', \hat{f}'',m'n',\eta'', s''}\tilde{V}^{(\hat{f}\eta;\hat{f}''\eta'')}_{mn;m'n'}(0; \vk,\vk') \Delta_{\hat{f}''m'\eta''s'';\hat{f}''n'\eta''s''}(\vk') \delta_{\hat{f}\hat{f}'} \delta_{\eta\eta'}\delta_{ss'} \label{eqn:Hartree_k} \\
	\mathcal{H}^{(F)}_{\hat{f}m\eta s,\hat{f}'n\eta's'}(\vk) =& -\frac{1}{\Omega_{\rm tot}}\sum_{\vk',m'n'}\tilde{V}^{(\hat{f}'\eta';\hat{f}\eta)}_{m'n;mn'}(\vk'-\vk;\vk,\vk') \Delta_{\hat{f}'m'\eta's'; \hat{f}n'\eta s}(\vk') \label{eqn:Fock_k}
\end{align}
Therefore the full mean field Hamiltonian is given by $\mathcal{H}^{HF}(\vk) = \mathcal{H}^{(0)}(\vk) + \mathcal{H}^{(H)}(\vk) + \mathcal{H}^{(F)}(\vk)$. The mean field Hamiltonian $\mathcal{H}^{HF}(\vk)$ is a $16\times 16$ matrix for each momentum. We use $\phi_{\hat{f}m\eta s, i}(\vk)$ and $E_i(\vk)$ to represent its eigenstates and eigenvalues, respectively:
\begin{equation}
	\sum_{\hat{f}',n,\eta',s'}\mathcal{H}^{HF}_{\hat{f}m\eta s, \hat{f}'n\eta's'}(\vk) \phi_{\hat{f}'n\eta's', i}(\vk) = E_i(\vk)\phi_{\hat{f}m\eta s, i}(\vk)\,.
\end{equation}
The eigenvalues $E_i(\vk)$ give us the Hartree Fock band structure, and the wavefunctions give us the self-consistent condition for the order parameter:
\begin{equation}\label{eqn:self-consistent-condition}
	\Delta_{\hat{f}m\eta s;\hat{f}'n\eta's'}(\vk) = \sum_{i\in{\rm occ}}\left(\phi^*_{\hat{f}m\eta s, i}(\vk)\phi_{\hat{f}'n\eta's', i}(\vk) - \frac12 \delta_{\hat{f}\hat{f}'}\delta_{mn}\delta_{\eta\eta'}\delta_{ss'}\right)\,,
\end{equation}
in which the $N = \nu\times N_M$ states with the lowest energies $E_i(\vk)$ are occupied. For each given value of filling factor $\nu$, we start the numerical calculation by various initial conditions of the order parameter, and then solve the mean field Hamiltonian $\mathcal{H}^{HF}(\vk)$ for the new order parameter using Eq.~(\ref{eqn:self-consistent-condition}) until convergence. The total energy of a solution is given by the following formula:
\begin{equation}\label{seq-EHF}
	E_{HF} = \Big{\langle} H_0 + \frac{1}{2}\left(H^{(H)} + H^{(F)}\right) \Big{\rangle}\,.
\end{equation}
And for each given parameter $w_0/w_1$ and $U$, we choose the state with the lowest energy.

Being an iterative method, the choice of the initial order parameter is crucial for the convergence of the HF algorithm. The HF order parameter could depend on the choice of initial condition, thus some initial conditions might lead to a local minimum. For that purpose, we have used several possible initial conditions for the each filling factor. We build the initial order parameter from the initial many-body wavefunction $|\Psi_0\rangle$. Defining the half filled Dirac cone wavefunction $|\phi_D\rangle$ as
\begin{equation}
	|\phi_D\rangle = \prod_{\vk,\eta,s}\hb^\dagger_{\vk,-1, \eta, s}|0\rangle\,,\label{app:eq:halffilleddirac}
\end{equation}
our initial many-body wavefunctions is built as the tensor product $|\Psi^{(0)}\rangle= |\phi^{(0)}_{\rm{TBG}}\rangle \otimes |\phi_D\rangle$. Here $|\phi^{(0)}_{\rm{TBG}}\rangle$ is a single Slater determinant many-body wavefunction with only the TBG electrons. As discussed in Sec.~\ref{sec:HF}, the Dirac fermion density $\delta \rho^{\hb}$ is small due to the large Fermi velocity, therefore we expect the ground state will be approximately given by the tensor product of TBG ground state and half filled Dirac fermion. For each filling factor, we choose several possible initial states $|\Psi^{(0)}\rangle$, motivated by the possible physics that could emerge at a given $\nu$, to build the initial order parameter $\Delta(\vk)$. The full list of these specific initial states can be found in Table~\ref{tab:init}. In addition to this list, we also tested a randomly generated HF order parameter $\Delta(\vk)$ for each HF calculation. The randomly generated initial conditions are no longer the tensor product between a half filled Dirac fermions and TBG states. Random initial condition is harder to converge which prevent its systematic usage. Nevertheless, we verified that random initial condition is able to obtain the phase diagram at $\nu=-3$. The resulting state with the lowest HF total energy is identified as the HF ground state.

\begin{table}[h]
\centering
\begin{tabular}{c||c||l}
\hline
$\nu$ & $|\Psi^{(0)}\rangle$ & description\\
\hline
\hline
$-3$ & $\displaystyle \prod_{\vk}\hat{d}^\dagger_{\vk,1,+\uparrow}|\phi_D\rangle$ & valley polarized Chern insulator\\
\hline
$-3$ & $\displaystyle \prod_{\vk}\frac{1}{\sqrt2}\left(\hat{d}^\dagger_{\vk, 1,+,\uparrow} + \hat{d}^\dagger_{\vk, 1,-,\uparrow}\right)|\phi_D\rangle$ & intervalley coherent Chern insulator\\
\hline
\hline
$-2$ & $\displaystyle \prod_{\vk,e_Y=\pm1}\frac{1}{\sqrt2}\left(\hat{d}^\dagger_{\vk,e_Y, +, \uparrow} + e_Y \hat{d}^\dagger_{\vk,e_Y,+,\uparrow}\right) |\phi_D\rangle$ & intervalley coherent state with zero Chern number\\
\hline
$-2$ & $\displaystyle \prod_{\vk}\hat{d}^\dagger_{\vk,1,+,\uparrow}\hat{d}^\dagger_{\vk,1,+,\downarrow}|\phi_D\rangle$ & valley polarized Chern insulator with zero spin\\
\hline
$-2$ & $\displaystyle \prod_{\vk}\hat{d}^\dagger_{\vk,1, +, \uparrow}\hat{d}^\dagger_{\vk, -1, +, \downarrow} |\phi_D\rangle$ & valley polarized state with zero Chern number and total spin\\
\hline
$-2$ & $\displaystyle \prod_{\vk,e_Y=\pm1}\hat{d}^\dagger_{\vk,e_Y, +,\uparrow}\hat{d}^\dagger_{\vk,e_Y,+,\uparrow} | \phi_D\rangle $ & fully polarized state\\
\hline
$-2$ & $\displaystyle \prod_{\vk,\eta=\pm}\hat{d}^\dagger_{\vk,1,\eta,\uparrow}|\phi_D\rangle$ & Chern insulator state with $N_v = 0$\\
\hline
$-2$ & $\displaystyle \prod_{\vk}\hat{d}^\dagger_{\vk,1,+,\uparrow}\hat{d}^\dagger_{\vk,-1,-,\uparrow}|\phi_D\rangle$ & spin polarized state with zero Chern number and $N_v$\\
\hline
\hline
$-1$ & $\displaystyle \prod_{\vk}\hat{d}^\dagger_{\vk, 1, +, \downarrow}\prod_{e_Y=\pm1}\frac{\left(\hat{d}^\dagger_{\vk,e_Y,+,\uparrow} + e_Y \hat{d}^\dagger_{\vk,e_Y,-,\uparrow}\right)}{\sqrt2} | \phi_D\rangle$ & two occupied intervalley coherent bands and a valley polarized band\\
\hline
$-1$ & $\displaystyle \prod_{\vk}\hat{d}^\dagger_{\vk,1,+,\uparrow}\hat{d}^\dagger_{\vk,-1,+,\uparrow}\hat{d}^\dagger_{\vk,1, +, \downarrow}|\phi_D\rangle$ & valley polarized Chern insulator state with $|\nu_C| = 1$\\
\hline
$-1$ & $\displaystyle \prod_{\vk} \hat{d}^\dagger_{\vk,1,+,\uparrow}\hat{d}^\dagger_{\vk,-1,+,\uparrow}\hat{d}^\dagger_{\vk,1,-,\uparrow} | \phi_D\rangle$ & spin polarized Chern insulator state with $|\nu_C| = 1$\\
\hline
$-1$ & $\displaystyle \prod_{\vk}\hat{d}^\dagger_{\vk,1,+,\uparrow}\hat{d}^\dagger_{\vk, 1, -, \uparrow}\hat{d}^\dagger_{\vk,1, +, \downarrow}|\phi_D\rangle$ & Chern insulator state with $|\nu_C| = 3$\\
\hline
\hline
$0$ & $\displaystyle \prod_{\vk,e_Y=\pm1,s}\frac{1}{\sqrt2}\left(\hat{d}^\dagger_{\vk,e_Y, +, s} + e_Y \hat{d}^\dagger_{\vk,e_Y,+,s}\right) |\phi_D\rangle$ & intervalley coherent state\\
\hline
$0$ & $\displaystyle \prod_{\vk,\eta, s}\hat{d}^\dagger_{\vk, 1, \eta, s}|\phi_D\rangle$ & spin valley unpolarized state with $|\nu_C| = 4$\\
\hline
$0$ & $\displaystyle \prod_{\vk,\eta}\hat{d}^\dagger_{\vk,1,\eta,\uparrow}\hat{d}^\dagger_{\vk,-1,\eta,\downarrow}|\phi_D\rangle$ & spin valley unpolarized state with $\nu_C = 0$ \\
\hline
\hline
\end{tabular}
\caption{The initial many-body wavefunctions $|\Psi^{(0)}\rangle$ that were used in the HF mean field calculations for each filling factor $\nu$. For the Dirac fermion sector, the initial states always assumed a half-filled Dirac cone given by $|\phi_D\rangle$ of Eq.~\ref{app:eq:halffilleddirac}. Note that in addition to these initial states, initial randomly generated HF order parameters were also considered for every filling factors.}
\label{tab:init}
\end{table}

\subsection{Hartree Fock band structure along high symmetry lines}\label{app_subsec:analysis}

Due to the difficulty of performing the numerical calculation (for example, it needs $1.7\rm\,s$ for one iteration of the self-consistent calculation on a $10\times 10$ momentum lattice with a single core $2.6\,\rm GHz$ Skylake CPU, and the convergence typically require around 1000 iterations), the momentum lattice that we use to discretize the MBZ cannot be dense enough to show a smooth dispersion of the Hartree Fock bands clearly along the high symmetry lines. As shown in Fig.~\ref{fig:examples_plots}(a), a three dimensional dispersion plot in $(k_x, k_y, E)$ space can be made easily for a given solution to the Hartree Fock Hamiltonian. However, the amount of the momentum points are not enough to obtain a continuous band structure plot along high symmetry lines. 

In order to visualize the Hartree Fock band structure, we calculate the approximate mean field Hamiltonian at an arbitrary momentum $\vk$ along high symmetry line by using the order parameter $\Delta(\vk)$ obtained on the discrete but rare MBZ lattice after the order parameter $\Delta(\vk)$ converges. The expressions for the Hartree and Fock terms $\mathcal{H}^{(H)}(\vk)$ and $\mathcal{H}^{(F)}(\vk)$ are still given by Eqs. (\ref{eqn:Hartree_k}) and (\ref{eqn:Fock_k}), but the momentum $\vk'$ which appear in the summations is constrained on the loose discrete lattice, while the momentum $\vk$ is a point on the high symmetry line. Therefore, in order to obtain the interaction matrix elements $\tilde{V}^{(\hat{f}\eta;\hat{f}''\eta'')}_{mn;m'n'}(0; \vk,\vk')$ and $\tilde{V}^{(\hat{f}'\eta';\hat{f}\eta)}_{m'n;mn'}(\vk'-\vk;\vk,\vk')$ which are required by the Hartree Fock terms $\mathcal{H}^{(H)}(\vk)$ and $\mathcal{H}^{(F)}(\vk)$, we only need the single-body wavefunctions on the 
rare discrete momentum lattice and along the high symmetry line, instead of a dense mesh. By diagonalizing the mean field Hamiltonian $\mathcal{H}^{HF}(\vk) = \mathcal{H}^{(0)}(\vk) + \mathcal{H}^{(H)}(\vk) + \mathcal{H}^{(F)}(\vk)$, we can obtain the band structure along the high symmetry line, as shown in Fig.~\ref{fig:examples_plots}(b).

\begin{figure}[!htbp]
	\centering
	\includegraphics[width=0.95\linewidth]{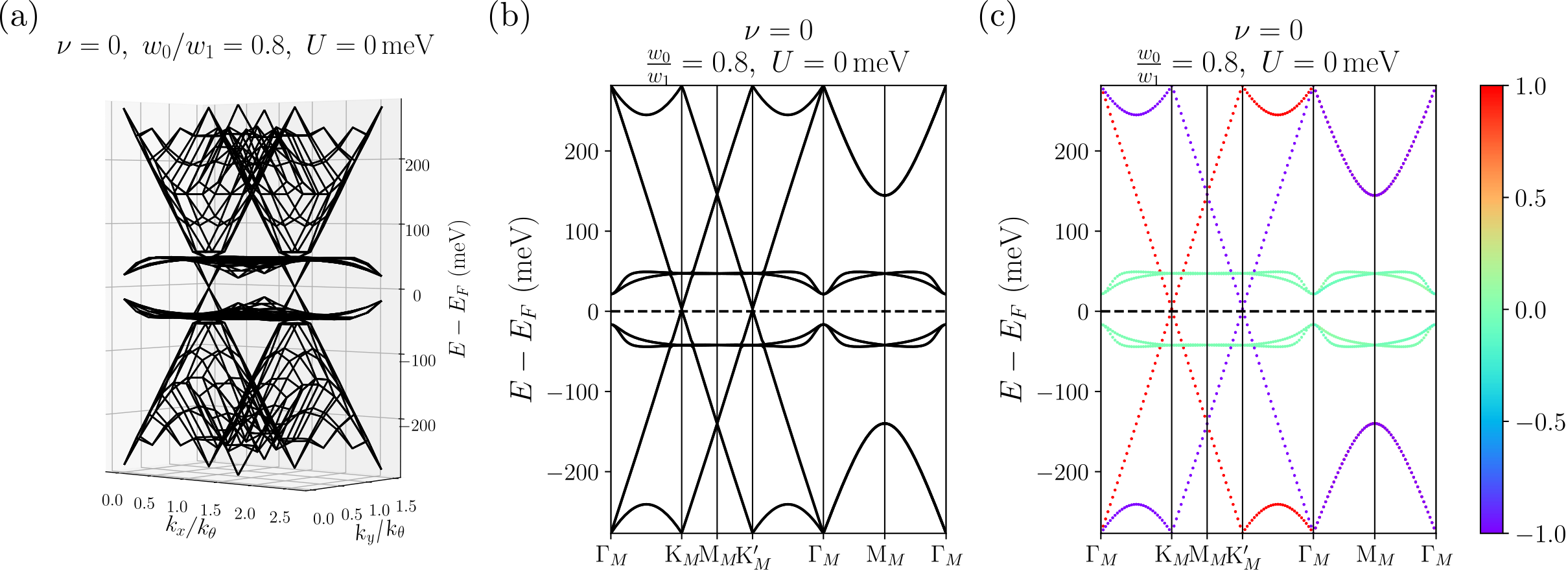}
	\caption{(a) The Hartree Fock band structure obtained from a $9\times9$ discrete momentum lattice, shown in a $(k_x, k_y, E)$ 3-dimensional space. (b) The Hartree Fock band structure plotted along the high symmetry lines in the moir\'e Brillouin zone. Note that the band structure in (b) is obtained from the numerical results (the density order parameter and the Hartree Fock Hamiltonians) obtained on the discrete lattice shown in (a). (c) Same as (b) but here we use the color scale to provide $v_i(\vk)$, i.e., the valley polarization for each Hartree Fock band.}
	\label{fig:examples_plots}
\end{figure}

Several quantites can also be shown for each point in the band structure plots, for example, the valley polarization of each single body state. By diagonalizing the Hartree Fock Hamiltonian $\mathcal{H}^{(HF)}(\vk)$ at a given $\vk$ along the high symmetry line, we can also obtain the corresponding wavefunction $\phi_{\hat{f}m\eta s, i}(\vk)$. For each given single body eigenstate of the Hartree Fock Hamiltonian, the valley polarization can be defined as follows:
\begin{equation}\label{eqn:def-valley-polarization}
	v_{i}(\vk) = \sum_{\hat{f},m,s, \eta, \eta'}\phi_{\hat{f}m\eta s, i}^*(\vk)(\tau_z)_{\eta\eta'}\phi_{\hat{f}m\eta's, i}(\vk)\,,
\end{equation}
in which $\tau_z$ is the Pauli matrix acting in valley space. This quantity measures the valley polarization, therefore $v_i(\vk)=-1$ or $1$ if the state is valley polarized, and $-1 < v_i(\vk) < 1$ if there is a superposition between the two valleys. Fig.~\ref{fig:examples_plots}(c) shows the corresponding results of valley polarization for each state $\phi_{\hf m\eta s, i}(\vk)$ along the high symmetry line, using the same order parameter $\Delta(\vk)$ as Figs.~\ref{fig:examples_plots}(a) and \ref{fig:examples_plots}(b). In this example, and using the color code visualization for $v_i(\vk)$, it can be seen clearly that the occupied flat bands are in intervalley coherent state at $\nu=0$ filling, as predicted in Refs. \cite{BUL20a,ZHA20,LIA20}.

\section{Additional Numerical Results}\label{app_sec:numerical_results}
\subsection{Numerical results at filling factor \texorpdfstring{$\nu=-3$}{nu=-3}}\label{app_subsec:-3}
In this appendix we provide additional HF results for various system sizes at filling factor $\nu=-3$. First in Fig.~\ref{app_fig:additional_nu_-3_phase} we give the phase diagrams in the $(w_0, U)$ plane with a color code representing valley polarization $N_v/N_M$ on several momentum lattices: $6\times 6$, $7\times 7$ and $8\times 8$. Note that Fig.~\ref{app_fig:additional_nu_-3_phase}(c) was already provided in Fig.~\ref{fig:nu=-3_phase}(a) and was added here for convenience. On the $7\times 7$ and $8\times 8$ momentum lattices, the HF calculation were performed at exact integer filling $\nu=-3$. For the $6\times 6$ momentum lattice (or any lattice of the size $3n \times 3n$ where $n$ is an integer), we removed four electrons from the exact integer filling, which is denoted by $\nu=-3 - 4\mathrm{e}^-$. Indeed, this lattice discretization exactly hits the Dirac points $K_M$ and $K'_M$, which induces degeneracy in the non-interacting band structure, plaguing the convergence of the HF self-consistent calculation. Removing four electrons improves the convergence by avoiding filling these degenerate states at Dirac points. As seen in these three phase diagrams, the positions of the three regions I, II and III do not strongly depend on the system size. We also notice that there is a small region with $w_0/w_1 \lesssim 0.3$ and around $U \approx 50\,\rm meV$ on $6\times 6$ momentum lattice located in region I, with small valley polarization, as opposed to the expected full polarization of region I. 
Such region does not show up in the other lattice discretizations. To test if this is a finite size effect or if this partially polarized region is induced by hitting exactly the $K_M$ and $K_M'$ points, we also provide the plot of the $U$ dependence of several quantities for fixed $w_0/w_1 = 0.2$ on $9\times 9$ momentum lattice at $\nu=-3$ filling with four electrons removed in Fig.~\ref{app_fig:additional_nu_-3_phase}(d) (note that a full phase diagram is computationally out of reach for this discretization).
Similar to the $6\times 6$ momentum lattice, the $9\times 9$ momentum lattice also has $K_M$ and $K'_M$. However, the valley polarization $N_v$ shown in Fig.~\ref{app_fig:additional_nu_-3_phase}(d) does not drop around $U \approx 50\,\rm meV$. This implies the region with small $N_v$ around $U\approx 50\,\rm meV$ on $6\times 6$ lattice is most probably due to size effects. We also provide the HF band structure obtained on $9\times 9$ lattice at filling factor $\nu=-3-4\mathrm{e}^-$ in Fig.~\ref{fig:additional_nu_-3_bands}. Similar to the HF band structures in Fig.~\ref{fig:nu=-3_band}, we obtain one occupied valley polarized flat band in region I as shown in Fig.~\ref{fig:additional_nu_-3_bands}(a), a gapless metal state without valley polarization in region II shown in Fig.~\ref{fig:additional_nu_-3_bands}(b). 
We also obtained a state in region III shown in Fig.~\ref{fig:additional_nu_-3_bands}(c) on $9\times 9$ momentum lattice. When compared with Fig.~\ref{fig:nu=-3_band}(c) on $10\times 10$ momentum lattice in the main text, the result on $9\times 9$ lattice also have $\nu_D \approx -1$ and $\nu_{\rm TBG} \approx -2$, although the occupied TBG flat bands have different valley polarization.

\begin{figure}[!htbp]
	\centering
	\includegraphics[width=\linewidth]{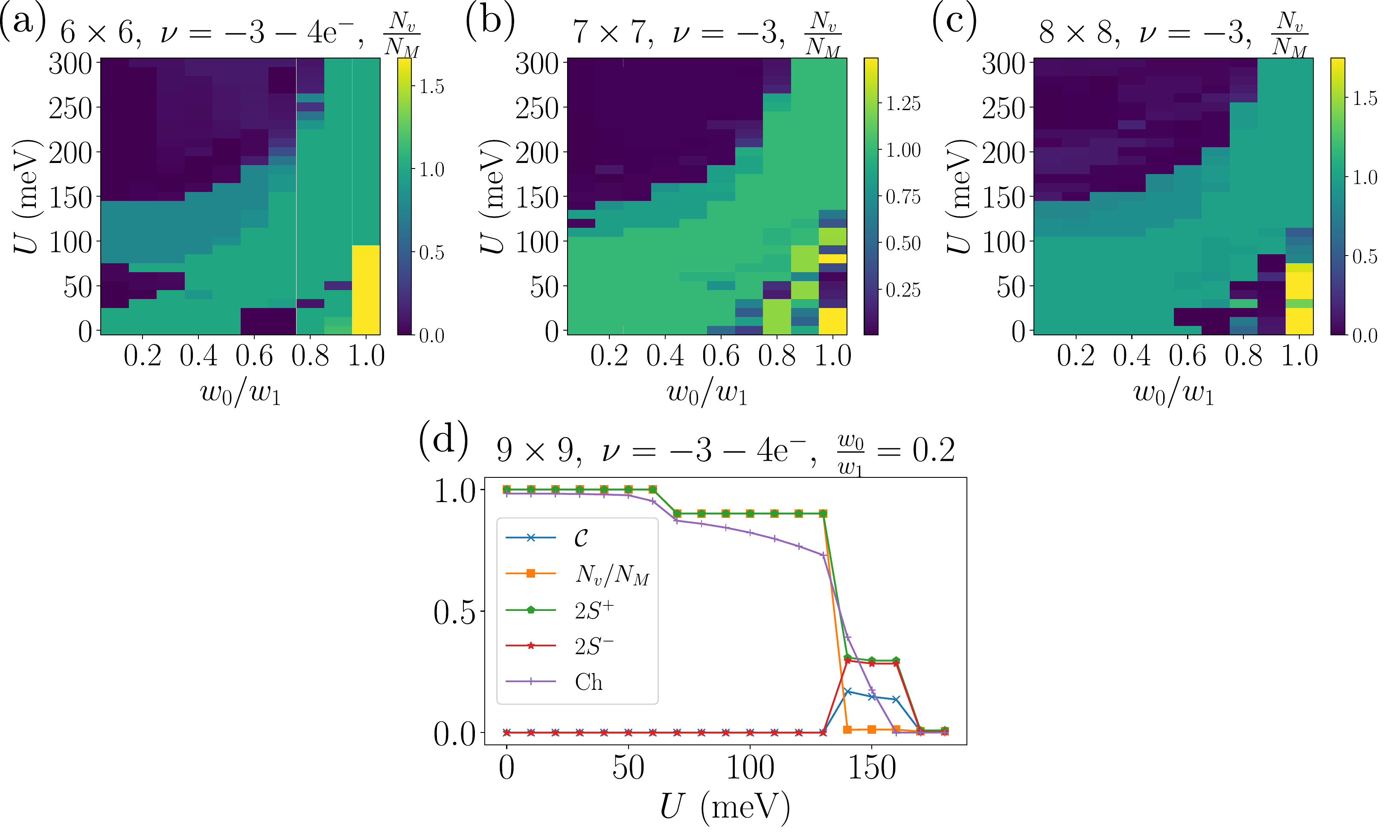}
	\caption{(a-c) Additional phase diagrams at filling factor $\nu=-3$ on different lattice sizes. (d) The displacement field dependence of the quantities $N_v$, $\mathcal{C}$, $\mathrm{Ch}$ and $S^\pm$ with fixed $w_0/w_1 = 0.2$ on $9\times 9$ lattice at filling factor $\nu=-3 - 4\mathrm{e}^-$.}
	\label{app_fig:additional_nu_-3_phase}
\end{figure}

\begin{figure}[!htbp]
	\centering
	\includegraphics[width=\linewidth]{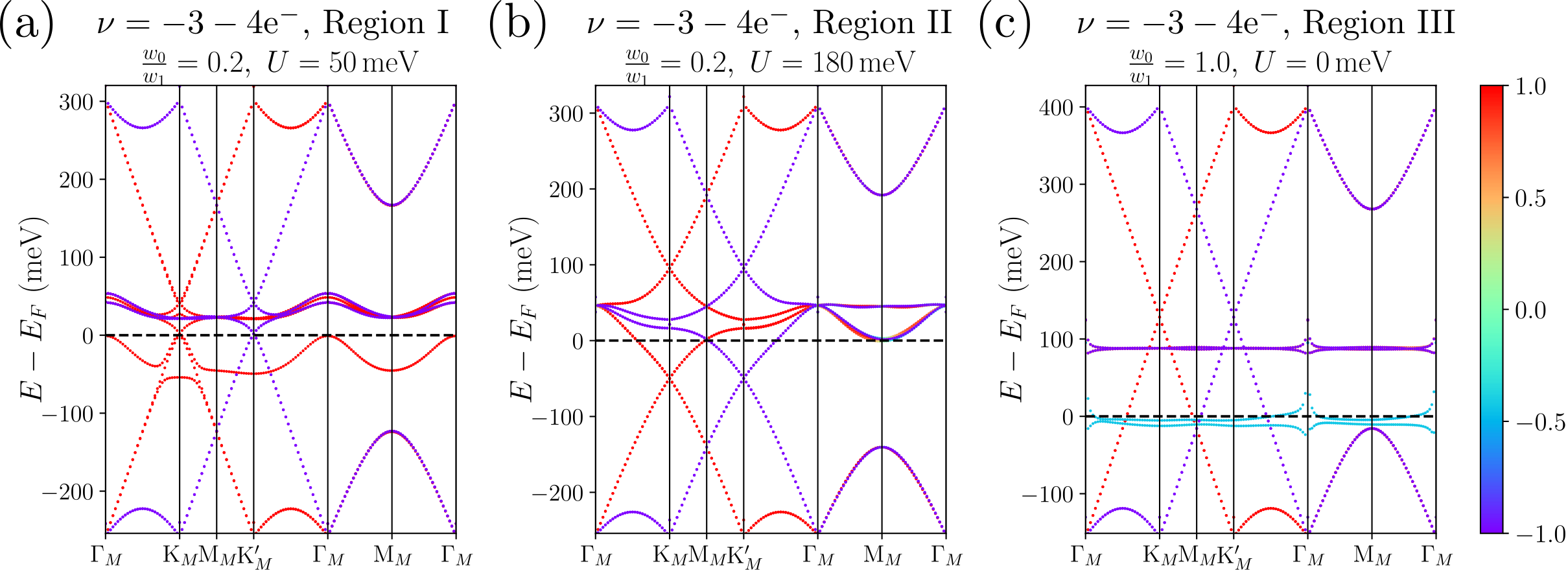}
	\caption{Some typical HF band structures illustrating the three regions of the phase diagram at filling factor $\nu=-3 - 4\mathrm{e}^-$ on $9\times 9$ momentum lattice. The color of each point represents the valley polarization $v_i(\vk)$ of each single body state, which is defined in Eq. (\ref{eqn:def_singlevp}). The parameters are the same as in Fig.~\ref{fig:additional_nu_-3_bands}.}
	\label{fig:additional_nu_-3_bands}
\end{figure}

To explain the regions I and III that are connected to the $U=0$ physics, we also provide the plots of $\nu_D$ and $\nu_{\rm TBG}$ as a function of $w_0/w_1$ at zero displacement field for several momentum lattice sizes in Fig.~\ref{app_fig:addtional_nuD_nuTBG}. Similar to the calculation on $6\times 6$ lattice, we also removed four electrons from the integer filling on $9\times 9$ lattice. It can be clearly seen that the electrons are moving from Dirac bands into TBG flat bands, when $w_0/w_1$ gets larger. The transition point and the shape of these plots are not exactly the same, while they share very similar trends. 

\begin{figure}[!htbp]
	\centering
	\includegraphics[width=\linewidth]{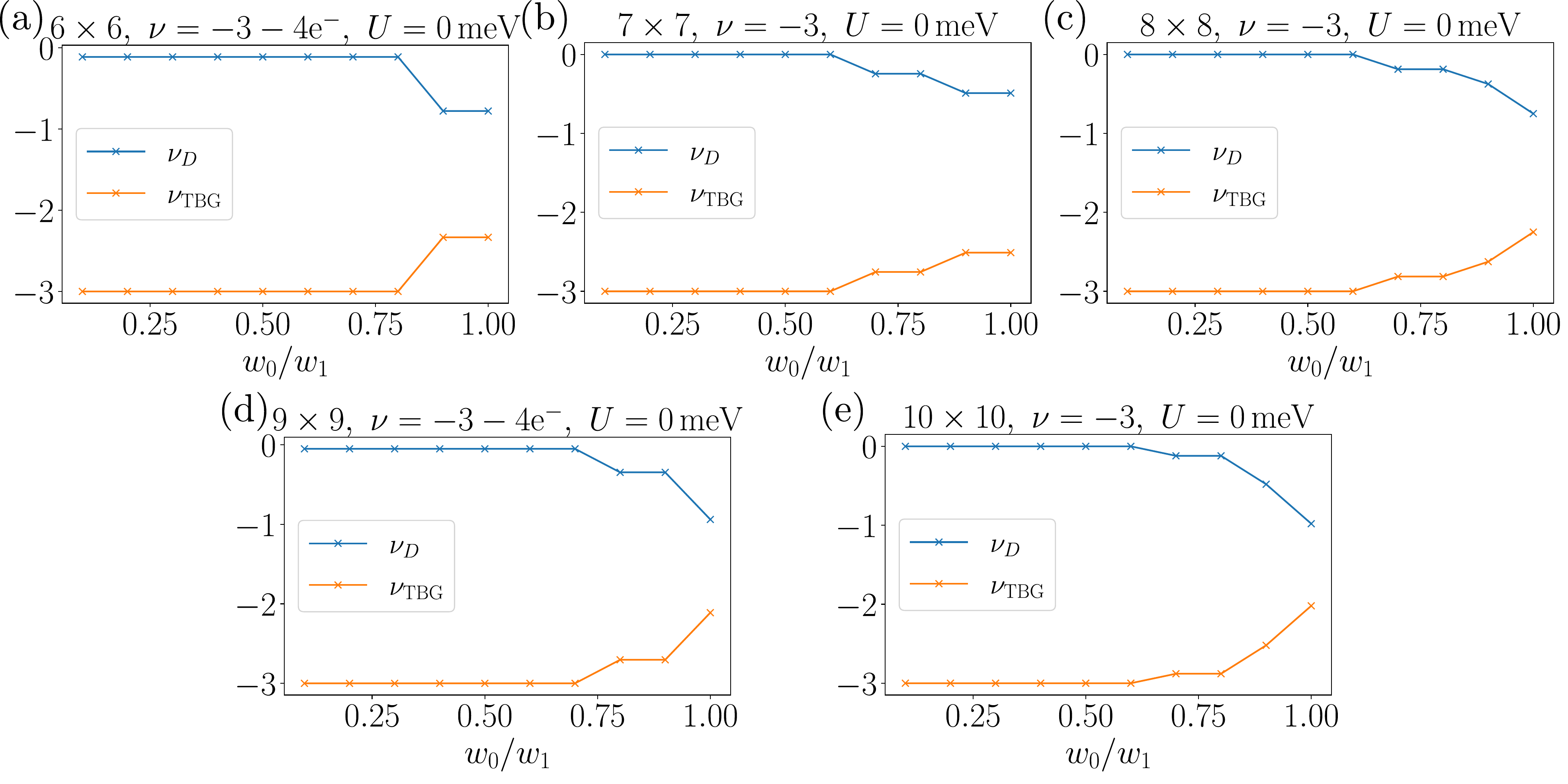}
	\caption{The filling factor of Dirac and TBG fermions as a function of $w_0/w_1$ when there is no displacement field ($U=0$) at filling factor $\nu=-3$ on various momentum lattices. Here on the $6\times 6$ and $9\times 9$ lattice we removed 4 electrons from the exact integer filling.}
	\label{app_fig:addtional_nuD_nuTBG}
\end{figure}

We now focus on the transition between regions III, I and II at fixed $w_0/w_1=0.8$ [see Fig.~\ref{fig:nu=-3_phase}(a) in the main text]. The displacement field dependence of several physical quantities, including $N_v$, $\mathcal{C}$, $\mathrm{Ch}$ and $S^\pm$, on different lattice sizes are shown in Fig.~\ref{app_fig:additional_nu_-3_Usweep}. In all these four diagrams, the system is in a valley and spin polarized state with one TBG flat band occupied when $60{\rm\, meV} \lesssim U \lesssim 250{\rm\, meV}$ (which is in region I), and the valley polarization $N_v$ vanishes when $U > 250\,\rm meV$ (which is likely to be a metal state in region II). However, when the displacement is smaller than $50\rm\,meV$ (which corresponds to region III), these four plots strongly change from one momentum lattice to another. 
Such a lattice size dependence could be related to the breakdown of the translation symmetry assumed for our HF order parameters (see Sec.~\ref{sec:HF}) as we have already argued in Sec.~\ref{sec:nu_-3}.

\begin{figure}[!htbp]
	\centering
	\includegraphics[width=0.8\linewidth]{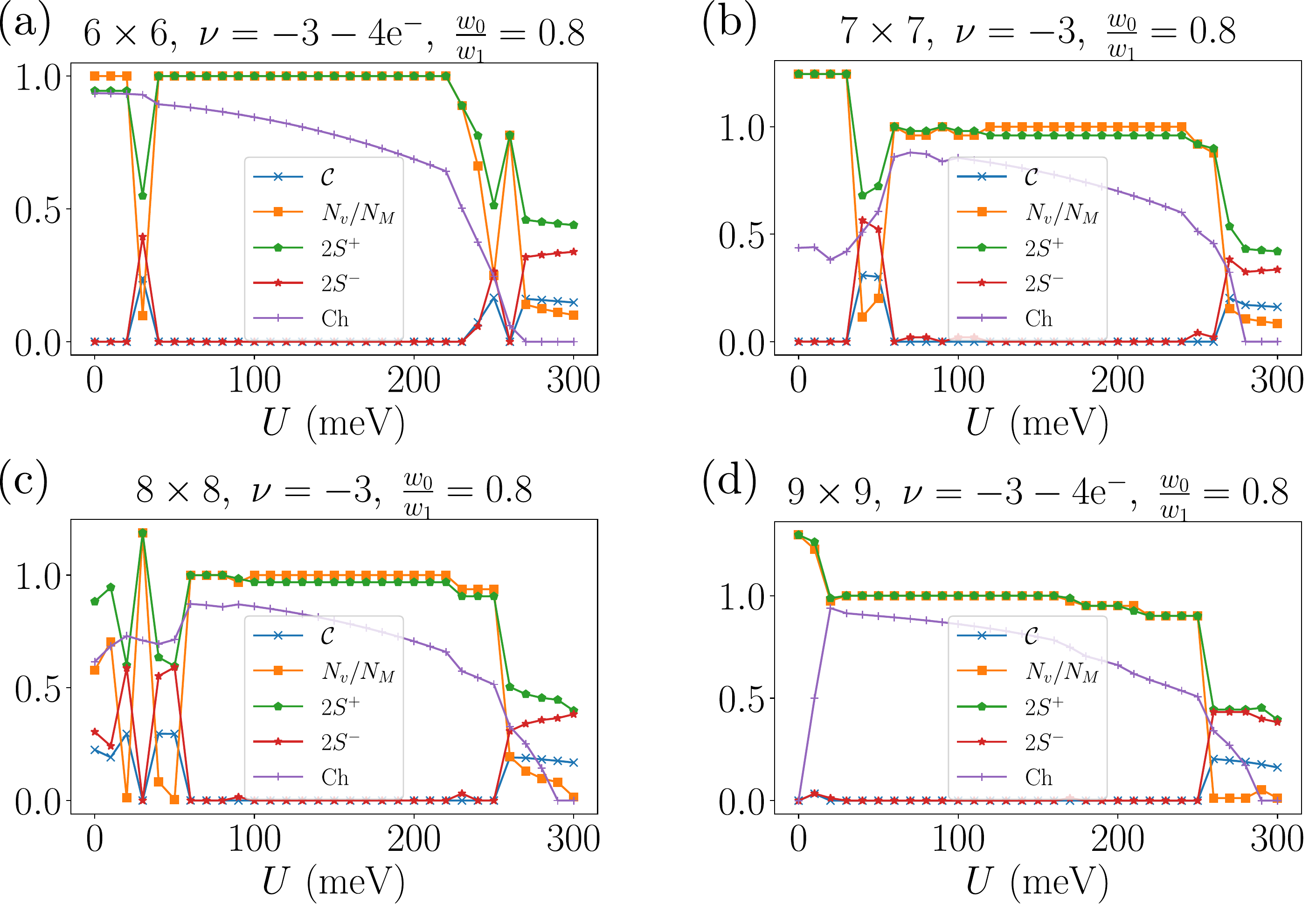}
	\caption{The displacement field dependence of the quantities $N_v$, $\mathcal{C}$ and $S^\pm$ on various system sizes. Similar to Fig.~\ref{app_fig:addtional_nuD_nuTBG}, we removed four electrons from integer filling on $6\times 6$ and $9\times 9$ lattices.}
	\label{app_fig:additional_nu_-3_Usweep}
\end{figure}

Finally, we address the question of translation symmetry breaking in region III. Using density matrix renormalization group, exact diagonalization, or HF, Refs.~\cite{KAN20a,SOE20,XIE20a} found that a period-$2$ stripe phase in which translation symmetry is broken at momentum $M_M$ (see Fig.~\ref{fig:mbz_dirac}(b) and (c)) becomes energetically competitive in ordinary magic-angle TBG for $w_0/w_1 \gtrsim 0.8$ and odd integer fillings.
While a full-fledged HF calculation allowing for general translation symmetry breaking is beyond the scope of this article, we compute a phase diagram of the system at $\nu = -3$, by allowing symmetry breaking with the wave vector corresponding to the $M_M$ point. Denoting $\mathbf{q}_{M_M}$ the moir\'e momentum at the $M_M$ point (see Fig.~\ref{fig:mbz_dirac}), we assume the HF order parameter of the system to be given by 
\begin{equation}
    \label{eq:ord_param_symbreak}
	\Delta_{\hat{f}m\eta s; \hat{f}'n\eta's'}(\vk_1,\vk_2) = \Big{\langle} \hf^\dagger_{\vk_1,m,\eta,s}\hf'_{\vk_2,n,\eta',s'} \left(\delta_{\vk_1,\vk_2} + \delta_{\vk_1,\vk_2 + \mathbf{q}_{M_M}} + \delta_{\vk_1,\vk_2 - \mathbf{q}_{M_M}} \right) - \frac12 \delta_{\vk_1,\vk_2}\delta_{\hat{f}\hat{f}'}\delta_{mn}\delta_{\eta\eta'}\delta_{ss'} \Big{\rangle}\,.
\end{equation}
Compared to Eq.~(\ref{eq:ord_param_no_symbreak}),  Eq.~(\ref{eq:ord_param_symbreak}) additionally allows for non-vanishing correlation between fermions whose momenta are separated by $\mathbf{q}_{M_M}$. To measure the degree of translation symmetry breaking, we define the quantity $\mathcal{T}$ which is based on the norm of the off-diagonal (in momentum space) order parameter matrix elements 
\begin{equation}
    \mathcal{T} = \frac{1}{2 N_M}\sum_{\vk_1 , \vk_2 \in{\rm MBZ}}\sum_{\substack{\hat{f}\hat{f}',mn, \\ ss',\eta\eta'}} \left(1-\delta_{\vk_1,\vk_2} \right) |\Delta_{\hat{f}m\eta s;\hat{f}'n\eta' s'}(\vk_1,\vk_2)|^2 \,.\label{eq:definitionT}
\end{equation}
$\mathcal{T}$ ranges from zero, when there is no translation symmetry breaking, to a maximum value of $\mathcal{T} = 5/2$ (the factor 5 is due to 5 projected bands at $\nu=-3$ in our calculation), when the fermions at $\vk$ and $\vk + \mathbf{q}_{M_M}$ are fully coherent.

In Fig.~\ref{app_fig:additional_nu_-3_translation}, we provide the phase diagram of the system in the $(w_0/w_1, U)$ plane for an $8 \times 8$ momentum lattice similar to Fig.~\ref{fig:nu=-3_phase} in the main text, allowing for translation symmetry breaking as defined in Eq.~(\ref{eq:ord_param_no_symbreak}). The translation symmetry breaking parameter $\mathcal{T}$ vanishes in regions I and II, but indicates that the HF ground-state breaks the translation symmetry in region III. However, even in the presence of this translation symmetry breaking, the valley polarization $N_v /N_M$ shows qualitatively the same features in region III as in the case where the translation symmetry is enforced (see Fig.~\ref{fig:nu=-3_phase}), namely electrons transfer from the Dirac valence bands into the TBG flat bands (as implied by $N_v/N_M > 1$).

We note that there are other possible translation symmetry breaking momenta (e.g., Kekul\'e order) which have been discussed in $\nu=-3$ TBG \cite{XIE20a,KWAN2021}, and the charge transfers between Dirac and flat bands in TSTG may further allow the translation symmetry breaking momenta to shift to other values. Therefore, our results here allowing only translation breaking at $M_M$ may not give the ground state, and a future study allowing more translation breaking momenta is needed.

\begin{figure}
    \centering
    \includegraphics[width=0.6\linewidth]{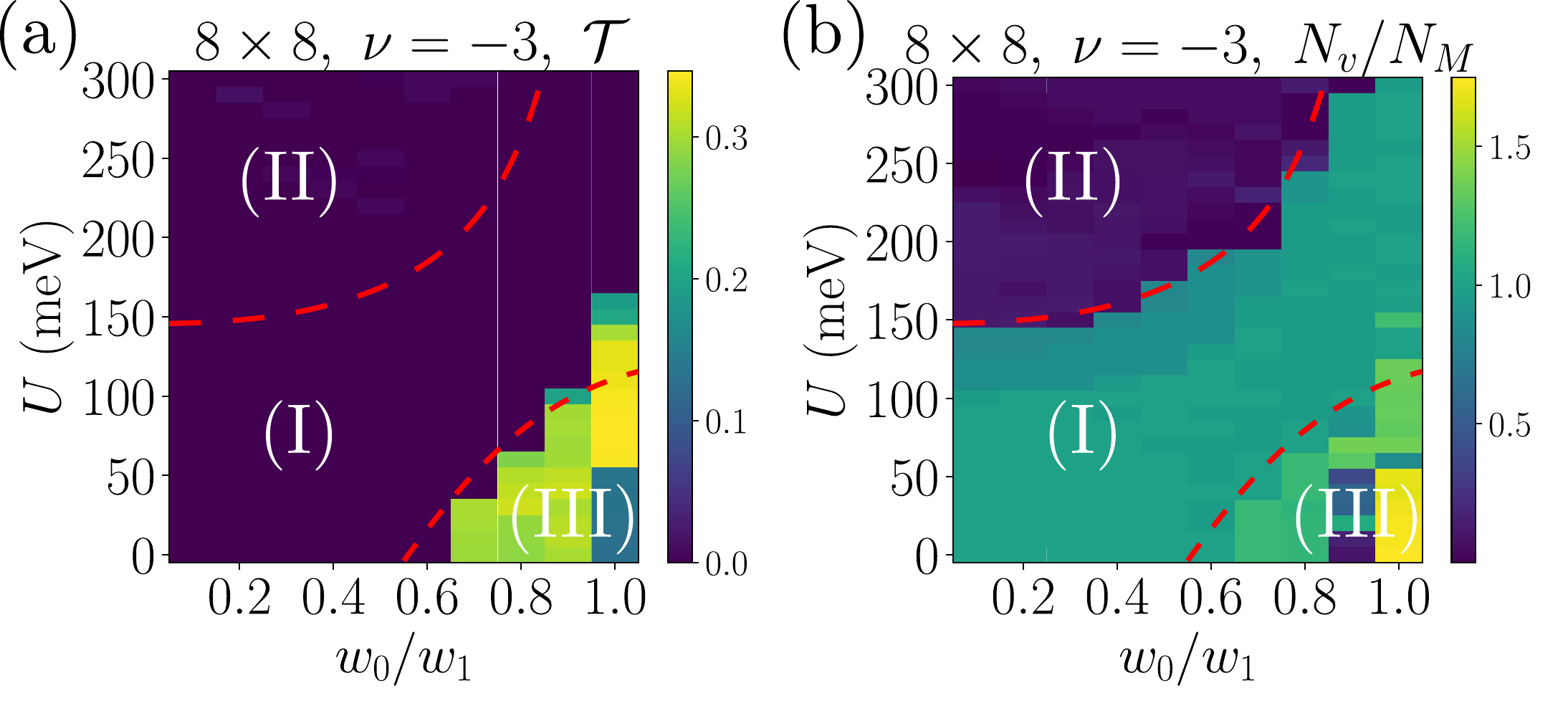}
    \caption{Propensity towards translation symmetry breaking in TSTG at $\nu=-3$. (a) Value of the translation symmetry breaking order parameter $\mathcal{T}$ as defined in Eq.~(\ref{eq:definitionT}), in the $(w_0/w_1, U)$ parameter space. As expected, the translation symmetry is only broken in region III. (b) Valley polarization of the HF ground state in the $(w_0/w_1, U)$ parameter space without assuming the translation symmetry. In both panels, we consider a $8 \times 8$ momentum lattice, similar to Fig.~\ref{fig:nu=-3_phase}. Note that the red lines delimiting the three regions are those of Fig.~\ref{fig:nu=-3_phase}(a).}
    \label{app_fig:additional_nu_-3_translation}
\end{figure}

\subsection{Numerical results at filling factor \texorpdfstring{$\nu=-2$}{nu=-2}}
In this appendix we present numerical results obtained on several momentum lattice sizes at integer filling $\nu=-2$. We start with the phase diagrams at integer filling $\nu=-2$. Figs.~\ref{fig:add_-2_6x6}(a) and \ref{fig:add_-2_7x7}(a) show the intervalley coherence $\mathcal{C}$ obtained on $6\times 6$ and $7\times 7$ momentum lattices. Other HF quantities on these system sizes at $w_0/w_1 = 0.2$ and $w_0/w_1 = 0.8$ are also shown in Figs.~\ref{fig:add_-2_6x6}(b-c) and \ref{fig:add_-2_7x7}(b-c). Clearly the phase diagrams on the $6\times 6$ lattice are noisy, but the phase boundary where intervalley coherence $\mathcal{C}$ vanishes is approximately the same as that of the larger sizes.

\begin{figure}
    \centering
    \includegraphics[width=0.8\linewidth]{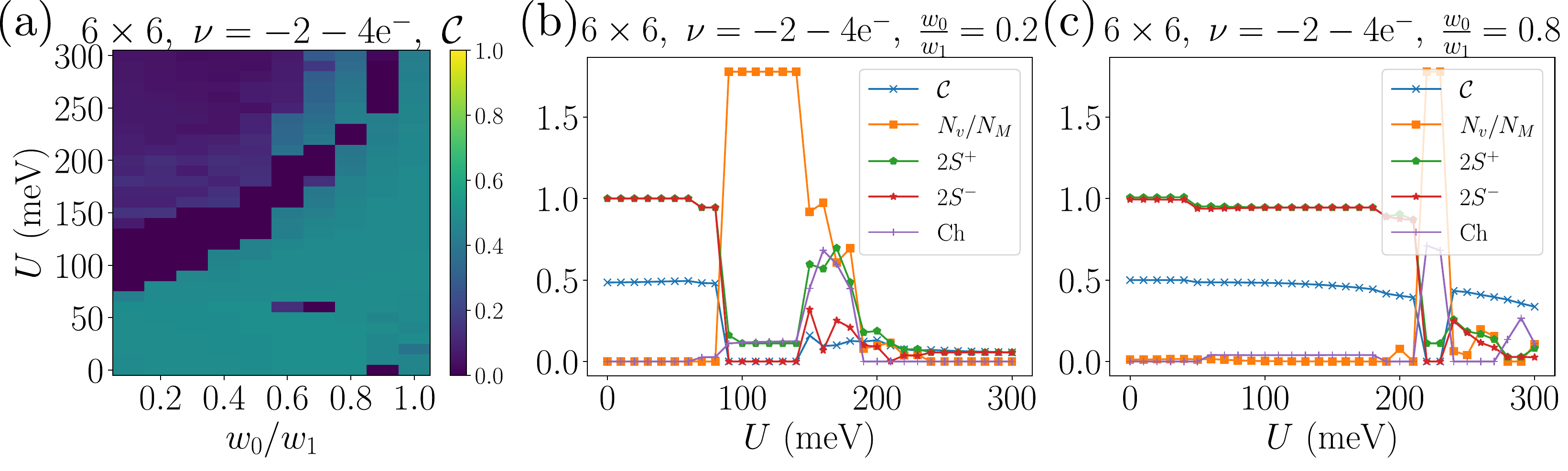}
    \caption{The phase diagrams at integer filling $\nu=-2-4\mathrm{e}^-$ on $6\times 6$ lattice.}
    \label{fig:add_-2_6x6}
\end{figure}

\begin{figure}
    \centering
    \includegraphics[width=0.8\linewidth]{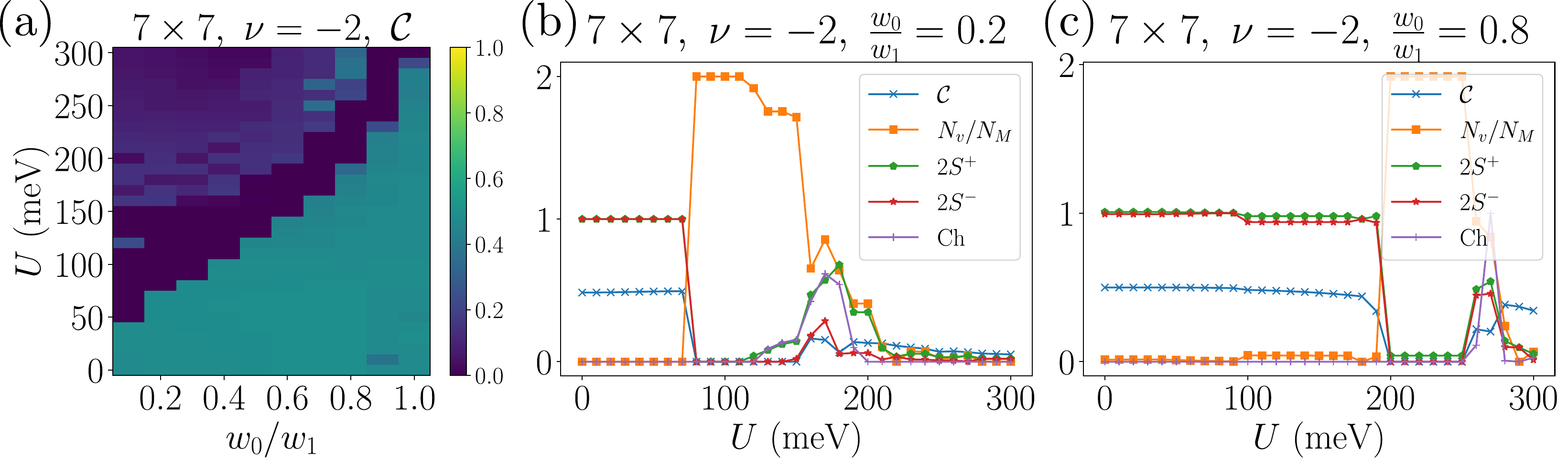}
    \caption{The phase diagrams at integer filling $\nu=-2$ on $7\times 7$ lattice.}
    \label{fig:add_-2_7x7}
\end{figure}

We also provide additional band structure plots in the metallic phase, namely the region III at $\nu=-2$. As shown in Fig.~\ref{fig:nu_-2_phase}(b), the intervalley coherence $\mathcal{C}$ can be a small but non-zero value in the metallic phase. We also observe that the intervalley coherence will decrease when $U$ becomes larger. This phenomenon can be observed in the HF band structures clearly. As shown in Fig.~\ref{fig:add_-2_band}, the HF bands at $w_0/w_1 = 0.2$ and $U = 180, 250, 300\,\rm meV$ are intervalley coherent around $M_M$ points indicated by the valley polarization $v_i(\vk)$ at each $\vk$. Numerically, we observe that the intervalley coherence values are $\mathcal{C} \approx 0.10$, $0.03$ and $0.02$, respectively. The HF band structure also becomes more similar to the non-interacting band structure when $U$ becomes larger.

\begin{figure}
	\centering
	\includegraphics[width=0.75\linewidth]{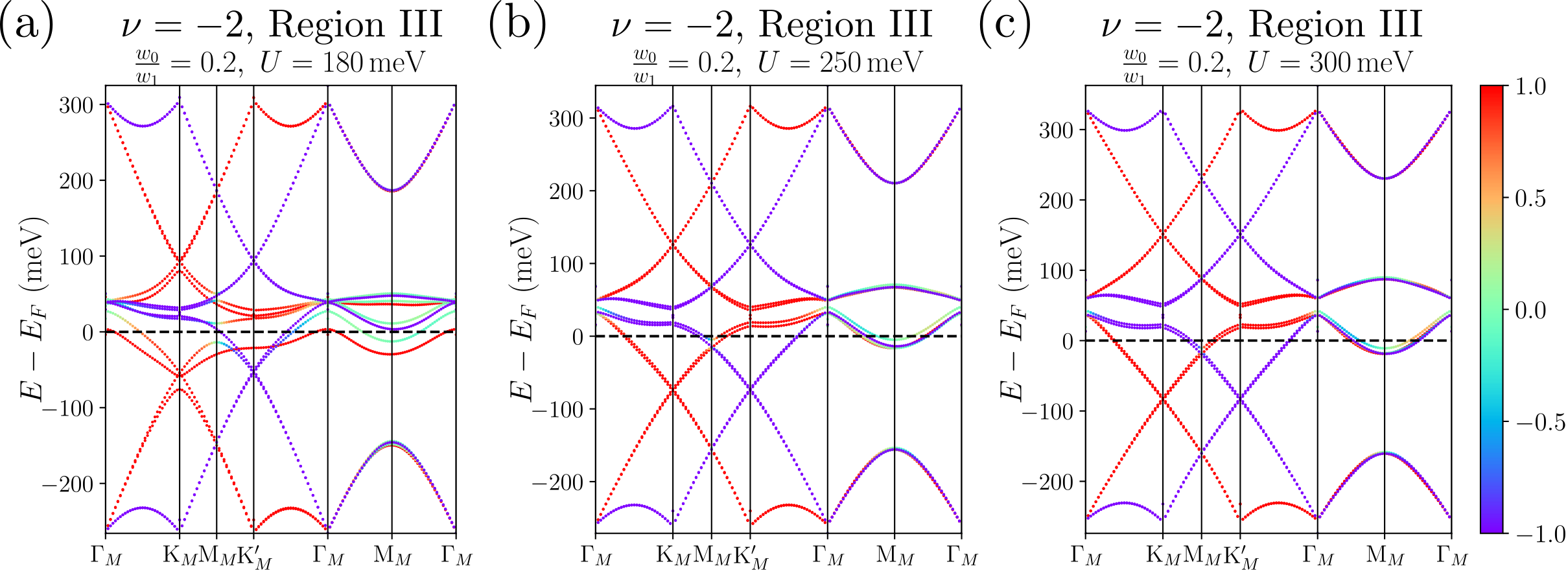}
	\caption{The HF band structure obtained at filling factor $\nu=-2$ on $10\times 10$ momentum lattice at $w_0/w_1 = 0.2$ for $U = 180\,\rm meV$ (a), $U=250\,\rm meV$ (b) and $U = 300\,\rm meV$ (c). The intervalley coherence values for these three cases are $\mathcal{C} \approx 0.10$, $0.03$ and $0.02$, respectively. The color code represents the valley polarization $v_i(\vk)$.}
	\label{fig:add_-2_band}
\end{figure}

\subsection{Numerical results at filling factor \texorpdfstring{$\nu=-1$}{nu=-1}}

Next, we provide some additional phase diagrams at $\nu=-1$ filling on $6\times 6$ and $7\times 7$ lattices, which can be found in Figs.~\ref{fig:add_-1_6x6_phase} and \ref{fig:add_-1_7x7_phase}. Similar to other filling factors, the result obtained on smaller lattices, especially on $6\times 6$ is more noisy than the result on $8\times 8$. However, the phase boundary where intervalley coherence disappears is not strongly affected.

\begin{figure}
    \centering
    \includegraphics[width=0.75\linewidth]{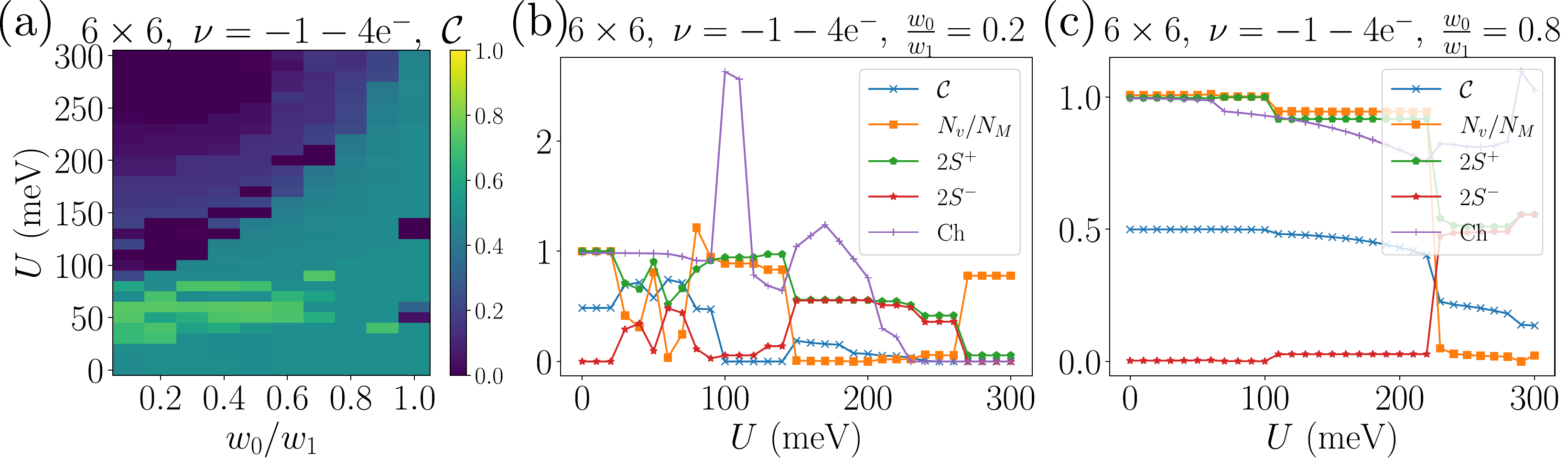}
    \caption{Phase diagrams at filling factor $\nu=-1 - 4\mathrm{e}^-$ on $6\times 6$ lattice.}
    \label{fig:add_-1_6x6_phase}
\end{figure}

\begin{figure}
    \centering
    \includegraphics[width=0.8\linewidth]{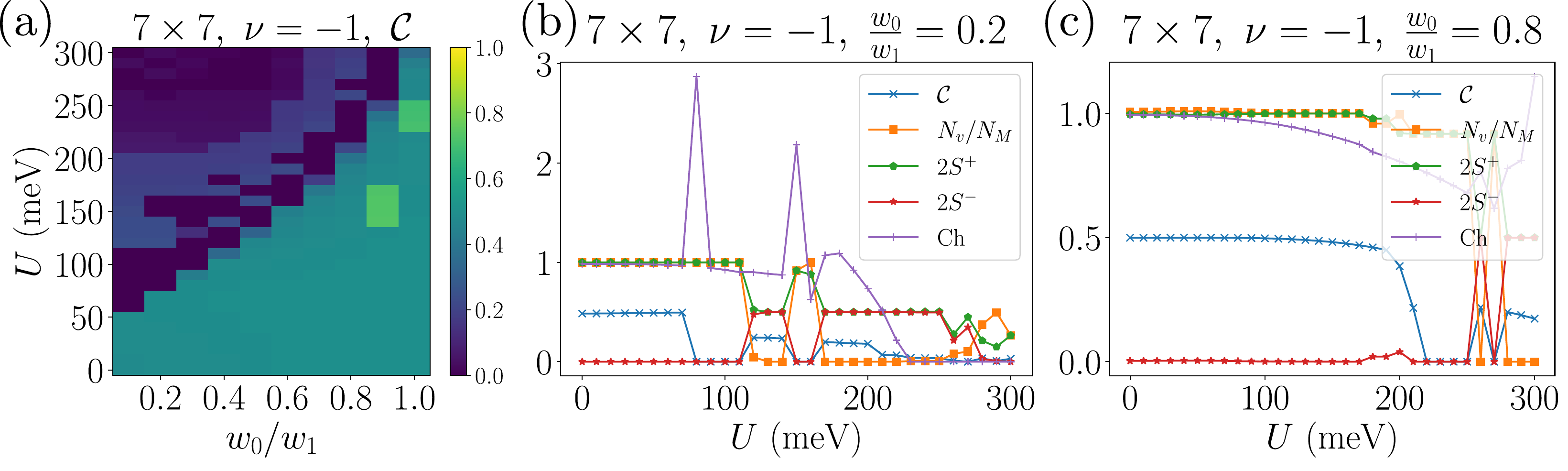}
    \caption{Phase diagrams at filling factor $\nu=-1$ on $7\times 7$ lattice.}
    \label{fig:add_-1_7x7_phase}
\end{figure}

\subsection{Numerical results at filling factor \texorpdfstring{$\nu=0$}{nu=0}}\label{app_subsec:0}

Similar to other fillings, we provide some phase diagrams obtained on $6\times 6$ and $7\times 7$ momentum lattices at filling factor $\nu=0$. As shown in Figs.~\ref{fig:add_0_phase_6x6}(a) and \ref{fig:add_0_phase_7x7}(a), these phase diagrams are similar to the results obtained on $8\times 8$ lattice, which has been discussed in the main text in Fig.~\ref{fig:phase_0}(a). However, as shown in Fig.~\ref{fig:add_0_phase_6x6}(b) and (c), the HF parameter $\mathrm{Ch}$ value on $6\times 6$ lattice at filling $\nu=0-4\mathrm{e}^-$ is no longer zero, and the curves become noisy. The noisy curves can also be observed in the results obtained on $9\times 9$ momentum lattice at $\nu=0-4\mathrm{e}^-$ filling in Fig.~\ref{fig:add_0_phase_9x9} at $w_0/w_1=0.2$ and $U\approx 150\,\rm meV$ and at $w_0/w_1 = 0.8$ and $U \approx 250\,\rm meV$. This observation means that there is a possible competing order with different Chern numbers in region II, as we mentioned in Sec.\ref{sec:nu_0}.

We also provide the Hartree-Fock band structure plots of two extra points in region III in Fig.~\ref{app_fig:add_0_band}. The discontinuous band structures at $K_M$, $K_M'$ and $\Gamma_M$ points can also be seen. Similar to Fig.~\ref{fig:band_0}(c), the HF bands at $K_M$ and $K_M'$ points are gapless, showing that region III is a semimetallic state at $\nu=0$ filling.

\begin{figure}
    \centering
    \includegraphics[width=0.75\linewidth]{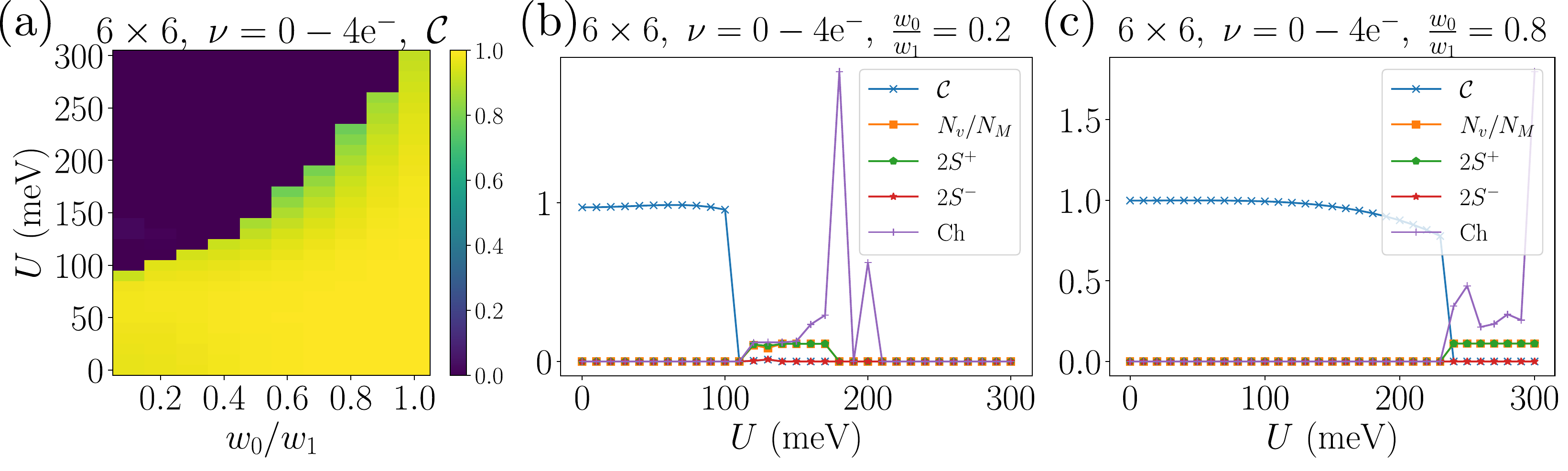}
    \caption{The phase diagrams obtained at filling factor $\nu=0-4\mathrm{e}^-$ on $6\times 6$ momentum lattice.}
    \label{fig:add_0_phase_6x6}
\end{figure}

\begin{figure}
    \centering
    \includegraphics[width=0.75\linewidth]{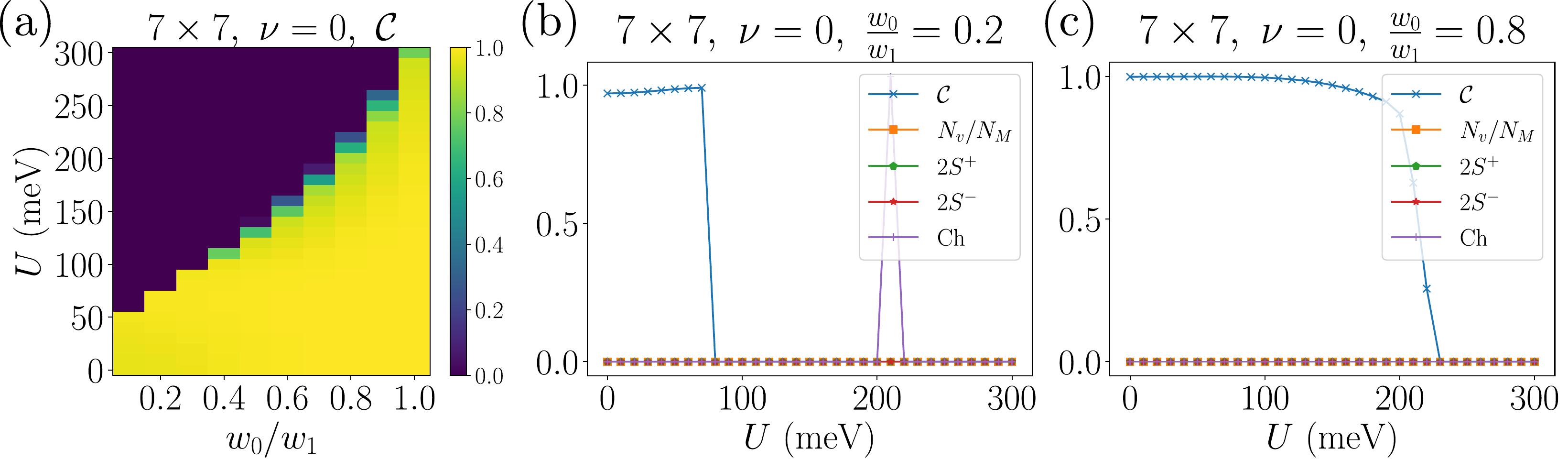}
    \caption{The phase diagrams obtained at filling factor $\nu=0$ on $7\times 7$ momentum lattice.}
    \label{fig:add_0_phase_7x7}
\end{figure}

\begin{figure}
    \centering
    \includegraphics[width=0.6\linewidth]{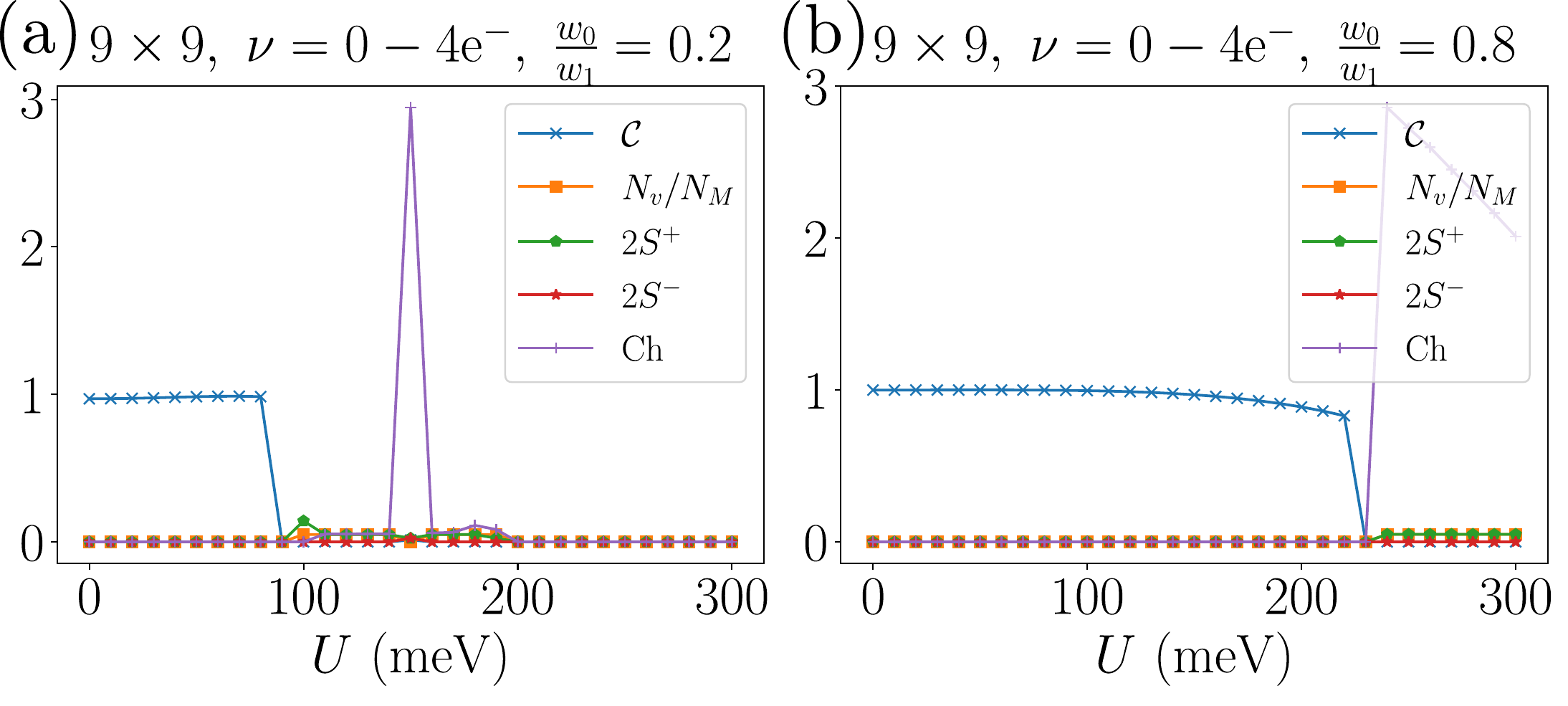}
    \caption{The HF parameters obtained at filling factor $\nu=0-4\mathrm{e}^-$ and $w_0/w_1 = 0.2$ (a) and $w_0/w_1 = 0.8$ (b) on $9\times 9$ lattice.}
    \label{fig:add_0_phase_9x9}
\end{figure}

\begin{figure}
    \centering
    \includegraphics[width=\linewidth]{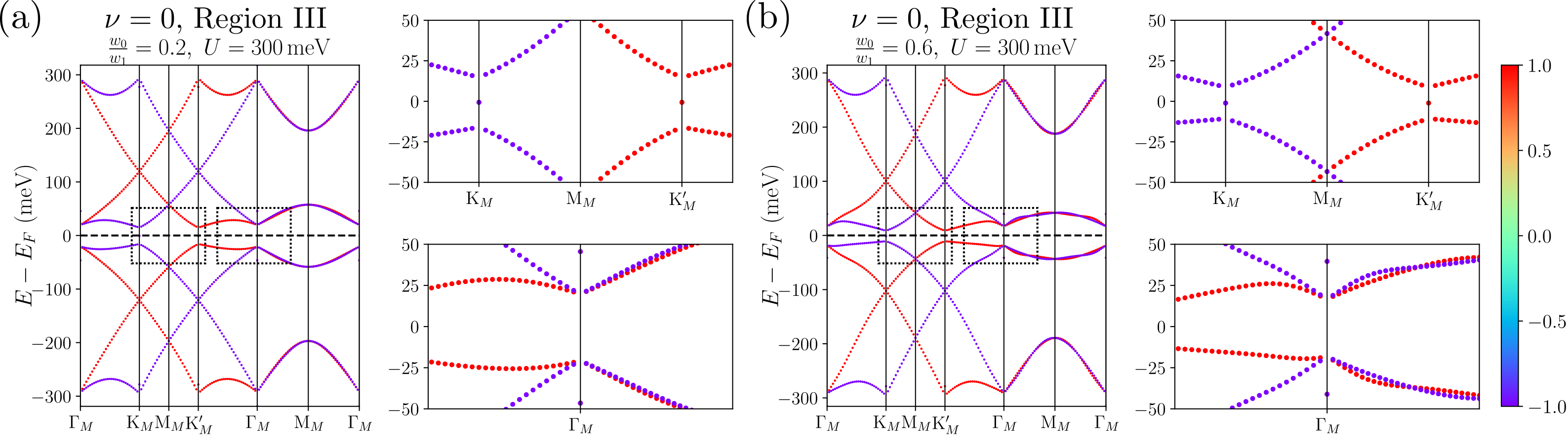}
    \caption{The Hartree-Fock band structures at filling factor $\nu=0$ on $10\times 10$ lattice at $w_0/w_1 = 0.2$ for $U = 300\,\rm meV$ (a) and at $w_0/w_1 = 0.6$ for $U = 300\,\rm meV$ (b).}
    \label{app_fig:add_0_band}
\end{figure}

\section{Understanding of the gaplessness of Dirac semimetal phases at all integer fillings at small \texorpdfstring{$U$}{U}}\label{sec:Ggammasymmetry}

Our HF calculation shows that at each integer filling $\nu=0,-1,-2,-3$, there is a Dirac semimetal phase at small $U$ (region I of the phase diagram at each filling).
Here we show that at even fillings $\nu=-2,0$, the gaplessness of the Dirac point in this Dirac semimetal phase is protected by a combined symmetry of the ground state given by $C_{2z}T$ combined with a valley $\hat{z}$ rotation. At odd fillings $\nu=-1,-3$, we show that the gapless Dirac point is protected by the $C_{2z}T$ symmetry within a $C_{2z}T$ invariant spin-valley flavor.

\subsection{\texorpdfstring{$U=0$}{U=0} at all integer fillings}\label{sec:Ggammasymmetry-1}

We first consider the $U=0$ case within region I of the phase diagram at each integer filling. At $U=0$, the Dirac fermion sector and the TBG sector do not hop with each other, so the filling of electrons in each sector is conserved, which we denote as $\nu_D$ and $\nu_{TBG}$, and the total filling is $\nu=\nu_D+\nu_{TBG}$. Therefore, the ground state at a given filling $\nu$ is generically a tensor product of states within the two sectors:
\begin{equation}\label{seq:prodU=0-general}
|\Psi_\nu\rangle=|\Psi_{TBG,\nu_{TBG}}\rangle\otimes|\Psi_{D,\nu_D}\rangle\ ,
\end{equation}
for certain $\nu_D=\nu-\nu_{TBG}$. 

We shall only discuss the case that the TSTG ground state has $\nu_D=0$ and $\nu_{TBG}=\nu$ at a given integer filling $\nu=0,-1,-2,-3$, and $|\Psi_{TBG,\nu}\rangle$ is given by the insulating ground state of TBG at filling $\nu$ (with no translation symmetry breaking) studied in Refs. \cite{BUL20,LIA20}. Under this assumption, the ground state at filling $\nu$ takes the form
\begin{equation}\label{seq:prodU=0}
|\Psi_\nu\rangle=|\Psi_{TBG,\nu}\rangle\otimes|\Psi_{D,0}\rangle\ .
\end{equation}
Our HF calculations showed that this is true (i.e., $\nu_D=0$, $\nu_{TBG}=\nu$, and the TBG sector is gapped) at $U=0$ for small enough $w_0/w_1$ in region I of all integer fillings. In particular, we have numerically checked that the HF Hamiltonian $\mathcal{H}^{HF}(\vk)$ at $U = 0$ has zero matrix elements hybridizing the Dirac fermions and TBG fermions, which confirms the direct product state nature of the ground state. Note that $\nu_D=0$ assumed here is required for the system to be strictly a Dirac semimetal with a point-like Fermi surface. At large $w_0/w_1$, small electron and hole pockets may arise due to overlapping of the HF dispersions of the conduction and valence flat bands in the TBG sector, in which case the system is a metal (thus gapless) and will not be considered here.

We now show that the Dirac sector state $|\Psi_{D,0}\rangle$ is a gapless Dirac semimetal. We first recall that the interaction between the TBG sector and the Dirac sector is (see Eq. (\ref{eq:proj-HI}))
\begin{equation}\label{seq:proj-HI-inter}
	H_I^{\hat{b}\hat{c}} = \frac{1}{N_M\Omega_{c}}\sum_{\vq,\mathbf{G}\in\mathcal{Q}_0}V(\vq + \mathbf{G})\overline{\delta\rho}^{\hat{b}}_{\vq + \mathbf{G}}\overline{\delta\rho}^{\hat{c}}_{-\vq - \mathbf{G}}\,,
\end{equation}
where $\hat{b}$ and $\hat{c}$ stand for fermions in the Dirac and TBG sectors, respectively. As shown in Ref.~\cite{LIA20}, in the flat-band limit at even fillings, or in the chiral-flat limit at any integer fillings $\nu$, the insulating TBG ground state satisfies $\overline{\delta\rho}^{\hat{c}}_{\vq + \mathbf{G}}|\Psi_{TBG,\nu}\rangle=N_MA_{\mathbf{G}}\delta_{\vq,\mathbf{0}}|\Psi_{TBG,\nu}\rangle$ for some constants $A_{\mathbf{G}}$. Away from the flat-band limit or chiral-flat limit, we do not have the above exact relation, but in the HF approximation (which is the subject of study of this paper), $\overline{\delta\rho}^{\hat{c}}_{\vq + \mathbf{G}}$ provides a Hartree mean field $\langle \overline{\delta\rho}^{\hat{c}}_{\vq + \mathbf{G}}\rangle=N_MA_{\mathbf{G}}\delta_{\vq,\mathbf{0}}$ with some constants $A_{\mathbf{G}}$ for the Dirac fermions, provided that the translation symmetry is unbroken. Note that in the original sublattice basis is $\overline{\delta\rho}^{\hat{b}}_{\vq}=\sum_\mathbf{k}\hb^\dag_{\vk+\vq,\eta,s}\hb_{\vk,\eta,s}$, where $\hb_{\vk,\eta,s}=(\hb_{\vk,\eta,s,A},\hb_{\vk,\eta,s,B})^T$ is the fermion basis in sublattices $A$ and $B$ ($\eta,s$ stand for valley and spin), and $\vk$ is restricted within the first moir\'e BZ (since we project into the lowest 2 Dirac bands). Thus, the inter-sector interaction in Eq. (\ref{seq:proj-HI-inter}) solely yields a chemical potential $\mu_H=\frac{V(\mathbf{0})A_\mathbf{0}}{\Omega_c}$ to the Dirac fermion, which will not affect the Dirac sector ground state $|\Psi_{D,0}\rangle$ at fixed filling $0$. 

Therefore, the Dirac fermion ground state $|\Psi_{D,0}\rangle$ is solely determined by interactions within the Dirac fermion sector. Since the Dirac fermion sector alone is no different from the model of monolayer graphene, we expect the Dirac ground state $|\Psi_{D,0}\rangle$ to be a gapless Dirac semimetal, in analogy to that of the monolayer graphene. Here we give a heuristic understanding for the gaplessness of such a Dirac semimetal. Consider the Hartree-Fock approximation for Dirac fermions with a single-particle Hamiltonian $H_D=\sum_\mathbf{k}\hb^\dag_{\vk,\eta,s} v_F\bm{\sigma}\cdot\mathbf{k} \hb_{\vk,\eta,s}$ and interaction $V(\vk)$. We now examine the possibility of an HF mean field order parameter of the form $\sum_\vk m_{HF}(\vk)\hb_{\vk,\eta,s}^\dag\sigma_z\hb_{\vk,\eta,s}$ around $\vk=\mathbf{0}$. At the charge neutrality, the self-consistent HF equation then yields (for fixed spin $s$ and valley $\eta$)
\begin{equation}\label{seq:mHFself}
m_{HF}(\vk)=-\int \frac{d^2\vk'}{(2\pi)^2}V(\vk'-\vk)\langle b_{\vk',\eta,s}^\dag \frac{\sigma_z}{2}b_{\vk',\eta,s}\rangle =\int \frac{d^2\vk'}{(2\pi)^2}V(\vk'-\vk)\frac{m_{HF}(\vk')}{2\sqrt{m_{HF}^2+v_F^2\vk'^2}}\ .
\end{equation}
Assume the maximal value of $|m_{HF}(\vk)|$ among all $\vk$ is $\overline{m}_{HF}$, and we take the interaction $V(\vq) = \pi\xi^2U_\xi\frac{\tanh(\xi q/2)}{\xi q/2}\ge0$ in Eq. (\ref{eq:Vq}), we have
\begin{equation}\label{eq:mHFupper}
|m_{HF}(\vk)|\le \int \frac{d^2\vk'}{(2\pi)^2}V(\vk'-\vk)\frac{\overline{m}_{HF}}{2v_F|\vk'|}\lesssim \overline{m}_{HF} \frac{ U_\xi}{2v_F \sqrt{\xi^{-2}+\vk^2}}\ln(1+\frac{\Lambda_k}{2\sqrt{\xi^{-2}+\vk^2}})  \ ,
\end{equation}
where $\Lambda_k$ is the UV momentum cutoff. Note that the above bound is quite lose, with all $m_{HF}(\vk')$ in the integral relaxed to $\overline{m}_{HF}$, while in fact $m_{HF}(\vk')$ decays at least as $\ln(|\vk'|)/|\vk'|$ at large $|\vk'|$ according to Eq. (\ref{eq:mHFupper}) above. 
Nevertheless, this yields a loose bounding condition
\begin{equation}
0\le \overline{m}_{HF} \le r_c \overline{m}_{HF}\ ,
\end{equation}
where $r_c= \frac{ \xi U_\xi}{2v_F}\ln(1+\frac{\xi\Lambda_k}{2})$. In this paper we have $\xi=10$nm, $U_\xi=24$meV, and $v_F\approx610$meV$\cdot$nm. 
For monolayer graphene, $\Lambda_k\approx 4\pi/3a_0\approx17$nm$^{-1}$ ($a_0=0.246$nm is the graphene lattice constant), which yields $r_c\approx 0.85$. In our practical calculation, we only keep the lowest two Dirac bands, which corresponds to $\Lambda_k\approx 0.5$nm$^{-1}$, and yields $r_c\approx0.25$. In either case, $r_c<1$, and we find the HF mass is bounded to $\overline{m}_{HF}=0$. So the Dirac ground state $|\Psi_{D,0}\rangle$ is gapless.

Therefore, we find the TSTG ground state is a gapless Dirac semimetal at $\nu=0$ and $U=0$.

\subsection{\texorpdfstring{$U>0$}{U>0} at even integer fillings}

In this case, the interlayer potential $U$ yields a hopping term $H_U$ between the Dirac fermion sector and the TBG sector as given in Eq. (\ref{eq-HU}). 

The TBG ground state at even fillings $\nu=0$ or $-2$ is a gapped intervalley coherent state given by the wavefunction \cite{BUL20,LIA20}
\begin{equation}
|\Psi_{TBG,\nu}\rangle=\prod_{\mathbf{k}} \prod_{s\in \mathcal{S}_\nu} \prod_{e_Y=\pm} \frac{e^{-i\gamma/2}\hd^\dag_{\mathbf{k},e_{Y},+,s}+ e^{i\gamma/2} e_{Y}\hd^\dag_{\mathbf{k},e_{Y},-,s}}{\sqrt{2}}|0\rangle\ ,
\end{equation}
where $\hd^\dag_{\mathbf{k},e_{Y},\eta,s}$ is the TBG Chern band basis defined in Eq.~(\ref{eq:chernbasis}), $\gamma$ is the intervalley coherent spontaneous symmetry breaking phase, and $\mathcal{S}_0=\{\uparrow,\downarrow\}$, $\mathcal{S}_{-2}=\{\uparrow\}$ are the set for the spin index summation. Noting that $C_{2z}T\hd^\dag_{\mathbf{k},e_{Y},\eta,s}(C_{2z}T)^{-1}=\hd^\dag_{\mathbf{k},-e_{Y},\eta,s}$ (under our $C_{2z}T$ gauge fixing), we see the TBG ground state has a remaining antiunitary symmetry $C_{2z}T$ combined with valley $z$ rotation:
\begin{equation}\label{seq:Ggamma}
\mathcal{G}_\gamma=C_{2z}T e^{i(\gamma+\frac{\pi}{2})S^{z0}}\ , \qquad \mathcal{G}_\gamma^2=1\ ,\qquad \mathcal{G}_\gamma \left(\hd^\dag_{\mathbf{k},e_{Y},\eta,s}, \hb^\dag_{\mathbf{k},\eta,s,\alpha}\right)\mathcal{G}_\gamma^{-1}=e^{-i\eta(\gamma+\frac{\pi}{2})}\left(\hd^\dag_{\mathbf{k},-e_{Y},\eta,s}, \hb^\dag_{\mathbf{k},\eta,s,-\alpha}\right)\ .
\end{equation}
where $S^{z0}=\sum_{\vk,\eta,s}\eta \left(\sum_{e_Y}\hd^\dag_{\mathbf{k},e_{Y},\eta,s}\hd_{\mathbf{k},e_{Y},\eta,s}+\sum_{\alpha}\hb^\dag_{\mathbf{k},\eta,s,\alpha}\hb_{\mathbf{k},\eta,s,\alpha}\right)$ is the valley $z$ rotation generator. Moreover, note that the absence of a mass term $\overline{m}_{HF}$ in the Dirac ground state $|\Psi_{D,0}\rangle$ in \ref{sec:Ggammasymmetry-1} indicates that the Dirac ground state $|\Psi_D\rangle$ also obeys the antiunitary symmetry $\mathcal{G}_\gamma$ in Eq. (\ref{seq:Ggamma}). So the even filling TSTG ground state at $U=0$ (Eq. (\ref{seq:prodU=0})), which is a Dirac semimetal tensor producted with an intervalley coherent TBG ground state, respects the $\mathcal{G}_\gamma$ symmetry.

When $U>0$, since $H_U$ in Eq. (\ref{eq-HU}) only contains intravalley hoppings and respect the $C_{2z}T$ symmetry, we know that $H_U$ also respects the antiunitary symmetry $\mathcal{G}_\gamma$ in Eq. (\ref{seq:Ggamma}). Therefore, provided $U$ is small enough, we expect the $\mathcal{G}_\gamma$ symmetry to remain respected by the TSTG ground state. The $\mathcal{G}_\gamma$ symmetry then protects the gaplessness of Dirac points, similar to the protection by $C_{2z}T$ symmetry (see for example Ref. \cite{SON20b}). This is in agreement with our findings in for region I of the $\nu=0,-2$ TSTG phase diagram, where the ground state is an intervalley coherent Dirac semimetal.

\subsection{\texorpdfstring{$U>0$}{U>0} at odd integer fillings}

At odd integer fillings $\nu=-1,-3$, the TBG ground state is shown in Ref.~\cite{LIA20} to be a Chern insulator of Chern number $\pm1$, which breaks the $C_{2z}T$ symmetry. However, the TBG ground state in this case always has at least one valley-spin flavor fully empty (which is exact, since the number of electrons in each spin-valley flavor ($\eta,s$) in TBG or TSTG is conserved). More concretely, the $\nu=-3$ TBG ground state is valley polarized and has one band in flavor ($+,\uparrow$) occupied and all the other bands empty; the $\nu=-1$ TBG ground state occupies 2 intervalley coherent bands in the spin $\downarrow$ sector, and one band in flavor ($+,\uparrow$). In both cases, the valley-spin flavor ($-,\uparrow$) is fully empty. Therefore, the $C_{2z}T$ symmetry is preserved within the valley-spin flavor ($-,\uparrow$) of TBG. 

On the other hand, at $U=0$, the Dirac ground state $|\Psi_{D,0}\rangle$ of the TSTG also preserves the $C_{2z}T$ symmetry because of the absence of a Dirac mass. Therefore, we conclude that the TSTG ground state in region I of the odd fillings $\nu=-1,-3$ at $U=0$ preserves the $C_{2z}T$ symmetry in the sector of valley-spin flavor ($-,\uparrow$).

When $U>0$, $H_U$ in Eq. (\ref{eq-HU}) respects the $C_{2z}T$ symmetry and only contains hoppings within each spin-valley flavor. Therefore, for small enough $U>0$, we expect the $C_{2z}T$ symmetry within the valley-spin flavor ($-,\uparrow$) to remain respected by the odd-filling TSTG ground state. The $C_{2z}T$ symmetry therefore at least protects the gapless Dirac nodes in the empty valley-spin flavor ($-,\uparrow$), ensuring the TSTG ground state in region I of odd fillings at $U>0$ to be a gapless Dirac semimetal.

\end{widetext}

\end{document}